\documentclass[11pt,notitlepage]{report}
\usepackage{amstex}
\usepackage{amssymb}
\makeatletter
\renewcommand{\@evenhead}{%
  \llap{\bfseries\thepage\quad}\slshape\leftmark\hfil}
\renewcommand{\@oddhead}{%
  \hfil{\slshape\rightmark\/}\rlap{\quad\bfseries\thepage}}%
\if@twoside
  \renewcommand{\ps@headings}{%
    \let\@oddfoot\@empty\let\@evenfoot\@empty
    \renewcommand{\@evenhead}{%
      \llap{\bfseries\thepage\quad}\slshape\leftmark\hfil}
    \renewcommand{\@oddhead}{%
      \hfil{\slshape\rightmark\/}\rlap{\quad\bfseries\thepage}}%
    \let\@mkboth\markboth
    \renewcommand{\chaptermark}[1]{%
      \markboth{\@chapapp\ \thechapter.\quad ##1}{}}%
    \renewcommand{\sectionmark}[1]{\markright{\thesection.\ ##1}}}
\else
  \renewcommand{\ps@headings}{%
    \let\@oddfoot\@empty
    \renewcommand{\@oddhead}{%
      \hfil{\slshape\rightmark\/}\rlap{\quad\bfseries\thepage}}%
    \let\@mkboth\markboth
    \renewcommand{\chaptermark}[1]{%
      \markright{\@chapapp\ \thechapter.\quad ##1}}}
\fi
\makeatother

\numberwithin{equation}{chapter}

\newcommand{\scri}{\mathcal{I}}
\newcommand{\acro}[1]{{\scshape\lowercase{#1}}}
\newcommand{\sss}[1]{{\scriptscriptstyle #1}}
\newcommand{\mb}[1]{\boldsymbol{#1}}
\newcommand{\half}{\tfrac{1}{2}}
\newcommand{\fourth}{\tfrac{1}{4}}
\newcommand{\real}{{\text{I\kern-.2em R}}}
\newcommand{\Lie}{\operatorname{\text{\pounds}}}
\newcommand{\nablas}{%
  \smash[b]{\mathord{\text{\raisebox{\depth}{$\bigtriangledown$}}}}}
\newcommand{\nablat}{\mathord{\bigtriangleup}}

\newcommand{\Tr}{\operatorname{Tr}}
\newcommand{\e}{{\mathrm{e}}}

\newcommand{\parenfrac}{\fracwithdelims{(}{)}}
\newcommand{\ads}{\textsc{a}{\scriptsize d}\textsc{s}}
\newcommand{\rnads}{\textsc{rn}\ads}

\providecommand{\eqref}[1]{(\ref{#1})}
\begin{document}
\pagestyle{plain}
\pagenumbering{roman}
\title{Gravitational Calorimetry}
\author{Jolien D. E. Creighton%
  \thanks{A thesis presented to the University of Waterloo
    in fulfillment of the thesis requirement
    for the degree of Doctor of Philosophy in Physics}}
\date{Waterloo, Ontario, Canada, 1996}
\maketitle
\thispagestyle{empty}
\begin{abstract}\noindent
I generalize the quasilocal formulation of thermodynamics of Brown
and York to include dilaton theories of gravity as well as Abelian
and Yang-Mills gauge
matter fields with possible dilaton couplings.  The resulting
formulation is applicable to a large
class of theories including two-dimensional toy models and the low
energy limit of string theory as well as to many types of matter
such as massless scalar fields, electromagnetism, Yang-Mills fields,
and matter
induced cosmological constants.  I use this formalism to evaluate
the thermodynamic variables for several black hole spacetimes.
I find that the formulation can handle black hole spacetimes that
are not asymptotically flat as well as rotating black hole spacetimes,
and black hole spacetimes possessing a dilaton field, an electric
charge, and a magnetic charge.
\vspace{\fill}
\begin{center}
  \copyright Jolien D. E. Creighton 1996
\end{center}
\vspace{\fill}
\end{abstract}

\chapter*{Acknowledgements}
It is my pleasure to thank my supervisor, Robert Mann, for his
guidance.  I have enjoyed the support of the Natural Sciences
and Engineering Research Council (NSERC) of Canada and the
hospitality of the Department of Applied Mathematics and Theoretical
Physics of the University of Cambridge.  Valuable criticisms of an
early draft of this work were provided by the members of my
examining committee: Valeri Frolov, George Leibbrandt, Raymond
McLenaghan, and Eric Poisson.  Finally, I would like
to thank Jean Giannakopoulou for her support and careful editing of
the manuscript.

\tableofcontents
\listoftables
\listoffigures

\clearpage
\pagestyle{headings}
\pagenumbering{arabic}

\chapter{Introduction}
\label{c:intro}

The connection between heat and mechanical energy was one of the
most interesting discoveries of thermodynamics.  This realization
provided the first clues about how phenomena at the macroscopic
level must arise from the statistical properties of the mechanics
of microscopic objects.  The failure of the classical statistical
mechanics\footnote{%
For example, the equipartition theorem suggested that the specific
heat capacity of a solid was given by the rule of Dulong and Petit
(1819)~$c_\sss{\text{V}}=3R$ where $R$ is the molar gas constant.
However, the experiments of Weber (1875) disagreed with this result
at low temperatures.  The resolution was only found in quantum
statistical mechanics by Einstein in 1907 and Debye in 1912.}
was the first indication of quantum mechanical effects.
The study of systems at the macroscopic scale yields insight into
the more fundamental theories of nature and has allowed physicists
to take the first steps to understand these theories.

Although quantum mechanics seems to be quite
sufficient for most applications, the theory suffers from serious
problems in its foundation, especially when one attempts to develop
a quantum theory of gravitation.  As an aid in understanding
quantum gravity, one needs a system in which the quantum and the
classical behaviour exist in juxtaposition.  The black hole is
such a system; one hopes to gain insight into the nature of quantum
gravity by studying the thermodynamics of black holes.

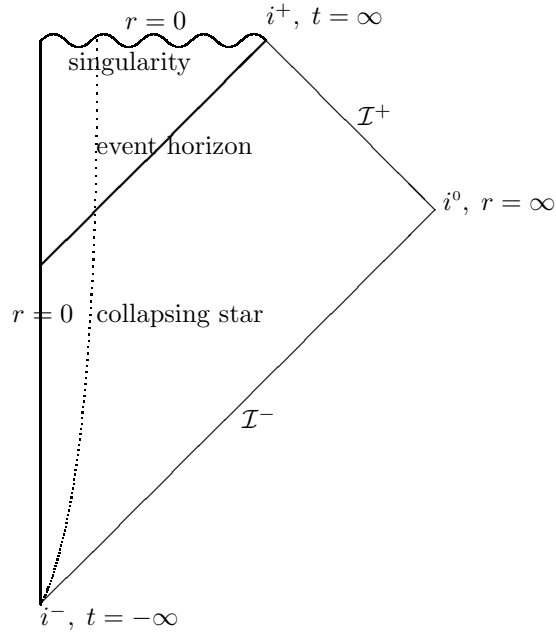
\begin{figure}[t]
\renewcommand{\baselinestretch}{1}\small
\begin{center}
\setlength{\unitlength}{0.75mm}
\begin{picture}(120,120)(-5,-5)
\put(0,0){\line(0,1){100}}
\put(0,0){\line(1,1){70}}
\put(40,100){\line(1,-1){30}}
\multiput(2.22,100)(8.88,0){5}{\qbezier(-2.22,0)(0,2.22)(2.22,0)}
\multiput(6.66,100)(8.88,0){4}{\qbezier(-2.22,0)(0,-2.22)(2.22,0)}
\qbezier[100](0,0)(10,20)(10,100)
\thicklines
\put(0,60){\line(1,1){40}}
\put(35,35){\raisebox{-\height}{$\,\scri^{-}$}}
\put(55,85){$\;\scri^{+}$}
\put(70,70){$\;i^\sss0,\;r=\infty$}
\put(5,98){\raisebox{-\height}{singularity}}
\put(15,102){\raisebox{\depth}{$r=0$}}
\put(40,102){\raisebox{\depth}{$i^{+},\;t=\infty$}}
\put(0,0){\raisebox{-\height}{$i^{-},\;t=-\infty$}}
\put(10,80){event horizon}
\put(10,50){collapsing star}
\put(-5,50){$r=0$}
\end{picture}
\end{center}
\begin{quote}\leavevmode
\caption[Black hole formed from spherical collapse]{\small
  Conformal diagram of a black hole formed from the spherical
  collapse of a star.  The points at infinity include timelike
  and spacelike infinities ($i^+$, $i^-$, and~$i^\sss0$), and
  null infinity ($\scri^+$ and~$\scri^-$).  Each point
  represents a sphere.  Radial null rays travel along lines inclined
  at~$45^\circ$.}
\label{fig:bh}
\end{quote}
\end{figure}

The black hole is an object that straddles the domains of classical
mechanics and quantum mechanics: of the macroscopic and the
microscopic.  Although the black hole has many interesting properties,
it is an extremely simple system: at the classical level it is vacuum.
The defining feature of a black hole is its
event horizon.\footnote{The standard definition of a black hole also
requires that the spacetime be asymptotically flat, but I shall consider
a more general class of objects that can arise in spacetimes that do
not possess asymptotic flatness.}
This surface separates events that are inside
the horizon from the ones that are outside the horizon;
the events inside the event horizon are never within
the causal past of the ones outside.  This feature can be seen in
figure~\ref{fig:bh}, which depicts the formation of a black hole
by the spherical collapse of a star.  The conformal transformation
that brings points at infinity to a finite distance preserves the
light-cone structure so that light rays travelling radially inwards
or outwards travel on lines inclined by~$45^\circ$; these rays are
called null.  Notice that the event horizon in figure~\ref{fig:bh}
is a null surface, so
the events within it are never in the past light-cone of events
outside of the horizon.

The presence of an event
horizon causes some unusual effects on quantum fields existing
in the black hole spacetime.  By acting as a one-way membrane,
the event horizon can trap one of the ``virtual'' particle-pairs
produced by quantum processes.%
\footnote{In quantum field theory, vacua are not empty: instead they
contain ground state quantum fields.  These fields can be thought of
as spontaneous creation and annihilation of virtual particles.  The
vacua of different spacetimes are not equivalent and this is why
the vacuum around a black hole has particle content when compared
to the vacuum in flat spacetime.}
The escaping particle (which is no longer virtual) appears to have
been radiated from the black
hole.  The radiation, known as Hawking radiation, has a thermal
spectrum (if one neglects scattering off of the gravitational field)
with a temperature proportional to the surface gravity
of the event horizon.
Although this picture is drawn from semi-classical quantum
field theory, it should be qualitatively correct since the
gravitational field at the event horizon of a black hole need
not be very strong.

Another strange property of black holes was realized by Wheeler:
one can forever hide information from the outside world by dropping
it into a black hole.  Indeed this property seems to pose a problem
with the second law of thermodynamics since a black hole will quickly
return to a very simple state even if an object with a large
amount of entropy is dropped into it.  Bekenstein~\cite{b:72,b:80}
resolved this
problem by speculating that the black hole itself is a thermodynamic
system with the area of the event horizon representing the entropy.
Since the area always increases when matter of positive energy
is added to a black hole, the problem with the second law of
thermodynamics can be resolved.  That the black hole is also
surrounded by quantum fields with a thermal spectrum seems to
support Bekenstein's speculation.  Various attempts to understand
the entropy of black holes in terms of the number of quantum states
contained within or near the event horizon have been made (two
recent reviews are given in references~\cite{b:94,f:95}).

Therefore, black holes are interesting systems to study, if only
theoretically, as their classical and semi-classical properties
may hint at the nature of a quantum theory of gravity.  Black
holes can be treated as a thermodynamic systems whose properties
must be reproduced in the statistics of the quantum fields.
However, the thermodynamic properties of black holes must first
be understood.

\section{The Laws of Black Hole Mechanics}
\label{s:intro laws}

Black holes have a variety of interesting properties.  For all black
holes that are stationary and axially symmetric with time-azimuth
reflection symmetry, the surface gravity of the event horizon is
constant on the event horizon~\cite{rw:96}.  The surface gravity of
the event horizon, $\varkappa_\sss{\text{H}}$, can be thought of as
the force required to hold a unit test mass on the event horizon in
place by an observer who is far from the black hole.\footnote{%
This is a crude definition that only holds for black holes
that are static, spherically symmetric, and asymptotically flat.}
Furthermore, Bardeen, Carter, and Hawking~\cite{bch:73} showed that,
for Ricci-flat, asymptotically flat black holes, the change in the
mass of the hole is given by
\begin{equation}
  \delta M = \frac{\varkappa_\sss{\text{H}}}{8\pi}\,\delta 
  A_\sss{\text{H}} + \varOmega_\sss{\text{H}}\,\delta J
  \label{first law: bch}
\end{equation}
for any process.  Here $M$ and~$J$ are the Arnowitt-Deser-Misner
(\acro{ADM}) definitions of mass and angular momentum of
the (asymptotically flat) black hole spacetime~\cite{adm:62},
$A_\sss{\text{H}}$~is
the area of the event horizon, and~$\varOmega_\sss{\text{H}}$
is the angular velocity of the event horizon.

An analogy can be made between the laws of black hole mechanics and
the laws of thermodynamics.  The constancy of the surface gravity is
similar to the zeroth law of thermodynamics, which states that the
temperature of a thermodynamic system in thermal equilibrium is
constant across the system;
furthermore equation~\eqref{first law: bch} is similar to the first
law of thermodynamics if the \acro{ADM} mass of the spacetime is
interpreted as the internal energy of the system and
the~$\varOmega_\sss{\text{H}}\,\delta J$ term is interpreted as some
work term.  According to equation~\eqref{first law: bch}, then, one
would expect that
the~$(8\pi)^{-1}\varkappa_\sss{\text{H}}\,\delta A_\sss{\text{H}}$
term is an expression for the heat exchange, and, if one were to
interpret the surface gravity as being proportional to the temperature
(in analogy with the zeroth law), then one would expect that the area
could be interpreted as being proportional to an entropy, and that a
second law of black hole
mechanics---that the area will not decrease---should exist.  In fact,
Hawking has proven that the area of the event horizon will never
decrease if the black hole spacetime satisfies certain positive
energy conditions~\cite{h:71}.  In addition to the zeroth, first, and
second laws, there appears to be an analog of the third law of
thermodynamics in the black hole mechanics:  it seems that it is
impossible to bring a black hole to a state of zero surface
gravity~\cite{w:74} in a finite number of processes.  These laws are
summarized in table~\ref{tab:laws}.
\begin{table}[b]
\renewcommand{\baselinestretch}{1}\small
\begin{quote}\leavevmode
\caption[Laws of black hole mechanics and thermodynamics]{\small
  A comparison of the four laws of black hole mechanics and the
  four laws of thermodynamics}
\label{tab:laws}
\end{quote}
\begin{center}
\begin{tabular}{lcc}
\hline
Law & Black Hole Mechanics & Thermodynamics \\
\hline\hline
Zeroth & $\varkappa_\sss{\text{H}}$ constant over event horizon &
  $T$ constant throughout system \\
First & $\delta M = (8\pi)^{-1}\varkappa_\sss{\text{H}}\,
  \delta A_\sss{\text{H}} + \varOmega_\sss{\text{H}}\,\delta J $ &
  $dE = T\,dS + \text{work}$ \\
Second & $\delta A_\sss{\text{H}}\ge0$ & $\delta S\ge0$ \\
Third & cannot reach $\varkappa_\sss{\text{H}}=0$ & cannot reach $T=0$ \\
\hline
\end{tabular}
\end{center}
\end{table}

The analogy with thermodynamics suffered from the following weakness:
because matter could only enter and never leave a black hole, a
black hole could never come into thermal equilibrium with a hot body
since the body would constantly radiate into the black hole.  However,
as mentioned earlier, if quantum fields are present around the event
horizon of a black hole, then the black hole appears to radiate as
a black body with a temperature proportional to the surface
gravity~\cite{h:75,w:95}.  Thus, a hot body can be brought into
thermal equilibrium with a black hole with quantum fields.  The 
analogy between the surface gravity and the temperature of the black
hole is thus given a physical justification.  Also, because the
quantum fields in the black hole spacetime do not satisfy the
positive energy condition, it is possible for the area of the black
hole to decrease, and thus heat may be extracted from the black
hole spacetime as expected.

At a thermodynamic level, a generalized second law of thermodynamics
can be formulated: the total entropy of the universe,
plus the total area of event horizons of black holes, will never
decrease.  This completely justifies the interpretation of the
area of a black hole as its entropy.  Nevertheless,
at the level of statistical mechanics, one would like to be able to
interpret the area of a black hole as some measure of the degeneracy
of states in a black hole.  An interesting dilemma arises.  If a
black hole were to radiate all its internal energy away and thus
``evaporate,'' what would be left of any information that has fallen
into the black hole?  One might expect to be able to recover any
information that had been lost to the black hole from correlations
in the not-quite thermal radiation.  However, such correlations do
not seem to be present.  If information is indeed lost, there would
be a loss of determinism from the universe because one would not be
able to propagate any final state backwards in time to before the
evaporation of the black hole.  This puzzle, known as the information
loss problem, is of considerable interest: any consistent quantum
theory of gravity must confront the information loss problem, and the
resolution to the problem is an indication of the qualitative
features of the theory.

It has been difficult to reconcile quantum field theory with General
Relativity.  Nevertheless, there are many candidate theories for a
quantum theory of gravity.  At present, it is not possible to
determine which approach, if any, is the correct one.  Part of the
problem seems to be due to the difficulty in performing realistic
calculations in four-dimensional spacetime.  Thus, many researchers
have been forced to deal with more tractable, if less realistic,
problems.  One may have to make modifications to the
theory of gravity or deal with ``unphysical'' or lower-dimensional
spacetimes.  Because puzzles such as the information loss problem
will yield \emph{qualitative} information about the nature of a
theory, it is hoped that insight into quantum gravity will be gained
even from somewhat unphysical problems.  For example, if one
were to determine the statistical origin of entropy of a black hole
in a three-dimensional black hole, then this would suggest that there
is a similar origin for four-dimensional black holes.

\section{Quasilocal Thermodynamics}
\label{s:intro thermo}

By considering black holes that arise in a variety
of ``toy theories,'' one hopes to be able to quantize a black hole
spacetime and thus gain insight into more realistic quantum theories
of gravity.  But first, one needs to know what
classical behaviour the toy theories exhibit.  In particular, one
needs to know if the black holes arising in the toy theory have
thermodynamic properties that are suitably similar to the ones
shown by black holes in General Relativity.  In order to understand
the thermodynamics of toy theories, the laws of black hole mechanics
must be generalized to these theories.

Although the constancy of the surface gravity over the event horizon
of the black hole and the increase in the area of the black hole
event horizon are quite general results, the remaining two laws of
black hole mechanics are not.  In particular,
equation~\eqref{first law: bch} holds only for four-dimensional
asymptotically flat spacetimes, and it only holds at spacelike
infinity.  Therefore, equation~\eqref{first law: bch} will not
necessarily hold for black holes in a toy theory of interest;
this equation represents an idealization even in the
context of General Relativity.  One is interested in a black
hole as some thermodynamic system in contact with the rest of
the universe, not as the whole universe itself.  One
would like to have the thermodynamic laws apply to a \emph{finite}
system that contains a black hole.  For these
reasons, one is motivated to adopt a new thermodynamic formalism
that will be adaptable to many different theories---without having
to do the laborious task of recomputing the laws of black hole
mechanics in each theory---and that applies on a finite compact
surface that surrounds the black hole.  As the surface surrounding
the black hole expands to infinity, one would expect to recover
equation~\eqref{first law: bch} for the asymptotically flat vacuum
solutions of General Relativity.

Therefore, one is interested in adopting \emph{quasilocal} definitions
for the thermodynamic variables.  By quasilocal, I mean that the
quantity is constructed from information that exists on the
\emph{boundary}
of a gravitating system alone.  Just as with Gauss' law, such
quasilocal quantities will yield information about the spacetime
contained within the system boundary.  Two advantages of using such
a quasilocal method are the following:  first, one is able to
effectively separate the gravitating system from the rest of the
universe;  second, the formalism does not depend on the particular
asymptotic behaviour of the system, so one can accommodate a wide class
of spacetimes with the same formalism.

The separation of the system from the rest of the universe has
important physical significance.  In order for a black hole to be
in equilibrium, the system must also contain a radiation field
of the same temperature as the black hole.  In general,
the self-gravitation of the radiation will cause the equilibrium
to be unstable unless the system is small enough~\cite{y:86}.
Thus, a finite system size is important in ensuring that the
system is stable.  Furthermore, if the black hole is rotating, then
the radiation field must rotate rigidly with the same angular
velocity.  However, the system must be small enough that the radiation
is not rotating faster than the speed of light~\cite{ft:89}.
Although an equilibrating radiation field is not considered in
this work, one must acknowledge its presence in principle; a system
of finite size is required for a stable system in thermodynamic
equilibrium.

There are certain sacrifices that one must make in order to
discuss thermodynamics of gravitating systems of finite size.
First, the zeroth law of thermodynamics must be partially
abandoned.  Because a gravitational field will redshift the
temperature of an object, the temperature on the quasilocal surface
will not generally be constant, even if the system is in a suitable
equilibrium.  Second, one must adjust the definitions of extensive
and intensive thermodynamic variables because a gravitating system
cannot be partitioned into self-similar subsystems.
Following Brown and York~\cite{by:93a},
an extensive variable is defined to be one that is a function of the
phase space variables alone.  Thus, extensive variables are
constructed out of fields that lie along spacelike hypersurfaces
of constant time.  An important consequence of this requirement
is that the thermodynamic variables will depend on the observer---%
as different observers may specify
different measures of time and, thus, different spacelike
hypersurfaces.  This is not a paradox: these different specifications
will all have perfectly self-consistent thermodynamics.

Recently, Brown and York~\cite{by:93a} developed a method for
calculating energy and other charges contained within a specified
surface surrounding a gravitating system.  Although their quasilocal
energy depends only on quantities defined on the boundary of the
gravitating system, it yields information about the total
gravitational energy contained within the boundary.  In an second
paper~\cite{by:93b}, Brown and York showed that this quasilocal
energy is the thermodynamic internal energy.
The quasilocal energy and momentum are obtained from the gravitational
action via a Hamilton-Jacobi analysis.  A review of the
Hamilton-Jacobi analysis for non-gravitating systems is presented
in appendix~\ref{a:action}.  Although the analysis of Brown and York
was restricted to General Relativity in four dimensions,
the thermodynamic variables were obtained from the action rather
than from the field equations, so it is adaptable to other theories
of gravitation such as dilaton gravity.

\section{Dilaton Gravity}
\label{s:intro dilaton}

One of the goals of this work is to generalize the analysis of
Brown and York~\cite{by:93a,by:93b} to include theories of dilaton
gravity.  A dilaton is a scalar field that couples to the
\emph{curvature} of spacetime.
There are several reasons why a dilaton field might be of interest.
Any scalar ``matter'' field that has a non-minimal coupling---that
is one that couples to the curvature of spacetime instead of just
the metric---can be considered to be a dilaton.  Such scalar fields
may be of interest because they will be conformally invariant provided
that a suitable choice is made for the curvature coupling.
Black holes endowed with such non-minimally coupled dilaton fields
will have peculiar thermodynamics that cannot be accommodated by the
original formalism of Brown and York.

Scalar-tensor theories of gravitation have been proposed as alternate
theories of gravity to General Relativity.  The scalar field, which
I call a dilaton, is interpreted as an additional dynamical
gravitational field.  One of the earlier scalar-tensor theories
was proposed by Brans and Dicke~\cite{bd:61} in an attempt to include
Mach's principle into gravity.  The Brans-Dicke theory and other
scalar-tensor theories have been used as foils in testing General
Relativity.  Recently, there has been a renewed interest in
scalar-tensor theories for two reasons.  First, the low-energy limit
of string theory in the string frame is a scalar-tensor theory of
gravity with additional matter terms~\cite{gm:88}.  Second, in
two-dimensional gravity, the additional dynamical scalar field is
necessary in order to have a non-trivial gravitational theory.

Whether the scalar field that couples to the curvature is considered
a matter field or a gravitational field is unimportant: the action
of the theory will include some coupling between the scalar field
and the curvature.  Such theories I will call dilaton theories of
gravity.  When such a coupling is present, the equivalence principle
will be violated.\footnote{%
When the dilaton is considered to be part of
the gravitational field, and there is no coupling between the dilaton
and any matter field, then it is the strong equivalence principle
that is violated.  However, if the dilaton is considered to be matter
or if there are couplings between the dilaton and matter fields, then
the Einstein equivalence principle is violated.}
Nevertheless, experiments have not excluded the possibility of
dilaton gravity~\cite{will:93} so one may be interested in dilaton
gravity as an
experimental foil to General Relativity.  Furthermore, if one
views string theory as \emph{the} theory of gravity, then dilaton
gravity should be adopted as the low energy gravitational theory.
Finally, black hole solutions to dilaton gravity in lower dimensions
may make useful toy models.  For these reasons, a generalization
of the Brown and York method to theories of dilaton gravity is of
interest.

\section{Overview}
\label{s:intro overview}

The generalization of the quasilocal formalism of Brown and York
is presented in chapter~\ref{c:quasi}.  I obtain the boundary
terms generated from the variation induced in the Lagrangian density
of dilaton gravity by the variations of the dilaton and metric fields.
I construct the quasilocal variables from the boundary terms that
appear on the quasilocal surface.  Some of these quasilocal
variables will diverge as the system size grows; I present
two procedures by which a reference spacetime can be chosen to
``renormalize'' these diverging variables.  The interpretation of
the quasilocal variables is aided by the Hamiltonian as well as
by analyzing conserved charges associated with spacetimes possessing
special symmetries.  I also compute the Noether charge associated
with the coordinate independence of the action, and I use this charge
to obtain the entropy and, from the entropy, the first law of
thermodynamics in terms of the quasilocal variables.

In chapter~\ref{c:matter}, I extend the quasilocal analysis
to include the contributions from matter gauge fields (with possible
couplings to the dilaton).  Once again
I construct quasilocal variables from the variation of the matter
Lagrangian, and I show how these variables are related to the
matter Hamiltonian and to conserved matter charges.  I also compute
the extra contribution to the Noether charge that is used to
obtain the additional matter work terms in the first law of
thermodynamics.  I consider the special cases of a
minimally-coupled scalar matter field, of an electromagnetic
two-form field strength, and of an $n$-form field strength that
resembles a cosmological constant in some ways.  I then consider
the non-Abelian Yang-Mills matter field.

Next I turn to the applications of the quasilocal theory.
In chapter~\ref{c:GR}, I consider black hole spacetimes that
are solutions to the field equations of General Relativity.  The
first such spacetime that I consider is the
Reissner-Nordstr{\"o}m-anti-de\thinspace Sitter spacetime, which
is a charged, static, four-dimensional solution to the Einstein-Maxwell
field equations with a negative cosmological constant.  The analysis
of this solution shows how the quasilocal formalism can accommodate
non-asymptotically flat spacetimes as well as both electric and
magnetic charged solutions.  In order to show how the quasilocal
formalism deals with rotating black holes, I consider the
rotating three-dimensional black hole solution, which is a solution
to the Einstein field equations with a negative cosmological constant.

In dilaton gravity, the issue of conformal invariance of physical
quantities such as the mass of a spacetime is of interest.  I begin
chapter~\ref{c:dilaton} with a discussion about how quasilocal
quantities change when they are computed in conformally related
theories of dilaton gravity.  Conformal invariance of the mass
critically depends on the manner in which the reference spacetime
is selected.  I illustrate this conformal invariance for the
Garfinkle-Horowitz-Strominger solution in dilaton gravity by
calculating the mass of the solution in both the ``Einstein frame''
and the ``string frame.''  I also examine the thermodynamics of
this solution as well as two two-dimensional solutions to dilaton
gravity theories.  Finally, in chapter~\ref{c:summary}, I summarize
the results.

\section{Notation}
\label{s:intro notation}

In this work, I use the sign conventions of Misner, Thorne,
and Wheeler~\cite{mtw:73} and of Wald~\cite{w:84}.  I adopt
``natural'' units in which the rationalized Planck constant, $\hbar$,
the speed of light, $c$, and the Boltzmann constant,
$k_\sss{\text{B}}$, are set to unity.  Later, in chapters
\ref{c:GR} and~\ref{c:dilaton}, Newton's constant, $G$, will
also be set to some dimensionless value.  Many tensors relating
to the geometry of the spacetime and its boundary are defined in
appendix~\ref{a:man}.

Tensor quantities will be represented in two ways.  For the most
part I will use the abstract index notation of Wald~\cite{w:84}.
In addition, I will often use bold-faced characters to represent
differential forms.  Sometimes I will use both of these notations if
a quantity is, for example, a vector valued $p$-form.  The operation
of contraction of a vector onto the first index of a differential
form will be denoted by a centered dot.  The ``pull-back'' of a
differential form defined on the tangent space of a manifold onto
the tangent space of an embedded submanifold will be denoted by
an overline.  If the differential form was proportional to the
volume form on the manifold, then this operation is roughly the
same as contraction of the form with the normal vector to the
submanifold.  If the differential form was tensor valued, then
the pull-back operation also includes the projection of all the
tensor indices onto the submanifold.  I will often need to
distinguish between the spacelike and timelike boundaries of
a manifold.  The pull-back onto the timelike boundary will then
be denoted by an overline while the pull-back onto the spacelike
boundary by an underline.  Both an overline and an underline are
used when a form is pulled-back onto the quasilocal surface which
is the intersection of the spacelike and timelike boundaries.
Appendix~\ref{a:man} has a detailed description of the boundary
manifolds and the tensors defined on them.

\chapter{The Quasilocal Formalism}
\label{c:quasi}

I present the quasilocal formalism for obtaining the
desired thermodynamic quantities from the gravitational action.
I choose the action for dilaton gravity; the case of General
Relativity can be thought to be a special case.  Initially I 
ignore the presence of any matter (except for the dilaton field),
but I will include the effects of various types of matter later
in chapter~\ref{c:matter}.  Also, I do not restrict the
dimensionality of the spacetime in this chapter.

The quasilocal formalism is based on a Hamilton-Jacobi analysis
of the action of the gravitating system.  A review of this analysis
as applied to a non-relativistic mechanical system is presented
in appendix~\ref{a:action}.  In this appendix, the method by
which thermodynamic variables are extracted from the boundary
terms induced in the variation of the action is described as well
as the way that the entropy of the system can be found using
path integral techniques.  This chapter generalizes the analysis
of appendix~\ref{a:action} for a theory of dilaton gravity.

In this chapter, I follow a similar procedure to the one
presented in appendix~\ref{a:action}.  I first analyze the boundary
terms obtained from the variation of the gravitational action
in section~\ref{s:quasi var}.  Because I deal with fields,
the Lagrangian of appendix~\ref{a:action} is replaced with a
Lagrangian density and the action is defined for the evolution
of an initial configuration of the fields on a spacelike hypersurface
to a final configuration.  The endpoints of the paths in
appendix~\ref{a:action} are generalized to the boundary of the
gravitating system, which is the initial and final spacelike
hypersurfaces as well as the timelike boundary corresponding to
the history of the boundaries of the spacelike hypersurfaces
(see appendix~\ref{a:man}).  Time has a ``many-fingered'' character
for a gravitating system; it is encoded in the geometry of the
timelike boundary.  Consequently, the energy of
appendix~\ref{a:action}, which was conjugate to changes in the
time between the initial and final endpoints, is generalized to a
momentum conjugate to the metric on the timelike boundary.  This
momentum contains information on the surface energy, momentum, and
stress densities for the gravitating system.  These are obtained
in section~\ref{s:quasi thermo}.

The gravitational action may contain an arbitrary functional on the
boundary of the system; the contributions of this functional are
considered in section~\ref{s:quasi ref}.  The action is written
in canonical form in section~\ref{s:quasi canon} and it is shown
that the energy and momentum densities are conjugate to the lapse
function and shift vector.  A generalization of Noether's theorem
is used to obtain conserved charges for systems possessing special
symmetries in section~\ref{s:quasi charge}.  The Noether charge
associated with the diffeomorphism invariance of the gravitational
action is used to define a microcanonical action in
section~\ref{s:quasi entropy}.  The entropy of the system is
obtained from the microcanonical action via a microcanonical
functional integral.  This procedure is similar to that of
the last section of appendix~\ref{a:action}.

The action for the gravitational sector of dilaton gravity has
the form
\begin{equation}
  I_\sss{\text{G}} = \int_{\mathcal{M}} \mb{L}_\sss{\text{G}}
  = \int_{\mathcal{M}} ( \mb{L}_\sss{\text{D}}
  + \mb{L}_\sss{\text{H}} + \mb{L}_\sss{\text{V}} )\,,
  \label{dilaton gravity action}
\end{equation}
where
\begin{align}
  \mb{L}_\sss{\text{D}} &= \mb{\epsilon}\,D(\phi)R[g]\,,
  \label{dilaton Lagrangian: curvature}\\
  \mb{L}_\sss{\text{H}} &= \mb{\epsilon}\,H(\phi)g^{ab}\nabla_a\phi
    \nabla_b\phi\,,
  \label{dilaton Lagrangian: kinetic}\\
  \intertext{and}
  \mb{L}_\sss{\text{V}} &= \mb{\epsilon}\,V(\phi)
  \label{dilaton Lagrangian: potential}
\end{align}
are the curvature, kinetic, and potential terms for the dilaton
field~$\phi$.  The functions~$D(\phi)$, $H(\phi)$, and~$V(\phi)$
are functions of the dilaton field alone (and no derivatives of
the dilaton) and are used to determine the particular dilaton
gravity theory.  Thus, the action of
equation~\eqref{dilaton gravity action} actually encompasses a large
class of dilaton gravity theories.  Also, when the function~$D(\phi)$
is a constant, the action of General Relativity (with a
minimally-coupled scalar field) is recovered.  The field equations
for the dilaton field and the metric are obtained from the Lagrangian
density via a variational principle.  These field equations as well
as the boundary terms encountered in the variation of the Lagrangian
density are discussed in the following section.

\section{Variational Boundary Terms}
\label{s:quasi var}

Under variation of the action~\eqref{dilaton gravity action}, 
the (sourceless) field equations for the dilaton and the metric fields
are generated.  In order for the variational
principle to be well posed, one must specify how the fields are
to behave on the boundary of the manifold~$\mathcal{M}$.  Usually
one is not interested in these boundary terms, but rather in the
field equations themselves.  However, since one is considering
a spacetime region of \emph{finite} size, the boundary terms are
crucial.  These boundary terms are just those quantities
that will be used to define the thermodynamic variables on the
boundary.  Thus, one must include all boundary contributions when
one considers variations of the gravitational
action~$I_\sss{\text{G}}$.  In this section, there will be no
restrictions on the matter
action~$I_\sss{\text{M}}=\int\mb{L}_\sss{\text{M}}$ except for the
restriction that it contains no couplings to derivatives of the
metric or of the dilaton.  The present analysis~\cite{cm:95c}
generalizes that of Burnett and Wald~\cite{bw:90} to dilaton gravity.

Under a one-parameter family of variations of the dilaton field and
the spacetime metric, the induced variation in the
Lagrangian density~$\mb{L}_\sss{\text{G}}$ is
\begin{equation}
  \delta\mb{L}_\sss{\text{G}} = (\mb{E}_g)_{ab}\,\delta g^{ab}
    + (\mb{E}_\phi)\,\delta\phi + \mb{d\rho} \,,
  \label{variation of gravitational Lagrangian}
\end{equation}
where
\begin{gather}
  \begin{split}
    (\mb{E}_g)_{ab} &= \mb{\epsilon}\,\Bigl(
      D(\phi)G_{ab}[g] + g_{ab}\nabla^2 D(\phi)
      -\nabla_a\nabla_b D(\phi) \\
    &\qquad + H(\phi)\bigl( \nabla_a\phi \nabla_b\phi
      -\half g_{ab} (\nabla\phi)^2 \bigr) - \half g_{ab} V(\phi)
      \Bigr)\,,
  \end{split}
  \label{metric eom} \\
  (\mb{E}_\phi) = \mb{\epsilon}\,\biggl( \frac{dD}{d\phi}\,R[g]
    - \frac{dH}{d\phi}\,(\nabla\phi)^2 - 2H(\phi)\nabla^2\phi
    + \frac{dV}{d\phi} \biggr)\,,
  \label{dilaton eom} \\
  \intertext{and $\mb{\rho}=\rho\cdot\mb{\epsilon}$ with}
  \begin{split}
    \rho^a &= D(\phi) \bigl( \nabla^a(g_{cd}\,\delta g^{cd})
      - \nabla_b\delta g^{ab} \bigr)
      - g_{cd}\,\delta g^{cd} \nabla^a D(\phi)
      + \delta g^{ab} \nabla_b D(\phi) \\
    &\qquad + \bigl(2H(\phi)\nabla^a\phi\bigr)\,\delta\phi \:.
  \end{split}
  \label{boundary terms}
\end{gather}
Notice that~$\mb{\rho}$ is a function of both the fields and their
variations.  However, since~$\mb{\rho}$ appears as an exact
differential in
equation~\eqref{variation of gravitational Lagrangian}, it contributes
terms on the boundary only.  If the variations of the fields
are fixed on the boundary in a such a manner as to eliminate the
contribution from~$\mb{\rho}$, then the field
equations $(\mb{E}_g)_{ab}=\half\mb{T}_{ab}$
and~$(\mb{E}_\phi)=\half\mb{U}$ are recovered,
where~$\mb{T}_{ab}$ is the stress
tensor density arising from couplings between the metric and matter
fields in the matter action, $I_\sss{\text{M}}$, and~$\mb{U}$ is an
analogous quantity for the dilaton field:
\begin{align}
  \mb{T}^{ab} &= 2\,\frac{\delta\mb{L}_\sss{\text{M}}}{\delta g_{ab}}
  \label{stress tensor}\\
  \mb{U} &= 2\,\frac{\delta\mb{L}_\sss{\text{M}}}{\delta\phi} \:.
  \label{dilaton source}
\end{align}
The equations \eqref{metric eom} and~\eqref{dilaton eom} are related
by~$\nabla^a(\mb{E}_g)_{ab}=-\half(\mb{E}_\phi)\nabla_b\phi$.
Therefore, if the field equation~$(\mb{E}_\phi)=0$ is imposed
(so that there is no coupling between the dilaton and matter),
the quantity~$(\mb{E}_g)_{ab}$ is divergenceless.  Hence, when
the metric field equation is satisfied, the stress tensor of the
matter must also be divergenceless.

The boundary term~$\overline{\mb{\rho}}$, which is the pull-back
of~$\mb{\rho}$ onto the boundary~$\partial\mathcal{M}$, needs to be
examined.  Consider an element of~$\partial\mathcal{M}$ with the
normal vector~$n^a$.  The induced metric on this boundary element
is defined as~$\gamma_{ab}=g_{ab}\mp n_an_b$ and the extrinsic
curvature~$\varTheta_{ab}$ is defined in appendix~\ref{a:man}, though
the notation in this section has been generalized so that~$n^a$ may be
either spacelike or timelike with~$g_{ab}n^an^b=\pm1$ respectively.
Assume that the boundary is fixed under the variations so that
the variations of the normal dual-vector on the boundary are
proportional to the normal dual-vector.  This implies that, on
the boundary element,
$\delta g^{ab}=\delta\gamma^{ab}\pm2n^{(a}\delta n^{b)}$ with
$\gamma^{ab}\delta n_b=0$ and~$\delta\gamma^{ab}n_b=0$.  Then, the
boundary term~$\overline{\mb{\rho}}$ can be written in the form
\begin{equation}
  \overline{\mb{\rho}} = \mb{\pi}^{ab}\,\delta\gamma_{ab}
  + \mb{\varpi}\,\delta\phi + \delta\mb{\alpha} + \mb{d\beta}\,,
  \label{decomposition of boundary term}
\end{equation}
where
\begin{gather}
  \mb{\pi}^{ab}=\overline{\mb{\epsilon}}\,
    \bigl(\gamma^{ab}n^c\nabla_c D(\phi)
    + D(\phi)(\varTheta^{ab}-\gamma^{ab}\varTheta)\bigr)\,,
  \label{metric momentum}\\
  \mb{\varpi}=-2\overline{\mb{\epsilon}}\,
    \biggl( \varTheta\,\frac{dD}{d\phi}
    - H(\phi)n^c\nabla_c\phi \biggr)\,,
  \label{dilaton momentum}\\
  \mb{\alpha}=2\overline{\mb{\epsilon}}\,D(\phi)\varTheta\,,
  \label{boundary term one}\\
  \intertext{and $\mb{\beta}=\beta\cdot\overline{\mb{\epsilon}}$ with}
  \beta^a = D(\phi)\gamma^a{}_c\delta n^c \:.
  \label{boundary term two}
\end{gather}
Because the last term in
equation~\eqref{decomposition of boundary term} is an exact
differential, it will not contribute to
equation~\eqref{variation of gravitational Lagrangian}; thus,
it can be ignored.

Were it not for the~$\delta\mb{\alpha}$ term in
equation~\eqref{decomposition of boundary term}, the
suitable boundary conditions for a well-posed variational principle
would be $\delta\gamma_{ab}=0$ and~$\delta\phi=0$.  Thus, one is
motivated to modify the action of the gravitational sector to
include a boundary term that will cancel the~$\delta\mb{\alpha}$ term
under field variations:
\begin{equation}
  I^1_\sss{\text{G}} = \int_{\mathcal{M}} \mb{L}_\sss{\text{G}}
  - \int_{\partial\mathcal{M}} \mb{\alpha}\:.
  \label{first order gravitational action}
\end{equation}
Such a boundary functional will not affect the field equations.
The action of equation~\eqref{first order gravitational action}
is suitable for the fixation of the metric and dilaton on the
boundary of~$\mathcal{M}$.  Alternately, the variation
of~$I^1_\sss{\text{G}}$ between ``nearby'' solutions of the
equations of motion contains only boundary terms.  It is from
these boundary terms that the thermodynamic variables are constructed.

\section{Thermodynamic Variables}
\label{s:quasi thermo}

One is now ready to define the thermodynamic variables arising from
the gravitational sector of the action.  The \emph{physical}
action is~$I_\sss{\text{G}} = I^1_\sss{\text{G}} - I^0$ where
the quantity~$I^1_\sss{\text{G}}$ was defined in
equation~\eqref{first order gravitational action} and the
quantity~$I^0$ is an arbitrary functional of the boundary fields.
Restrictions on the form of~$I^0$ are made in
section~\ref{s:quasi ref}; it is shown that $I^0$ defines a
``reference spacetime,'' which can be
used as a reference point for the thermodynamic variables.
In the notation of appendix~\ref{a:man}, the action has the
following form:
\begin{equation}
  I_\sss{\text{G}} = \int_{\mathcal{M}} \mb{L}_\sss{\text{G}}
  + 2\int_{\varSigma} \underline{\mb{\epsilon}} D(\phi) K
  - 2\int_{\mathcal{T}} \overline{\mb{\epsilon}} D(\phi) \varTheta
  - I^0\:.
  \label{physical gravitational action}
\end{equation}
In equation~\eqref{physical gravitational action}, only the final
spacelike boundary contribution has been included; an initial boundary
contribution would have the same form as the negative of the final
spacelike boundary contribution (the negative arising because of the
inward orientation of the timelike normal vector on the initial
hypersurface).  Under variations
of the metric and the dilaton, the sourceless equations of motion
for these fields are obtained when the boundary fields are fixed.

The momenta conjugate to the boundary field configurations can be
obtained by taking functional derivatives of the action with respect
to the boundary fields and evaluating this derivative ``on-shell''
(i.e., when the field equations are satisfied).  The momenta
conjugate to the boundary metrics are
$\mb{p}^{ab}=(\delta I^1_\sss{\text{G}}/\delta h_{ab})_{c\ell}$
for the spacelike boundary and~$\mb{\pi}^{ab}=(\delta
I^1_\sss{\text{G}}/\delta\gamma_{ab})_{c\ell}$ for the timelike
boundary.  The subscripted ``$c\ell$'' indicates that the functional
derivatives are evaluated on a classical solution.  From
equation~\eqref{metric momentum}, these momenta are found to be
\begin{align}
  \mb{p}^{ab}&=-\underline{\mb{\epsilon}}\,
    \bigl(h^{ab}u^c\nabla_c D(\phi) + D(\phi)(K^{ab}-h^{ab}K)\bigr)
  \label{spacelike metric momentum}\\
  \intertext{and}
  \mb{\pi}^{ab}&=\overline{\mb{\epsilon}}\,
    \bigl(\gamma^{ab}n^c\nabla_c D(\phi)
    + D(\phi)(\varTheta^{ab}-\gamma^{ab}\varTheta)\bigr)\:.
  \label{timelike metric momentum}
\end{align}
Similarly, the momenta conjugate to the dilaton field are
$\mb{\wp}=(\delta I^1_\sss{\text{G}}/\delta\phi)_{c\ell}$
evaluated on the spacelike boundary
and~$\mb{\varpi}=(\delta I^1_\sss{\text{G}}/\delta\phi)_{c\ell}$
evaluated on the timelike boundary.
From equation~\eqref{dilaton momentum}, one finds
\begin{align}
  \mb{\wp} &= -\underline{\mb{\epsilon}}\, \biggl(
    2H(\phi)u^a\nabla_a \phi - 2 \frac{dD}{d\phi}\,K \biggr)
  \label{spacelike dilaton momentum}\\
  \intertext{and}
  \mb{\varpi} &= \overline{\mb{\epsilon}}\, \biggl( 2H(\phi)
    n^a\nabla_a \phi - 2 \frac{dD}{d\phi}\,\varTheta \biggr)\:.
  \label{timelike dilaton momentum}
\end{align}
The physical momenta conjugate to these fields are obtained from
the above quantities but with an additional contribution from the
background action functional~$I^0$;  $\mb{p}^{ab}_\sss0$,
$\mb{\pi}^{ab}_\sss0$, $\mb{\wp}_\sss0$, and~$\mb{\varpi}_\sss0$
are defined according to the same functional derivatives but
with~$I^0$ replacing~$I^1_\sss{\text{G}}$.

In appendix~\ref{a:action}, it was shown that the energy of a
system is conjugate to the ``time'' between the initial and final
point of the motion of the system.  However, for relativistic
systems, the ``time'' can evolve at different rates for different
parts of the boundary of the system.  The analogous quantity to
the time of the non-relativistic system is the metric induced on
the boundary~$\mathcal{T}$.  This metric contains more information
than just the lapse of time on the surface: it also describes the
shift between the leaves of the foliation of the
boundary~$\mathcal{T}$ as well as the geometry of these leaves.
Thus, the momentum conjugate to the
metric on the boundary~$\mathcal{T}$ yields the stress, energy,
and momentum of the quasilocal surface.

Suppose one induces a variation that takes a solution of
the field equation to a nearby solution.  The change in the action
arising from the change in the fields on the timelike boundary is
\begin{equation}
  \delta I_\sss{\text{G}} = \int_{\mathcal{T}} \bigl(
  (\mb{\pi}^{ab}-\mb{\pi}^{ab}_\sss0)\,\delta\gamma_{ab}
  + (\mb{\varpi}-\mb{\varpi}_\sss0)\,\delta\phi \bigr)\:.
  \label{variation on spacelike boundary}
\end{equation}
The variation of the metric~$\gamma_{ab}$ can be written in the
form of equation~\eqref{decomposition of timelike metric} on any
quasilocal surface; then
equation~\eqref{variation on spacelike boundary} becomes
\begin{equation}
  \delta I_\sss{\text{G}} = \int \mb{d}t\,\int_{\mathcal{B}} (
  -\mb{\mathcal{E}}\,\delta N + \mb{\mathcal{J}}_a\,\delta N^a
  +N\mb{\mathcal{S}}^{ab}\,\delta\sigma_{ab} +N\mb{\mathcal{\mu}}\,
  \delta\phi )
  \label{variation on history of quasilocal surface}
\end{equation}
with
\begin{align}
  \mb{\mathcal{E}} &= 2 \underline{u_a u_b ( \mb{\pi}^{ab}
  - \mb{\pi}^{ab}_\sss0 )}
  \label{definition of quasilocal energy}\\
  \mb{\mathcal{J}}^a &= -2 \underline{ u_b ( \mb{\pi}^{ab}
  - \mb{\pi}^{ab}_\sss0 )}
  \label{definition of quasilocal momentum}\\
  \mb{\mathcal{S}}^{ab} &= 
  ( \underline{\mb{\pi}}^{ab} - \underline{\mb{\pi}}^{ab}_\sss0 )
  \label{definition of quasilocal stress}\\
  \mb{\mathcal{\mu}} &=\underline{\mb{\varpi}}
  - \underline{\mb{\varpi}}_\sss0
  \label{definition of quasilocal dilaton potential}
\end{align}
and where the
relation~$\overline{\mb{\epsilon}}=-N\mb{d}t\wedge
\underline{\overline{\mb{\epsilon}}}$ has been used.  Therefore,
$\mb{\mathcal{E}}$,
$\mb{\mathcal{J}}_a$, $\mb{\mathcal{S}}^{ab}$, and~$\mb{\mu}$ are
forms on the quasilocal surface~$\mathcal{B}$ conjugate to the
variations of the lapse function, the shift vector, the metric
on~$\mathcal{B}$, and the dilaton field on~$\mathcal{B}$
respectively.  Since the lapse function is a measure of the rate
of time change for observers on the quasilocal surface, the conjugate
quantity, $\mb{\mathcal{E}}$, can be viewed as an energy density on
the quasilocal surface%
\footnote{I show later that this quantity has a more precise
definition as the thermodynamic internal energy.};
it is called the \emph{quasilocal surface energy density}.  Similarly
the shift vector can measure rotation of the quasilocal surface, and
thus the quantity~$\mb{\mathcal{J}}_a$ is called the \emph{quasilocal
surface momentum density}.  The quantity~$\mb{\mathcal{S}}^{ab}$ is
conjugate to geometry of the quasilocal surface; it is called the
\emph{quasilocal surface stress density}.  Finally, the
quantity~$\mb{\mu}$ is a \emph{dilaton potential density} conjugate
to the dilaton configuration on the quasilocal surface.

The quasilocal quantities defined in
equations~\eqref{definition of quasilocal energy}--%
\eqref{definition of quasilocal dilaton potential} can be evaluated
using equations 
\eqref{timelike metric momentum} and~\eqref{timelike dilaton momentum}
as well as equation~\eqref{extrinsic curvature relationship}.
The quasilocal quantities are found to be
\begin{gather}
  \mb{\mathcal{E}} = -2\underline{\overline{\mb{\epsilon}}}\,\bigl(
    n^a\nabla_a D(\phi) - D(\phi)k \bigr)
    - \mb{\mathcal{E}}_\sss0 \,,
  \label{quasilocal energy density}\\
  \begin{split}
    \mb{\mathcal{J}}^c &= 2\underline{\overline{\mb{\epsilon}}}\,
      D(\phi) n_a \sigma^c{}_b K^{ab} - (\mb{\mathcal{J}}_\sss0)^c \\
      &= -2 \overline{n_a \mb{p}^{ac}}
      - (\mb{\mathcal{J}}_\sss0)^c\,,
  \end{split}
  \label{quasilocal momentum density}\\
  \mb{\mathcal{S}}^{cd} = \underline{\overline{\mb{\epsilon}}}\,
    \Bigl( \sigma^{cd} n^a\nabla_a D(\phi) - D(\phi) \bigl( k^{cd}
    - \sigma^{cd}(k-n^a a_a) \bigr) \Bigr)
    - (\mb{\mathcal{S}}_\sss0)^{cd} \,,
  \label{quasilocal stress density}\\
  \intertext{and}
  \mb{\mu} = 2\underline{\overline{\mb{\epsilon}}}\,\biggl(
    H(\phi) n^a\nabla_a \phi - \frac{dD}{d\phi}(k - n^a a_a) \biggr)
    - \mb{\mu}_\sss0 \,,
  \label{quasilocal dilaton potential density}
\end{gather}
where $\mb{\mathcal{E}}_\sss0$, $(\mb{\mathcal{J}}_\sss0)^c$,
$(\mb{\mathcal{S}}_\sss0)^{cd}$, and~$\mb{\mu}_\sss0$ arise from
$\mb{\pi}_\sss0^{ab}$ and~$\mb{\varpi}_\sss0$.  It will also be
useful to decompose the variation of the metric of the quasilocal
surface as in equation~\eqref{decomposition of quasilocal metric},
i.e., into a ``shape preserving'' piece and a ``size preserving''
piece.  A similar decomposition of the quasilocal stress density
yields
\begin{equation}
  \mb{\mathcal{S}}^{ab}\,\delta\sigma_{ab} = \mathcal{S}\,
  \delta\underline{\overline{\mb{\epsilon}}}
  + \mb{\eta}^{ab}\,\delta\varsigma_{ab}\:.
  \label{decomposition of surface stress density}
\end{equation}
Here, $\mathcal{S}$ is the \emph{surface tension} field on the
quasilocal surface conjugate to changes in the size of the quasilocal
surface and $\mb{\eta}^{ab}$ is the \emph{shear density} conjugate
to changes in the shape of the quasilocal surface.  These are given by
\begin{gather}
  \mathcal{S} = 2 n^a\nabla_a D(\phi) + 2D(\phi) \Biggl( n^a a_a
  - k\biggl(\frac{n-3}{n-2}\biggr) \Biggr) - \mathcal{S}_\sss0
  \label{surface tension} \\
  \intertext{and}
  \mb{\eta}^{ab} = (\sqrt{\sigma})^{2/(n-2)}\mb{\mathcal{S}}^{ab}\:.
  \label{surface shear density}
\end{gather}

The \emph{quasilocal energy} is defined as the integral of the
quasilocal surface energy density over the quasilocal surface:
\begin{equation}
  E = \int_{\mathcal{B}} \mb{\mathcal{E}}\:.
  \label{quasilocal energy}
\end{equation}
This quantity has several useful properties, one of which is
additivity~\cite{by:93a}, although it is not necessarily positive
definite~\cite{h:94}.  The quasilocal energy is observer dependent:
even if there
is a natural way to choose the boundary~$\mathcal{T}$, the value
of the quasilocal energy depends on how this boundary is foliated
into quasilocal surfaces~$\mathcal{B}$.  Although this may appear
to be a defect in using equation~\eqref{quasilocal energy} as a
viable notion of energy, the observer dependence
is a necessary property of the thermodynamic internal energy.
In section~\ref{s:quasi charge}, I show how observer-independent
quantities can be obtained.

\section{Reference Spacetime}
\label{s:quasi ref}

In defining the thermodynamic quantities in the previous section,
I have made few restrictions on the background action
functional~$I^0$ and its contribution to the various momenta.
Since~$I^0$ does not contribute to the field equations, it would
seem to be rather unimportant.  However, it is important in the
definition of the quasilocal quantities; in particular, it is
an essential ``counterterm'' to make the quasilocal energy finite
for quasilocal surfaces at spacelike infinity in asymptotically
flat spacetimes.  In the present analysis, it is sufficient
to consider~$I^0$ to be a functional on the boundary~$\mathcal{T}$
alone; furthermore, I assume that $I^0$ is a functional of the
metric~$\gamma_{ab}$ and the dilaton configuration on the
boundary~$\mathcal{T}$.

In the absence of the contributions from the reference
action functional~$I^0$, the quasilocal energy and momentum densities
are constructed out of the phase-space
data~$\{(\mb{p}^{ab},h_{ab}),(\mb{\wp},\phi)\}$ alone.  (The
quasilocal stress and dilaton potential densities do not share this
property because they depend on the acceleration of the unit normal,
$a^a$, which is not a function of the phase-space data.)  I make
a further requirement on the form of the reference action functional:
the quantities $\mb{\mathcal{E}}_\sss0$
and~$(\mb{\mathcal{J}}_\sss0)_a$ must also be functionals of the
phase-space data.  Under this restriction, the quasilocal energy
and momentum densities are functions of the phase-space data.
Such variables are called \emph{extensive}.  The \emph{intensive}
variables, on the contrary, depend on the way the canonical
data evolve with time, i.e., they depend on the lapse and
shift functions (which are not part of the phase-space variables).

The above restrictions on the form of the background action
functional allow one to write~$I^0$ as follows:
\begin{equation}
  I^0 = \int \mb{d}t \, \int_{\mathcal{B}} \bigl(
  N \mb{\mathcal{E}}_\sss0 - N^a (\mb{\mathcal{J}}_\sss0)_a \bigr)\:.
  \label{bacground action functional}
\end{equation}
Here, $\mb{\mathcal{E}}_\sss0$ and~$(\mb{\mathcal{J}}_\sss0)_a$ are
functions of the phase-space variables and are thus extensive.
When $\mb{\mathcal{E}}_\sss0$ and~$(\mb{\mathcal{J}}_\sss0)_a$ are
known functions of $\sigma_{ab}$ and~$\phi$, one can compute
$(\mb{\mathcal{S}}_\sss0)^{ab}$ and~$\mb{\mu}_\sss0$ through the
relationship
\begin{equation}
  \int_{\mathcal{B}} N \bigl(
  (\mb{\mathcal{S}}_\sss0)^{ab}\,\delta\sigma_{ab} + \mb{\mu}_\sss0
  \,\delta\phi \bigr) = 
  - \int_{\mathcal{B}}  \bigl( N\,
  \delta\mb{\mathcal{E}}_\sss0 - N^a\,\delta(\mb{\mathcal{J}}_\sss0)_a
  \bigr) \:.
  \label{reference functional relationship}
\end{equation}
Even with the above restrictions, there is considerable freedom
in the way the reference action functional is chosen.  Of course, the
simplest choice would be to set~$I^0=0$ and to ignore the problem;
however, the contributions from the reference action functional
potentially cancel divergent terms in the quasilocal energy.
I present two prescriptions for constructing the background
action function.
\begin{enumerate}
  \item Consider a physically appropriate reference solution
  with metric~$(g_\sss0)_{ab}$ and dilaton~$\phi_\sss0$.  For example,
  when dealing with asymptotically flat reference solutions,
  one could choose Minkowski spacetime with a dilaton field;
  if one is interested in asymptotically anti-de\thinspace Sitter
  spacetimes, one could take anti-de\thinspace Sitter spacetime
  as a reference.  Suppose that it is possible to embed the
  quasilocal surface~$(\mathcal{B},\sigma_{ab})$ in some spacelike
  slice of the reference spacetime, and suppose that the pull back
  of the reference dilaton and matter fields have the same value on
  this surface as on the quasilocal surface of the original
  solution.  Then compute the trace of the extrinsic curvature,
  $k_\sss0$, of
  the quasilocal surface embedded in the reference spacetime.
  Then the reference action functional is
  \begin{equation}
    I^0 = 2\int \mb{d}t\, \int_{\mathcal{B}} D(\phi)Nk_\sss0\,
    \underline{\overline{\mb{\epsilon}}} \:;
    \label{boundary action functional 1}
  \end{equation}
  thus $\mb{\mathcal{E}}_\sss0=-2D(\phi)k_\sss0\,
  \underline{\overline{\mb{\epsilon}}}$
  and~$(\mb{\mathcal{J}}_\sss0)_a=0$.
  \item Another approach~\cite{hh:95} is to remove the restriction
  that the reference action functional, $I^0$, be a functional only
  on the boundary.  Instead let~$I^0$ be of the same form as the
  action~$I^1$; the fields $(g_\sss0)_{ab}$ and~$\phi_\sss0$ are
  independent of the fields $g_{ab}$ and~$\phi$ except on the
  boundary~$\mathcal{T}$ where the pull-backs of the fields are
  identified.  A particular solution to the field equations generated
  from~$I^0$ is taken to be the reference spacetime, and the
  contributions to the quasilocal quantities are found by applying
  the formul{\ae}~\eqref{quasilocal energy density}--%
  \eqref{quasilocal dilaton potential density} but with the
  pulled-back quantities (e.g., the dilaton and the lapse) identified
  with the original spacetime.
\end{enumerate}
The boundary action functional~$I^0$ may also contain contributions
from the matter fields; their inclusion would be a straightforward
generalization of the analysis above.  However, one is more often
interested in having a reference spacetime that is a \emph{vacuum}
solution (except for the dilaton field, which is viewed as part of
gravity), so I ignore these matter contributions.

Although there is considerable freedom in the choice of the
background action functional, this should not be viewed as a defect
in the definitions of the quasilocal quantities that depend on
this action.  A physical interpretation of the arbitrariness would
be the freedom an observer has in calibrating the measurements of
the quasilocal quantities.  Even classically the energy of a system
would have an arbitrary ``zero point''; the equations of motion
always depend on derivatives of the energy which eliminate the
ambiguity.  Similarly, it shall be shown that the ambiguity in the
choice of the reference action functional does not arise in the
first law of thermodynamics since one is only concerned with
changes in the quasilocal quantities.

\section{Canonical Decomposition of the Action}
\label{s:quasi canon}

The action~$I_\sss{\text{G}}$ can be written in canonical form as
I shall show in this section.  Often the boundary terms of the
Hamiltonian are used to obtain the internal energy of a gravitating
system and ultimately the first law of thermodynamics.  This
approach is suitable when the boundary terms are evaluated at
spacelike infinity for asymptotically flat spacetimes, but I am
interested in more general circumstances.  Nevertheless, the
boundary terms of the Hamiltonian are useful in interpreting the
various quasilocal boundary quantities that have been found.

The \emph{Hamiltonian density} for the gravitational sector is
defined as follows:
\begin{equation}
  \mb{d}t\wedge\mb{H}_\sss{\text{G}} = \mb{d}t\wedge(
  \mb{p}^{ab}\Lie_t h_{ab}
  + \mb{\wp}\Lie_t\phi) - \mb{L}_\sss{\text{G}} \:.
  \label{Hamiltonian density defined}
\end{equation}
For simplicity, I have ignored the contribution from the reference
action functional though its inclusion is straightforward.
Notice that I have used the Lie derivative with respect to the vector
field~$t^a$ as the ``time derivative'' of a field.  According to
equation~\eqref{extrinsic curvature as velocity} the
extrinsic curvature~$K_{ab}$ can be viewed as the ``velocity'' of the
induced
metric~$h_{ab}$; thus, one finds the following expression for the
first term of equation~\eqref{Hamiltonian density defined}:
\begin{equation}
  \mb{p}^{ab}\Lie_t h_{ab} = -2N\mb{p}^{ab}K_{ab} - 2N_a\nablas_b
  \mb{p}^{ab} + 2\nablas_b(N_a\mb{p}^{ab}) \:.
  \label{decomposition of hdot term}
\end{equation}
The second term of equation~\eqref{Hamiltonian density defined} can
be evaluated by splitting the vector~$t^a$ into a term proportional
to~$u^a$ and a term proportional to~$N^a$; one finds
\begin{equation}
  \mb{\wp}\Lie_t\phi = N\mb{\wp}
  \overset{\circ}{\raisebox{0pt}[1ex]{$\phi$}}
  + N^a\mb{\wp} \nablas_a\phi
  \label{decomposition of phidot term}
\end{equation}
where the symbol~$\overset{\circ}{\raisebox{0pt}[1ex]{$\phi$}}$
is shorthand
for~$u^a\partial_a\phi$.  The Lagrangian density can be decomposed
with the aid of equation~\eqref{Gauss Codacci a} along with the
relationship
\begin{equation}
  (\nabla\phi)^2 = (\nablas\phi)^2 - 
  \overset{\circ}{\raisebox{0pt}[1ex]{$\phi$}}{}^2 \:.
  \label{decomposition of nablaphi}
\end{equation}
A \emph{zero-vorticity} observer, that is, an observer who has a
velocity $n$-vector equal to the unit normal~$u^a$, experiences
an acceleration equal to~$a_a=N^{-1}\nablas_aN$.  Using these
expressions, one can evaluate the Hamiltonian density of the
gravitational sector:
\begin{equation}
  \begin{split}
    \mb{d}t\wedge\mb{H}_\sss{\text{G}} &= \mb{d}t\wedge \Bigl(
      N \mb{\mathcal{H}}_\sss{\text{G}}
      + N^a (\mb{\mathcal{H}}_\sss{\text{G}})_a
      + 2\nablas_b\bigl(N_a\mb{p}^{ab}-\underline{\mb{\epsilon}}
      N \nablas^b D(\phi)\bigr)  \Bigr)\\
      &\qquad+2\mb{\epsilon}\nabla_b\bigl(D(\phi)(u^bK+a^b)\bigr)\:.
  \end{split}
  \label{Hamiltonian density}
\end{equation}
Notice that the Hamiltonian density, apart from terms that contribute
on the boundary, is constrained with the lapse and the
shift acting as Lagrange multipliers (since they are not functions
on phase-space).
Here the \emph{Hamiltonian constraint},
$\mb{\mathcal{H}}_\sss{\text{G}}$, and the \emph{momentum constraint},
$(\mb{\mathcal{H}}_\sss{\text{G}})_a$, are given by
\begin{gather}
  \begin{split}
    \mb{\mathcal{H}}_\sss{\text{G}} &= - 2\mb{p}^{ab}K_{ab}
    + \mb{\wp}\overset{\circ}{\raisebox{0pt}[1ex]{$\phi$}} \\
    &\qquad - \underline{\mb{\epsilon}}\, \biggl(
    D(\phi) ( R[h] + K^{ab}K_{ab} - K^2 )
    +2K\frac{dD}{d\phi}\,\overset{\circ}{\raisebox{0pt}[1ex]{$\phi$}}
    -2 \nablas{}^2 D(\phi) \\
    &\qquad + H(\phi)(\nablas\phi)^2 - H(\phi)
    \overset{\circ}{\raisebox{0pt}[1ex]{$\phi$}}{}^2 + V(\phi)\biggr)
  \end{split}
  \label{Hamiltonian constraint}\\
  \intertext{and}
  (\mb{\mathcal{H}}_\sss{\text{G}})_a = -2 \nablas_b \mb{p}^b{}_a
  + \mb{\wp}\nablas_a\phi
  \label{momentum constraint}
\end{gather}
respectively.  The Hamiltonian constraint is associated with the
arbitrariness of the choice of the time function and, thus, with the
way in which~$\mathcal{M}$ is foliated into leaves~$\varSigma_t$,
while the
momentum constraint arises from the gauge invariance of the
metric~$h_{ab}$ to diffeomorphisms on~$\varSigma$.

The action for the gravitational sector~$I^1_\sss{\text{G}}$ can
be written in canonical form:
\begin{equation}
  \begin{split}
    I^1_\sss{\text{G}} &= \int_{\mathcal{M}}\mb{d}t \wedge(
    \mb{p}^{ab}
    \Lie_t h_{ab} + \mb{\wp} \Lie_t \phi - \mb{H}_\sss{\text{G}} )
    +2\int_{\varSigma}\underline{\mb{\epsilon}}\,D(\phi)K
    -2\int_{\mathcal{T}}\overline{\mb{\epsilon}}\,D(\phi)\varTheta\\
    &= \int \mb{d}t \Biggl( \int_{\varSigma} \bigl( \mb{p}^{ab}
    \Lie_t h_{ab} + \mb{\wp} \Lie_t \phi 
    - N \mb{\mathcal{H}}_\sss{\text{G}}
    - N^a (\mb{\mathcal{H}}_\sss{\text{G}})_a \bigr)\\
    &\qquad- \int_{\mathcal{B}} ( N\mb{\mathcal{E}}
    - N^a\mb{\mathcal{J}}_a ) \biggr)\:.
  \end{split}
  \label{canonical form of gravitational action}
\end{equation}
The gravitational \emph{Hamiltonian}, $H_\sss{\text{G}}$, is defined
in terms of the action by
\begin{equation}
  I^1_\sss{\text{G}} = \int \mb{d}t \biggl( \int_{\varSigma}
    ( \mb{p}^{ab} \Lie_t h_{ab} + \mb{\wp} \Lie_t \phi )
    - H_\sss{\text{G}} \biggr)\:;
  \label{Hamiltonian defined}
\end{equation}
thus,
\begin{equation}
  H_\sss{\text{G}} = \int_{\varSigma} \bigl(
  N \mb{\mathcal{H}}_\sss{\text{G}}
  + N^a (\mb{\mathcal{H}}_\sss{\text{G}})_a \bigr)
  + \int_{\mathcal{B}} (N\mb{\mathcal{E}}-N_a\mb{\mathcal{J}}^a)\:.
  \label{Hamiltonian}
\end{equation}
When the
constraint equations hold, the only contribution to the Hamiltonian
is from the boundary~$\mathcal{B}$.  Thus, the quasilocal
energy is the value of the ``on-shell'' Hamiltonian when the lapse
is unity and the shift vanishes on the quasilocal surface.  In
general, matter terms will also contribute to the boundary term
of the Hamiltonian; in this case, the quasilocal energy is recovered
when the Lagrange multipliers of these fields also vanish on the
quasilocal surface.
Notice that the on-shell value of the gravitational action for
stationary solutions (so that the Lie derivatives vanish) is given
by the $\mathcal{T}$-boundary contribution; this value is equal to
the negative of the on-shell Hamiltonian times the
time period between the initial and final spacelike hypersurfaces.

Regge and Teitelboim~\cite{rt:74} interpreted the energy of an
asymptotically flat spacetime as the value of the gravitational
Hamiltonian of General Relativity at spacelike infinity.  Iyer and
Wald~\cite{iw:94} used a generalization of this scheme to define
the energy for general gravitational theories.  In the case of
asymptotically flat solutions to the field equations of General
Relativity, it can be shown that the the value of the Hamiltonian
is equal to the \acro{adm} mass~\cite{iw:94,hh:95}.  The \acro{adm}
mass~\cite{adm:62} is essentially the surface term obtained when
the time-time component of the Einstein field equation is integrated
over a spacelike hypersurface~$\varSigma$ under asymptotically flat
field fall-off conditions.  A straightforward generalization shows
that the value of the on-shell Hamiltonian~\eqref{Hamiltonian}
of an asymptotically flat spacetime in dilaton gravity is
equal to the time-time component of the metric field
equation~\eqref{metric eom} integrated over a spacelike
hypersurface~$\varSigma$, where the boundary~$\mathcal{B}$ is taken
to be at spacelike infinity.  Although this method gives a
perfectly consistent thermodynamics~\cite{iw:94,hh:95}, I do not
adopt it here as I am interested in finite-sized quasilocal
surfaces and spacetimes that are not necessarily asymptotically flat.

\section{Conserved Charges}
\label{s:quasi charge}

Although the quasilocal energy is foliation-dependent, it is possible
to construct quantities on the timelike boundary~$\mathcal{T}$ that
are foliation-independent.  Since various classes of observers that
exist on~$\mathcal{T}$ will agree on the value of these quantities,
I shall call them conserved; in particular, the quantities will
not change with time for the observers on~$\mathcal{T}$.
The spacetime, however, must satisfy certain conditions to define such
conserved quantities.  The first requirement is that the spacetime
must possess a vector field~$\xi^a$ such that the Lie derivative
of all fields along this vector field vanish.  Such a vector field
may provide a natural choice for the boundary~$\mathcal{T}$---namely
the boundary that contains a congruence of these vectors.  Note
that~$\xi^a$ need not be defined over the entire manifold; it just
needs to be defined on~$\mathcal{T}$.

To obtain the conserved charge, I consider the equation of
motion~$2(\mb{E}_g)_{ab}=\mb{T}_{ab}$ and project the first
component normal to~$\mathcal{T}$ and the second onto~$\mathcal{T}$.
Using the analog of the Gauss Codacci
relation~\eqref{Gauss Codacci b} on the timelike
boundary~$\mathcal{T}$, one finds
\begin{equation}
  2\nablat_a(\mb{\pi}^{ab}-\mb{\pi}_\sss0^{ab}) =
  (\mb{\varpi}-\mb{\varpi}_\sss0)\nablat^b\phi - \overline{n_a
  \mb{T}^{ab}} \:.
  \label{divergence of metric momentum}
\end{equation}
The divergence of the momentum conjugate to the $\mathcal{T}$-boundary
metric has a source term.  However, if one contracts
equation~\eqref{divergence of metric momentum} with the Killing
vector~$\xi_b$, one obtains
\begin{equation}
  2\nablat_a\bigl(\xi_b(\mb{\pi}^{ab}-\mb{\pi}_\sss0^{ab})\bigr) =
  - \overline{n_a \xi_b \mb{T}^{ab}}\,,
  \label{contracted divergence of metric momentum}
\end{equation}
where the Killing properties of~$\xi^a$ have been used.  When 
this expression is integrated over~$\mathcal{T}$, the left hand side
has contributions only from the boundary~$\partial\mathcal{T}$.
One finds
\begin{gather}
  Q_\xi(\partial\mathcal{T}_{\text{final}})
  -Q_\xi(\partial\mathcal{T}_{\text{initial}})
  = \int_{\mathcal{T}} \overline{n_a\xi_b\mb{T}^{ab}}
  \label{conservation of charge}\\
  \intertext{where}
  Q_\xi(\mathcal{B}) = -2 \int_{\mathcal{B}}
  \underline{u_a\xi_b(\mb{\pi}^{ab}-\mb{\pi}_\sss0^{ab})} \:.
  \label{conserved charge}
\end{gather}
When the right hand side of equation~\eqref{conservation of charge}
vanishes, the value of~$Q_\xi$ becomes
independent of the particular cut of~$\mathcal{T}$ on which it is
evaluated; thus, it represents a conserved charge.  An additional
condition for the existence of a conserved charge, then, is the
vanishing of the right hand side of
equation~\eqref{conservation of charge}.  This may occur if the
boundary~$\mathcal{T}$ is chosen to be outside of the matter
distribution (so that~$\mb{T}_{ab}=0$), or if the
projection~$n^a\xi^b\mb{T}_{ab}$ vanishes for the specific type
of matter of interest.

Suppose that~$\varphi^a$ is a spacelike azimuthal Killing vector;
define the \emph{angular momentum}, $J=Q_\varphi$,
to be the conserved charge associated with this Killing vector.
If the quasilocal surface~$\mathcal{B}$ is taken to contain the
orbits of the Killing vector, then
\begin{equation}
  J = \int_{\mathcal{B}} \varphi^a\mb{\mathcal{J}}_a \:.
  \label{angular momentum}
\end{equation}
Thus, the $\varphi$-component of the quasilocal momentum
density can be interpreted as an angular momentum density that yields
a conserved angular momentum when~$\mathcal{B}$ contains the orbit
of~$\varphi$.  From equation~\eqref{Hamiltonian}, one sees that the
angular momentum is just the value of the on-shell Hamiltonian when
the lapse vanishes and the shift vector is~$\varphi^a$.

Alternately, when $\xi^a$ is timelike, 
a conserved \emph{mass}, $M=-Q_\xi$, can be defined.  For a static
spacetime, $\xi^a$~is surface forming and the quasilocal
surface~$\mathcal{B}$ can be chosen such that the Killing vector is
proportional to the timelike normal.  Then the mass can be written as
\begin{equation}
  M = \int_{\mathcal{B}} N\mb{\mathcal{E}} \:.
  \label{mass}
\end{equation}
By comparing equation~\eqref{mass} with~\eqref{quasilocal energy},
one sees that the quasilocal energy is not the same as the conserved
mass.  However, for asymptotically flat spacetimes, the quasilocal
energy and mass have the same value at spacelike infinity since the
lapse function becomes unity at this point.  More generally, when
the quasilocal surface is chosen to contain the orbits of the shift
vector, one finds the mass to be
\begin{equation}
  M = \int_{\mathcal{B}} (N\mb{\mathcal{E}}-N^a\mb{\mathcal{J}}_a)\:.
  \label{mass two}
\end{equation}
By comparing equation~\eqref{mass two} with~\eqref{Hamiltonian},
one sees that the mass is the same as the value of the on-shell
Hamiltonian.

There is another way of constructing conserved quantities on the
quasilocal surface.  The Lagrangian density for the gravitational
sector, $\mb{L}_\sss{\text{G}}$, is covariant under arbitrary
diffeomorphisms.  As was shown by Wald~\cite{w:93}, 
a conserved Noether current that is associated with this covariance
of the Lagrangian can be constructed.  From this conserved current,
one can then
construct a Noether charge; the value of this charge on the event
horizon is closely related to the entropy of a black hole.
I will follow Wald's procedure~\cite{w:93} in obtaining the Noether
charge arising from the Lagrangian density for dilaton gravity.

Consider a variation of the Lagrangian
density~$\mb{L}_\sss{\text{G}}$ that is associated with
diffeomorphisms along some vector field~$\xi^a$.  The variations
of the metric and the dilaton are given by
\begin{gather}
  \delta g_{ab} = \Lie_\xi g_{ab} = 2\nabla_{(a}\xi_{b)}
  \label{metric diffeo-variation}\\
  \intertext{and}
  \delta\phi = \Lie_\xi \phi = \xi^a \nabla_a\phi
  \label{dilaton diffeo-variation}
\end{gather}
respectively.  Because the Lagrangian density is covariant under
diffeomorphisms, the induced change in the Lagrangian density
by the variations \eqref{metric diffeo-variation}
and~\eqref{dilaton diffeo-variation} is equal to the Lie derivative
of the Lagrangian density along the vector~$\xi^a$.  Using Cartan's
identity,
\begin{equation}
  \Lie_\xi\mb{\varLambda} = \xi\cdot\mb{d\varLambda}
    + \mb{d}(\xi\cdot\mb{\varLambda})\,,
  \label{Cartan's identity}
\end{equation}
along with the fact that the Lagrangian density is an $n$-form, so
its exterior derivative must vanish, one finds that
\begin{equation}
  \delta\mb{L}_\sss{\text{G}} = \Lie_\xi \mb{L}_\sss{\text{G}}
    = \mb{d}(\xi\cdot\mb{L}_\sss{\text{G}})\:.
  \label{Lagrangian diffeo-variation}
\end{equation}
Therefore, one can define the Noether current,
\begin{equation}
  \mb{j}[\xi] = \mb{\rho}(g,\phi,\Lie_\xi g,\Lie_\xi\phi)
    - \xi\cdot\mb{L}_\sss{\text{G}}\,,
  \label{definition of Noether current}
\end{equation}
which is closed when the field equations of motion
($(\mb{E}_g)_{ab}=0$
and~$(\mb{E}_\phi)=0$) hold.  One can then construct an $(n-2)$-form,
$\mb{q}[\xi]$, that
satisfies~$\mb{j}[\xi]=\mb{dq}[\xi]$.
This \emph{Noether charge density} can be integrated over any closed
surface~$\mathcal{B}$ to give a conserved \emph{Noether charge}.

I use the expression for~$\rho^a$ given in
equation~\eqref{boundary terms}, as well as equations
\eqref{metric diffeo-variation} and~\eqref{dilaton diffeo-variation},
to obtain an expression for the Noether current:
\begin{equation}
  j^a[\xi] = -2\nabla_b \bigl( 2 \xi^{[a} \nabla^{b]} D(\phi)
    + D(\phi) \nabla^{[a}\xi^{b]} \bigr) + 2 \xi_b (E_g)^{ab}
  \label{Noether current}
\end{equation}
where $\mb{j}=j\cdot\mb{\epsilon}$
and~$(\mb{E}_g)_{ab}=\mb{\epsilon}\,(E_g)_{ab}$.  When the
(sourceless) field equation~$(E_g)_{ab}=0$ holds, the Noether
current~$j^a$ is divergenceless, so one can construct the
Noether charge density, $\mb{q}[\xi]$, on a closed spacelike
$(n-2)$-dimensional hypersurface.  The Noether charge
density is
\begin{equation}
  \mb{q}[\xi] = \underline{\overline{\mb{\epsilon}}}\, n^{ab}
  \bigl( 2\xi_a\nabla_b D(\phi) + D(\phi)\nabla_a \xi_b \bigr)
  \label{Noether charge density}
\end{equation}
where $n^{ab}$ is the bi-normal to the $(n-2)$-surface.  The
Noether charge is the integral of this quantity over the
$(n-2)$-surface.  When matter fields are present, both the
Noether current and the Noether charge will have additional
contributions.  I illustrate the effect of matter fields
in the next chapter.

\section{Entropy of Stationary Black Hole Spacetimes}
\label{s:quasi entropy}

The presence of an event horizon within the thermodynamic system
gives rise to an entropy for the system; in the absence of any
matter fields, the event horizons are the only source of entropy.
Here I explore the contributions to the entropy of thermodynamic
systems containing a stationary black hole.  For simplicity, I will
ignore possible matter fields and their contribution to the entropy.

I obtain the entropy from Euclidean path integral techniques
using a ``microcanonical'' action~\cite{by:93b}.  By
``microcanonical'' I mean the action for which the extensive
variables (alone) must be fixed on the boundary in order to have
a well-posed variational principle.  From
equation~\eqref{variation on history of quasilocal surface} one
sees that the action~$I_\sss{\text{G}}$ is \emph{not} a microcanonical
action because both extensive and intensive variables must be fixed
on the quasilocal boundary.  Following a suggestion of Iyer and
Wald~\cite{iw:95}, I consider the action
\begin{equation}
  I_{\text{m}} = \int_{\mathcal{M}} \mb{L}_\sss{\text{G}}
  - \int_{\mathcal{T}} \mb{d}t \wedge \mb{q}[t]
  \label{definition of microcanonical action}
\end{equation}
where only the boundary element~$\mathcal{T}$ is of interest.
Here, $\mb{q}[t]$ is the Noether charge density associated with
the diffeomorphism covariance of the gravitational Lagrangian
density along the timelike Killing vector field~$t^a$.

I first show that the action~$I_{\text{m}}$ is the desired
microcanonical action.  The Noether charge
density of equation~\eqref{Noether charge density} associated with
the vector field~$t^a$ must be evaluated.  One finds the first term in
equation~\eqref{Noether charge density} is~$-2Nn^a\nabla_aD(\phi)$
while a calculation of the second term yields (see~\cite{iw:95})
$2D(\phi)(Nu^au^b\varTheta_{ab}+N^au^b\varTheta_{ab})$.  Using
equation~\eqref{extrinsic curvature relationship} as well as the
definitions of the quasilocal energy and momentum densities,
equations \eqref{definition of quasilocal energy}
and~\eqref{definition of quasilocal momentum}, one finds
\begin{equation}
  -\mb{d}t\wedge\mb{q}[t] = -\mb{d}t\wedge( N\mb{\mathcal{E}}
  - N^a\mb{\mathcal{J}}_a) - \mb{\alpha}
  \label{Noether charge on quasilocal surface}
\end{equation}
where~$\mb{\alpha}$ is given in equation~\eqref{boundary term one}.
Therefore, the boundary contribution to the variation of the
action~$I_{\text{m}}$ is
\begin{equation}
  \begin{split}
     \delta I_{\text{m}} &= \int_{\mathcal{T}} ( -\mb{d}t\wedge
     \delta\mb{q}[t] + \overline{\mb{\rho}} ) \\
     &= \int_{\mathcal{T}} \bigl( -\mb{d}t \wedge \delta(
     N\mb{\mathcal{E}} - N^a\mb{\mathcal{J}}_a)
     + (\overline{\mb{\rho}} - \delta\mb{\alpha}) \bigr) \\
     &= \int \mb{d}t \int_{\mathcal{B}} \bigl(
     N\,\delta\mb{\mathcal{E}} - N^a\,\delta\mb{\mathcal{J}}_a
     +N\mb{\mathcal{S}}^{ab}\,\delta\sigma_{ab}+\mb{\mu}\,\delta\phi
     \bigr)
  \end{split}
  \label{variation of microcanonical action}
\end{equation}
where the second term in the second line has been calculated in
equation~\eqref{variation on history of quasilocal surface}.
Clearly it is the extensive variables that must be held fixed on the
quasilocal boundary in order for the variations of the
action~$I_{\text{m}}$ to yield the field equations; thus,
$I_{\text{m}}$ is a microcanonical action.

In order to calculate the entropy of the thermodynamic system,
I will use the method of Brown and York~\cite{by:93b}.  Consider the
``Euclidean section'' of the black hole spacetime obtained by the
Wick rotation~$t\to\tau=it$.  The extensive variables are invariant
under this transformation while the intensive variables
become imaginary.  Under such a transformation, the metric becomes
complex in general, so the term ``Euclidean'' is misleading.
In a coordinate system that becomes co-rotating near the event
horizon, however, the imaginary components of the metric
become small and the metric is approximately real with a Euclidean
signature close to the event horizon.  The time variable is periodic
near the bifurcation point;
the period can be fixed by the condition that there be no conical
singularity at the bifurcation point.  The topology of the Euclidean
manifold, $\hat{\mathcal{M}}$,
is the direct product of the topology of the quasilocal
surface~$\mathcal{B}$ with a disk where the origin of the disk is
the bifurcation point of the black hole spacetime; the interior of
the black hole is not present in the Euclidean section.  One can
view the Euclidean section as the identification of the initial and
final spacelike hypersurface with a period of imaginary time equal
to~$\Delta\tau=2\pi/\varkappa_\sss{\text{H}}$.  Here,
$\varkappa_\sss{\text{H}}^2=h^{ab}(\partial_aN)(\partial_bN)$~%
(evaluated on the event horizon) is the \emph{surface gravity}.
A diagram depicting the Lorentzian and Euclidean sections
of the Schwarzschild spacetime is presented in figure~\ref{fig:eucl}.
\begin{figure}[t]
\renewcommand{\baselinestretch}{1}\small
\begin{center}
\setlength{\unitlength}{1mm}
\begin{picture}(110,120)
\put(5,70){%
  \begin{picture}(100,50)(0,25)
  \put(10,50){\line(1,1){20}}
  \put(10,50){\line(1,-1){20}}
  \put(50,50){\line(-1,1){20}}
  \put(50,50){\line(-1,-1){20}}
  \put(50,50){\circle*{2}}
  \put(50,50){\line(1,1){20}}
  \put(50,50){\line(1,-1){20}}
  \put(90,50){\line(-1,1){20}}
  \put(90,50){\line(-1,-1){20}}
  \multiput(32.22,70)(8.88,0){5}{\qbezier(-2.22,0)(0,2.22)(2.22,0)}
  \multiput(36.66,70)(8.88,0){4}{\qbezier(-2.22,0)(0,-2.22)(2.22,0)}
  \multiput(32.22,30)(8.88,0){5}{\qbezier(-2.22,0)(0,-2.22)(2.22,0)}
  \multiput(36.66,30)(8.88,0){4}{\qbezier(-2.22,0)(0,2.22)(2.22,0)}
  \linethickness{1pt}
  \qbezier(50,50)(60,60)(70,60)
  \qbezier(50,50)(60,40)(70,40)
  \put(70,40){\line(0,1){20}}
  \put(45,49){$\mathcal{H}$}
  \put(60,43){$\varSigma_0$}
  \put(60,55){$\varSigma_{\Delta t}$}
  \put(71,49){$\mathcal{T}$}
  \put(60,49){$\mathcal{M}$}
  \put(70,40){\circle*{2}}
  \put(71,39){$\mathcal{B}$}
  \end{picture}}
\put(40,65){(a) Lorentzian Section}
\put(5,5){%
  \begin{picture}(100,50)(0,25)
  \qbezier(10,50)(10,70)(80,70)
  \qbezier(10,50)(10,30)(80,30)
  \linethickness{1pt}
  \qbezier(80,70)(75,70)(70,60)
  \qbezier(70,60)(65,50)(70,40)
  \qbezier(70,40)(75,30)(80,30)
  \qbezier(80,30)(85,30)(90,40)
  \qbezier(90,40)(95,50)(90,60)
  \qbezier(90,60)(85,70)(80,70)
  \qbezier(10,50)(10,40)(70,40)
  \put(70,40){\circle*{2}}
  \put(10,50){\circle*{2}}
  \put(35,43){$\varSigma_0\equiv\varSigma_{\Delta\tau}$}
  \put(72,39){$\mathcal{B}$}
  \put(5,49){$\mathcal{H}$}
  \put(72,59){$\hat{\mathcal{T}}$}
  \put(55,60){$\hat{\mathcal{M}}$}
  \end{picture}}
\put(40,0){(b) Euclidean Section}
\end{picture}
\end{center}
\begin{quote}\leavevmode
\caption[Lorentzian and Euclidean sections of Schwarzschild spacetime]%
  {\small A depiction of the manifold region~$\mathcal{M}$ in the
  extended Schwarzschild spacetime in (a) the Lorentzian section,
  and (b) the corresponding manifold region~$\hat{\mathcal{M}}$ in
  the Euclidean section.  Each point on these diagrams represents a
  spacelike sphere.  In the Euclidean section, the analytic
  extension of the Schwarzschild solution is not present.  In order
  for regularity at the ``point''~$\mathcal{H}$, the
  identification of~$\varSigma_0$ with~$\varSigma_{\Delta\tau}$
  must have the period~$\Delta\tau=2\pi/\varkappa_\sss{\text{H}}$.}
\label{fig:eucl}
\end{quote}
\end{figure}

I follow the argument of Iyer and Wald~\cite{iw:95} to calculate
the entropy.  The microcanonical density matrix is given by the
formal expression
\begin{equation}
  \begin{split}
    \nu \Bigl[
    \raisebox{-.5\height+1ex}{\shortstack{extensive\\variables}}
    \Bigr]
    &= \int \Bigl[
    \raisebox{-.5\height+1ex}{\shortstack{periodic\\histories}}
    \Bigr]
    \exp \biggl( \hat{I}_{\text{m}} \Bigl[
    \raisebox{-.5\height+1ex}{\shortstack{extensive\\variables}}
    \Bigr]
    \biggr) \\
    &\approx \exp\bigl( \hat{I}_{\text{m}} |_{c\ell} \bigr)
    \quad \text{(zeroth order)}\,,
  \end{split}
  \label{microcanonical density matrix}
\end{equation}
where $\hat{I}_{\text{m}}$ is the Euclideanized microcanonical
action---the usual microcanonical action calculated on the
manifold~$\hat{\mathcal{M}}$---and the subscripted~``$c\ell$''
indicates evaluation on the analytic continuation to imaginary time
of the classical solution.  The second line in
equation~\eqref{microcanonical density matrix} gives the evaluation
of the path integral to the ``zeroth order'' in quantum corrections.
The entropy, $S$, is the logarithm of the microcanonical density
matrix; thus, $S\approx\hat{I}_{\text{m}}$.  Then the entropy is
\begin{equation}
  \begin{split}
    S &\approx \Biggl[ i\int_{\hat{\mathcal{M}}}
    \mb{L}_\sss{\text{G}} - \int_{\hat{\mathcal{T}}}
    \mb{d}\tau \wedge \mb{q}[t] \Biggr]_{c\ell} \\
    &=\Delta\tau \Biggl[ \int_{\varSigma} t\cdot\mb{L}_\sss{\text{G}}
    + \int_{\mathcal{B}} \mb{q}[t] \Biggr]_{c\ell}\:.
  \end{split}
  \label{derivation of entropy}
\end{equation}
From equation~\eqref{definition of Noether current}, one finds
that~$t\cdot\mb{L}_\sss{\text{G}}=-\mb{dq}[t]$ when evaluated
on a classical solution.  Here, $\mb{\rho}$ vanishes because the
fields are stationary, so the Lie derivatives of the fields along the
vector~$t^a$ vanish.  The integrand of the first integral on
the second equality of equation~\eqref{derivation of entropy}
contributes two surface terms: one on the quasilocal
surface~$\mathcal{B}$ that cancels the second integral, and one
on the event horizon~$\mathcal{H}$.  Therefore,
\begin{equation}
  S \approx \frac{2\pi}{\varkappa_\sss{\text{H}}}\,
  \int_{\mathcal{H}} \mb{q}[t] \qquad \text{(zeroth order).}
  \label{definition of entropy}
\end{equation}

One can evaluate equation~\eqref{definition of entropy} using
equation~\eqref{Noether charge density}.  On the event
horizon~$\mathcal{H}$, one has $t_a=0$
and~$\nabla_at_b=-\kappa_\sss{\text{H}}n_{ab}$.  One finds
\begin{equation}
  S \approx 4\pi \int_{\mathcal{H}}
  \underline{\overline{\mb{\epsilon}}}\,D(\phi) \qquad
  \text{(zeroth order).}
  \label{entropy}
\end{equation}
For General Relativity, $D=(16\pi)^{-1}$, so the entropy is
one-quarter of the area of the event horizon, which is the standard
result.  In general, however, the dilaton field contributes to the
entropy.

Using equation~\eqref{variation of microcanonical action},
one can calculate the variation of entropy amongst classical
solutions.  Define
\begin{equation}
  \beta = \int N\mb{d}\tau \quad\text{and}\quad
  \omega^a = \beta^{-1}\int N^a\mb{d}\tau
  \label{temperature and velocity}
\end{equation}
to be the \emph{inverse temperature} and the \emph{velocity}
of the quasilocal surface.
Since~$\delta S\approx\delta\hat{I}_{\text{m}}$, one finds that
\begin{equation}
  \delta S \approx \int_{\mathcal{B}} \beta (
  \delta\mb{\mathcal{E}} - \omega^a\,\delta\mb{\mathcal{J}}_a
  + \mb{\mathcal{S}}^{ab}\,\delta\sigma_{ab} + \mb{\mu}\,\delta\phi
  )\:.
  \label{first law of thermodynamics}
\end{equation}
This is the integral form of the \emph{first law of thermodynamics}
for a system containing a black hole to the zeroth order in quantum
corrections.  Unfortunately, one cannot integrate this equation in
general.  As I have argued before, the temperature will not always
be constant on the quasilocal surface because of the redshift factor;
even if one chooses the quasilocal surface to be an isotherm, the
angular velocity of this surface will not necessarily be constant.
With static spherically symmetric spacetimes, however, one will be
able to integrate equation~\eqref{first law of thermodynamics}
to obtain the usual differential form of the first law of
thermodynamics.

\chapter{Matter Fields}
\label{c:matter}

Having defined the basic quasilocal quantities for the gravitational
part of the action of dilaton gravity, I now proceed to
analyze the contributions of matter fields.  For the present, I
only consider matter fields that are minimally coupled to
the metric and the dilaton, i.e., I do not allow the matter
fields to couple to derivatives of the metric or the dilaton.
I shall show that such minimally coupled matter fields will not
contribute explicitly to the entropy at the classical level
although they will provide
extra work terms in the first law of thermodynamics.

For most of this chapter I consider Abelian gauge fields only;
I defer a discussion of the non-Abelian Yang-Mills fields until
section~\ref{s:matter Yang-Mills}.  The class of matter
fields that I consider includes electromagnetism as well as matter
fields with larger gauge groups.  Most of my analysis will be general
enough to include massless scalar fields even though these are not
gauge fields.  Consider the matter action
\begin{equation}
  I_\sss{\text{M}} = -\int_{\mathcal{M}} \frac{1}{2}\, W(\phi)
  \mb{\frak{F}} \wedge \ast \mb{\frak{F}} = - \int_{\mathcal{M}}
  \mb{\epsilon}\,\frac{1}{2p!}\,W(\phi)\frak{F}^{a_1\cdots a_p}
  \frak{F}_{a_1\cdots a_p}
  \label{matter action}
\end{equation}
where $\mb{\frak{F}}=\mb{d\frak{A}}$ is a $p$-form field strength
obtained from the $(p-1)$-form potential~$\mb{\frak{A}}$.  The field
strength---and thus the matter action---is invariant under the gauge
transformation~$\mb{\frak{A}}\to\mb{\frak{A}}+\mb{d\frak{x}}$ for any
$(p-2)$-form~$\mb{\frak{x}}$.  The function~$W(\phi)$ controls the
coupling of the matter field with the dilaton.  The Hodge dual
on the manifold~$\mathcal{M}$ is represented by the symbol~$\ast$;
for any $q$-form~$\mb{\varLambda}$ one has
\begin{equation}
  (\ast\varLambda)_{b_1\cdots b_{n-q}} = \frac{1}{q!}\,
  \varLambda^{a_1\cdots a_q}
  \epsilon_{a_1\cdots a_q b_1\cdots b_{n-q}} \:.
  \label{Hodge dual}
\end{equation}
Therefore, the Hodge dual converts a $q$-form into an $(n-q)$-form.

One can decompose the field strength~$\mb{\frak{F}}$ into electric
and magnetic components on any spacelike hypersurface~$\varSigma$.  
Let~$u^a$ be the normal vector to the spacelike hypersurface;
the \emph{electric field} $(p-1)$-form on~$\varSigma$ is defined
as~$\mb{\frak{E}}=u\cdot\mb{\frak{F}}$.  Similarly, define
the \emph{magnetic field} $(n-p-1)$-form on~$\varSigma$
as~$\mb{\frak{B}}=(-)^pu\cdot(\ast\mb{\frak{F}})=\star\mb{\frak{F}}$
where $\star$~is the Hodge dual on the spacelike hypersurface, i.e.,
a Hodge dual defined just as in equation~\eqref{Hodge dual} but
with the volume form~$\underline{\mb{\epsilon}}=u\cdot\mb{\epsilon}$
on the spacelike hypersurface.  The kinetic energy of the field
strength can then be decomposed into electric and magnetic components
as follows:
\begin{equation}
  \frak{F}^{a_1\cdots a_p}\frak{F}_{a_1\cdots a_p} = 
  \frac{p!}{(n-p-1)!} \frak{B}^{a_1\cdots a_{n-p-1}}
  \frak{B}_{a_1\cdots a_{n-p-1}} - p\, \frak{E}^{a_1\cdots a_{p-1}}
  \frak{E}_{a_1\cdots a_{p-1}} \:.
  \label{decomposition of gauge kinetic energy}
\end{equation}

To obtain the equations of motion for the matter fields as well
as the desired quasilocal quantities, one must analyze the variation
of the matter Lagrangian, $\mb{L}_\sss{\text{M}}$.
I do this in the next section.

\section{Variations of the Matter Lagrangian}
\label{s:matter var}
  
The variation of the matter action yields the equations of motion
for the matter fields as well as the stress energy tensor and dilaton
source arising from the couplings between the matter fields and
the metric and dilaton.  In addition, boundary terms will
be obtained and, as before, I construct useful quasilocal
quantities from these boundary terms.  Once again this analysis
is a generalization of Burnett and Wald~\cite{bw:90}.

I introduce a one-parameter family of variations of the metric,
the dilaton, and the field potential~$\mb{\frak{A}}$.  Notice that
I choose to vary the potential rather than the field strength.
The induced variation in the matter Lagrangian
density~$\mb{L}_\sss{\text{M}}$ is found to be
\begin{equation}
  \delta\mb{L}_\sss{\text{M}} = -\half \mb{T}_{ab}\,\delta g^{ab}
  -\half \mb{U}\,\delta\phi
  + (\mb{E}_{\frak{A}})^{a_1\cdots a_{p-1}}\,
  \delta\frak{A}_{a_1\cdots a_{p-1}} + \mb{d\varrho}
  \label{variation of matter Lagrangian}
\end{equation}
where $\mb{T}_{ab}$ and~$\mb{U}$ are the stress tensor and dilaton
source densities defined by equations \eqref{stress tensor}
and~\eqref{dilaton source}.  The stress tensor is given by
\begin{gather}
  \mb{T}_{ab} = \mb{\epsilon}\,\frac{1}{p!}\,W(\phi) \bigl(
  p\, \frak{F}_a{}^{c_2\cdots c_p} \frak{F}_{bc_2\cdots c_p}
  -\half g_{ab}\frak{F}^{c_1\cdots c_p}\frak{F}_{c_1\cdots c_p}\bigr)
  \label{matter stress tensor} \\
  \intertext{and the dilaton source by}
  \mb{U} = \mb{\epsilon}\,\frac{1}{p!}\,\frac{dW}{d\phi}
  \frak{F}^{a_1\cdots a_p}\frak{F}_{a_1\cdots a_p} \:.
  \label{matter dilaton source}
\end{gather}
The (sourceless) equation of motion for the matter field
is~$(\mb{E}_{\frak{A}})^{a_1\cdots a_{p-1}}=0$ with
\begin{equation}
  (\mb{E}_{\frak{A}})^{a_1\cdots a_{p-1}} = \mb{\epsilon}\,
  \frac{1}{(p-1)!}\,\nabla_b \bigl( W(\phi)
  \frak{F}^{ba_1\cdots a_{p-1}} \bigr)
  \label{matter eom}
\end{equation}
while the boundary term~$\mb{\varrho}=\varrho\cdot\mb{\epsilon}$
is given by
\begin{equation}
  \varrho^a = - \frac{1}{(p-1)!}\,W(\phi)
  \frak{F}^{ab_1\cdots b_{p-1}}\,
  \delta\frak{A}_{b_1\cdots b_{p-1}}\:.
  \label{matter boundary term}
\end{equation}

Consider the contribution of the boundary term~$\mb{\varrho}$
to the timelike boundary~$\mathcal{T}$.  To do so one must pull-back
this form onto~$\mathcal{T}$:
\begin{gather}
  \overline{\mb{\varrho}} = \mb{\varPi}^{a_1\cdots a_{p-1}}\,
  \overline{\delta\frak{A}}_{a_1\cdots a_{p-1}}
  \label{pull-back of boundary term onto T}\\
  \intertext{with}
  \mb{\varPi}^{a_1\cdots a_{p-1}} = - \overline{\mb{\epsilon}}\,
  \frac{1}{(p-1)!}\,W(\phi) n_b \frak{F}^{ba_1\cdots a_{p-1}}
  \label{timelike matter momentum density}
\end{gather}
being the momentum density conjugate to the matter field on the
timelike boundary~$\mathcal{T}$.  One can further decompose the
pull-back of the variation of the matter field potential as
follows: define the potential $(p-2)$-form, $\mb{\frak{V}}$,
on the quasilocal surface~$\mathcal{B}$
by~$\mb{\frak{V}}=\underline{u\cdot\overline{\mb{\frak{A}}}}$, and the
potential $(p-1)$-form, $\mb{\frak{W}}$, on the quasilocal surface
by~$\mb{\frak{W}}=\underline{\overline{\mb{\frak{A}}}}$.  One finds
\begin{equation}
  \begin{split}
    \overline{\delta\frak{A}}_{a_1a_2\cdots a_{p-1}} &=
    - \frac{(p-1)}{N}\,u_{[a_1} \bigl(
    \delta(N\frak{V}_{a_2\cdots a_{p-1}]}) + \delta N^b
    \frak{W}_{|b|a_2\cdots a_{p-1}]} \bigr)\\
    &\qquad +\delta\frak{W}_{a_1a_2\cdots a_{p-1}}\:.
  \end{split}
  \label{decomposition of variation of matter potential}
\end{equation}
Using equation~\eqref{decomposition of variation of matter potential},
one can further decompose
equation~\eqref{pull-back of boundary term onto T} as follows:
\begin{equation}
  \begin{split}
    \mb{\varPi}^{a_1\cdots a_{p-1}}\,
    \overline{\delta\frak{A}}_{a_1\cdots a_{p-1}} &= -\mb{d}t\wedge
    \bigl( - \mb{\frak{Q}}^{a_1\cdots a_{p-2}} \,\delta( N
    \frak{V}_{a_1\cdots a_{p-2}}) + \mb{\frak{J}}_a\,\delta N^a \\
    &\qquad + N \mb{\frak{K}}^{a_1\cdots a_{p-1}} \,\delta
    \frak{W}_{a_1\cdots a_{p-1}} \bigr)\:.
  \end{split}
  \label{decomposition of timelike matter momentum}
\end{equation}
Here, $\mb{\frak{Q}}_{a_1\cdots a_{p-2}}$ is a $(p-2)$-form-valued
density on the quasilocal surface~$\mathcal{B}$ that is called
the \emph{quasilocal matter charge density}.  This density is given by
\begin{equation}
  \mb{\frak{Q}}^{a_1\cdots a_{p-2}} =
  \underline{\overline{\mb{\epsilon}}}\,\frac{1}{(p-2)!}\,W(\phi)
  n_b \frak{E}^{ba_1\cdots a_{p-2}}\:.
  \label{quasilocal matter charge density}
\end{equation}
Similarly, $\mb{\frak{J}}_a$ represents an \emph{electromotive force
\emph{(\acro{emf})} density} on the quasilocal surface.\footnote{%
The \acro{emf} density is the potential created by changes in the
electromotive force of the system~\cite{bmy:91,tpmsz:86}.}
It is defined by
\begin{equation}
  \mb{\frak{J}}_a = -\mb{\frak{Q}}^{b_1\cdots b_{p-2}}
  \frak{W}_{ab_1\cdots b_{p-2}}\:.
  \label{quasilocal EMF density}
\end{equation}
Finally, the \emph{surface current density} is the
$(p-1)$-form-valued density on the quasilocal surface:
\begin{equation}
  \mb{\frak{K}}^{a_1\cdots a_{p-1}} =
  -\underline{\overline{\mb{\epsilon}}}\,\frac{1}{(p-1)!}\,W(\phi)
  n_b \underline{\frak{F}}^{ba_1\cdots a_{p-1}}\:.
  \label{quasilocal surface current density}
\end{equation}
Because the field strength is pulled-back onto the spacelike
hypersurface in equation~\eqref{quasilocal surface current density},
the surface current density is defined
on the quasilocal surface~$\mathcal{B}$.
Equations \eqref{quasilocal matter charge density}--%
\eqref{quasilocal surface current density} define all the quasilocal
quantities that are needed for the work terms in the first law
of thermodynamics.

\section{Canonical Form of the Matter Action}
\label{s:matter canon}

The matter action~$I_\sss{\text{M}}$ can be written in canonical
form and from this one can obtain the Hamiltonian.  The
Hamiltonian is useful in interpreting the quasilocal quantities that
have just been constructed.

The first step in the canonical decomposition is to obtain the
momentum conjugate to the field configuration on the spacelike
boundary.  This momentum is obtained in the same manner as before:
here,
however, one must pull-back the boundary contribution~$\mb{\varrho}$
onto the spacelike boundary~$\varSigma$.  One finds
\begin{gather}
  \underline{\mb{\varrho}} = \mb{P}^{a_1\cdots a_{p-1}}\,
  \underline{\delta\frak{A}}_{a_1\cdots a_{p-1}}
  \label{pull-back of boundary term onto S}\\
  \intertext{with}
  \mb{P}^{a_1\cdots a_{p-1}} = \underline{\mb{\epsilon}}\,
  \frac{1}{(p-1)!}\,W(\phi) \frak{E}^{a_1\cdots a_{p-1}}\:.
  \label{spacelike matter momentum density}
\end{gather}
Notice that the momentum conjugate to the matter potential on the
spacelike hypersurface, $\mb{P}^{a_1\cdots a_{p-1}}$,
is proportional to the electric field strength.

Next, the Hamiltonian density for the matter sector is computed:
\begin{equation}
  \mb{d}t\wedge\mb{H}_\sss{\text{M}} = \mb{d}t\wedge
  \mb{P}^{a_1\cdots a_{p-1}}\,
  \Lie_t \frak{A}_{a_1\cdots a_{p-1}} - \mb{L}_\sss{\text{M}} \:.
  \label{matter Hamiltonian density defined}
\end{equation}
The Lie derivative in the first term can be evaluated using
Cartan's identity~\eqref{Cartan's identity}.  The first term
will contribute a total derivative as well as terms proportional
to the lapse, the shift, and the Lagrange
multiplier~$t\cdot\mb{\frak{A}}$.  The Lagrangian density has already
been decomposed into fields on the spacelike boundary in
equation~\eqref{decomposition of gauge kinetic energy}.
Thus, the matter Hamiltonian is
\begin{equation}
\begin{split}
    H_\sss{\text{M}} &= \int_{\varSigma} \mb{H}_\sss{\text{M}} \\
      &= \int_{\varSigma} \bigl( N \mb{\mathcal{H}}_\sss{\text{M}}
      + N^a (\mb{\mathcal{H}}_\sss{\text{M}})_a
      + t^b \frak{A}_{ba_1\cdots a_{p-2}}
      \mb{\mathcal{G}}^{a_1\cdots a_{p-2}} \bigr) \\
      &\qquad + \int_{\mathcal{B}} ( N \frak{V}_{a_1\cdots a_{p-2}}
      \mb{\frak{Q}}^{a_1\cdots a_{p-2}} - N^a \mb{\frak{J}}_a ) \:.
  \end{split}
  \label{matter Hamiltonian}
\end{equation}
The matter fields contribute to the Hamiltonian and the momentum
constraints; their contribution is given by
\begin{gather}
  \begin{split}
  \mb{\mathcal{H}}_\sss{\text{M}} &= \underline{\mb{\epsilon}}\,
  W(\phi) \biggl( \frac{1}{2(n-p-1)!}\,\frak{B}^{a_1\cdots a_{n-p-1}}
  \frak{B}_{a_1\cdots a_{n-p-1}} \\ &\qquad+ \frac{1}{2(p-1)!}
  \frak{E}^{a_1\cdots a_{p-1}} \frak{E}_{a_1\cdots a_{p-1}} \biggr)
  \end{split}
  \label{matter Hamiltonian constraint}\\
  \intertext{and}
  (\mb{\mathcal{H}}_\sss{\text{M}})_a = \frak{F}_{ab_1\cdots b_{p-1}}
  \mb{P}^{b_1\cdots b_{p-1}}
  \label{matter momentum constraint}
\end{gather}
respectively.  In addition, there is a Gauss law constraint that
arises from the gauge invariance of the Abelian group.  The Gauss
law constraint is
\begin{equation}
  \mb{\mathcal{G}}^{a_1\cdots a_{p-2}} = -(p-1) \nablas_b
  \mb{P}^{ba_1\cdots a_{p-2}}\:.
  \label{Gauss law constraint}
\end{equation}
The value of the on-shell Hamiltonian arises from the integral on
the quasilocal surface.  The two components on the quasilocal surface
come from the quasilocal surface charge density and the quasilocal
surface \acro{emf} density.
These components are conjugate to the
potential, $\mb{\frak{V}}$, blue-shifted from its value on the
event horizon, and the shift function respectively.  Finally,
one can write the matter action in the canonical form:
\begin{equation}
  \begin{split}
    I_\sss{\text{M}} &= \int_{\mathcal{M}}\mb{d}t \wedge(
    \mb{P}^{a_1\cdots a_{p-1}}\Lie_t \frak{A}_{a_1\cdots a_{p-1}}
    - \mb{H}_\sss{\text{M}} )\\
    &= \int \mb{d}t \biggl( \int_{\varSigma} \bigr(
    \mb{P}^{a_1\cdots a_{p-1}}\Lie_t \frak{A}_{a_1\cdots a_{p-1}}
    - N \mb{\mathcal{H}}_\sss{\text{M}}
    - N^a (\mb{\mathcal{H}}_\sss{\text{M}})_a \\
    &\qquad - t^b \frak{A}_{ba_1\cdots a_{p-2}}
    \mb{\mathcal{G}}^{a_1\cdots a_{p-2}} \bigr)
    -\int_{\mathcal{B}} ( N \frak{V}_{a_1\cdots a_{p-2}}
    \mb{\frak{Q}}^{a_1\cdots a_{p-2}}-N^a \mb{\frak{J}}_a )\biggr)\:.
  \end{split}
  \label{canonical form of matter action}
\end{equation}
The on-shell value of the matter action, therefore, is proportional
to the negative of the on-shell value of the Hamiltonian for a
stationary solution.

\section[Conserved Charges Arising from the Matter Action]%
  {Conserved Charges Arising from the\\ Matter Action}
\label{s:matter charge}

The presence of a Gauss law constraint in the Hamiltonian indicates
that one should be able to construct a charge on the quasilocal
surface.  To obtain a conserved quantity, consider the
equation of motion for the gauge field:
\begin{equation}
  \frac{1}{(p-1)!}\, \nabla_b \bigl( W(\phi)
  \frak{F}^{ba_1\cdots c_{p-1}} \bigr) =
  \frak{j}^{a_1\cdots a_{p-1}}
  \label{matter eom with source}
\end{equation}
where the $(p-1)$-form~$\mb{\frak{j}}$ represents a source for the
gauge field.  The source is divergenceless as can be seen from
equation~\eqref{matter eom with source}.
Furthermore, for any $(p-3)$-form~$\mb{\frak{w}}$, one has
\begin{equation}
  0 = \nabla_b ( \frak{j}^{bca_1\cdots a_{p-3}} \nabla_c
  \frak{w}_{a_1\cdots a_{p-3}} )
  \label{divergenceless matter source}
\end{equation}
and, thus, the
quantity~$\mb{\frak{q}}[\frak{w}]=
\frak{q}[\frak{w}]\cdot\mb{\epsilon}$ with
\begin{equation}
  \frak{q}^a[\frak{w}] = (p-1) \frak{j}^{abc_1\cdots c_{p-3}}
  \nabla_b \frak{w}_{c_1\cdots c_{p-3}}
  \label{definition of conserved charge density}
\end{equation}
is closed:
\begin{equation}
  0 = \int_{\mathcal{M}}\mb{d\frak{q}} =
  \int_{\varSigma_{\text{i}}}^{\varSigma_{\text{f}}}
  \underline{\mb{\frak{q}}}[\frak{w}] + \int_{\mathcal{T}}
  \overline{\mb{\frak{q}}}[\frak{w}] \:.
  \label{conservation of matter charge}
\end{equation}
The last term in the above equation is proportional to the normal
component of the matter source on the boundary~$\mathcal{T}$.
I assume that this term will vanish.  This will certainly
be the case if the boundary~$\mathcal{T}$ is chosen to be outside of
the matter source distribution.  When this is the case, the charge
\begin{equation}
  \int_{\varSigma_{\text{i}}} \underline{\mb{\frak{q}}}[\frak{w}] =
  \int_{\varSigma_{\text{f}}} \underline{\mb{\frak{q}}}[\frak{w}] = 
  \frak{Q}[\frak{w}]
  \label{definition of conserved charge}
\end{equation}
is conserved because its value does not depend on the
slice of the foliation on which it is calculated.

One can evaluate the conserved charge using its definition,
equation~\eqref{definition of conserved charge}, along with
equation~\eqref{definition of conserved charge density} and the
equation of motion~\eqref{matter eom with source}:
\begin{equation}
  \begin{split}
    \frak{Q}[\frak{w}] &=  \frac{1}{(p-2)!}\, \int_{\varSigma}
    \underline{\mb{\epsilon}} u_a \nabla_b \bigl( W(\phi)
    \frak{F}^{abcd_1\cdots d_{p-3}} \bigr) \nabla_c
    \frak{w}_{d_1\cdots d_{p-3}} \\
    &= \frac{1}{(p-2)!}\, \int_{\varSigma}\underline{\mb{\epsilon}}
    \nablas_b \bigl( W(\phi) \frak{E}^{bcd_1\cdots d_{p-3}} \bigr)
    \nabla_c \frak{w}_{d_1\cdots d_{p-3}} \bigr) \\
    &= \int_{\mathcal{B}} \mb{\frak{Q}}^{ab_1\cdots b_{p-3}} \nabla_a
    \frak{w}_{b_1\cdots b_{p-3}} \:.
  \end{split}
  \label{conserved matter charge}
\end{equation}
In obtaining the second equality, I have used the fact
that~$u_a=-N\partial_at$ as well as the antisymmetry of the
gauge field~$\mb{\frak{F}}$ to pull the~$\partial_at$ part
inside the covariant
derivative.  The lapse function can also be pulled inside, but
it converts the covariant derivative associated with
the metric~$g_{ab}$ into a covariant derivative associated with
the metric~$h_{ab}$.  The antisymmetric derivative of $\mb{\frak{w}}$
is similarly pulled inside the covariant derivative and one is
left with a total divergence.  Thus, one obtains the third equality.

Note that the second line of equation~\eqref{conserved matter charge}
can be re-written as
\begin{equation}
  \frak{Q}[\frak{w}] = - \int_{\varSigma}
  \mb{\mathcal{G}}^{ab_1\cdots b_{p-3}} \nabla_a
  \frak{w}_{b_1\cdots b_{p-3}} \,,
  \label{conserved charge from Gauss law}
\end{equation}
which shows that the conserved charge is directly associated
with the Gauss constraint.  Furthermore,
equation~\eqref{conserved matter charge} justifies the interpretation
of the quantity~$\mb{\frak{Q}}^{a_1\cdots a_{p-2}}$ as a quasilocal
surface charge density.  Notice that for every closed $(p-2)$-form,
there is an associated charge.  (I have written such a form as
the exact form~$\mb{d\frak{w}}$ above.)

One can also find the contribution to the Noether charge associated
with the diffeomorphism covariance of the matter Lagrangian density.
First, define the contribution to the Noether current from
the matter Lagrangian associated with diffeomorphism generated by
the vector field~$\xi^a$ as
\begin{equation}
  \mb{j}_{\text{extra}}[\xi] = \mb{\varrho}_\xi - \xi \cdot
  \mb{L}_\sss{\text{M}}
  \label{definition of matter Noether current}
\end{equation}
where the variations have been replaced with Lie derivatives
along~$\xi^a$ in~$\mb{\varrho}_\xi$.  With the aid of
equation~\eqref{matter boundary term}, one can evaluate
equation~\eqref{definition of matter Noether current}:
\begin{equation}
  \begin{split}
    j_{\text{extra}}{}^a[\xi] &= - \xi_b T^{ab} - (p-1)
    (E_{\frak{A}})^{ac_1\cdots c_{p-2}} \xi^b
    \frak{A}_{bc_1\cdots c_{p-2}}\\
    &\qquad - \frac{1}{(p-2)!} \nabla_b \bigl(
    W(\phi) \frak{F}^{abc_1\cdots c_{p-2}} \xi^d
    \frak{A}_{dc_1\cdots c_{p-2}} \bigr)
  \end{split}
  \label{matter Noether current}
\end{equation}
where $\mb{j}_{\text{extra}}=j_{\text{extra}}\cdot\mb{\epsilon}$,
$\mb{T}^{ab} = \mb{\epsilon}T^{ab}$
and~$(\mb{E}_{\frak{A}})^{a_1\cdots a_{p-1}}=\mb{\epsilon}
(E_{\frak{A}})^{a_1\cdots a_{p-1}}$.  When the sourceless matter
equation of motion holds, the second term in
equation~\eqref{matter Noether current} vanishes.  When the metric
equation of motion holds, the first term of
equation~\eqref{matter Noether current} cancels the last term of
equation~\eqref{Noether current}, and the total Noether current
becomes a total divergence.  The matter contribution to the
Noether current when the equations of motion hold is
\begin{equation}
  \mb{j}_{\text{extra}}[\xi] = \mb{d}
  \mb{q}_{\text{extra}}[\xi]
  = \mb{d} ( \mb{\frak{Q}}^{a_1\cdots a_{p-2}}
  \xi^b \frak{A}_{ba_1\cdots a_{p-2}} )\,,
  \label{on-shell matter Noether current}
\end{equation}
and the total Noether charge density on a closed
$(n-2)$-dimensional hypersurface is given by
equation~\eqref{Noether charge density} with the contribution
from equation~\eqref{on-shell matter Noether current}:
\begin{equation}
  \mb{q}[\xi] = \underline{\overline{\mb{\epsilon}}}\, n^{ab}
  \bigl( 2\xi_a\nabla_b D(\phi) + D(\phi)\nabla_a \xi_b \bigr)
  + \mb{\frak{Q}}^{a_1\cdots a_{p-2}} \xi^b
  \frak{A}_{ba_1\cdots a_{p-2}} \:.
  \label{total Noether charge density}
\end{equation}

\section{Matter Contributions to the Thermodynamics}
\label{s:matter thermo}

Now I obtain the contributions of the matter fields to the
thermodynamics and to the entropy of the system.  I assume that
the system contains a stationary black hole.  Once again I define
the microcanonical action by
equation~\eqref{definition of microcanonical action}, but now I
include the extra piece
\begin{equation}
  I_{\text{extra}} = \int_{\mathcal{M}} \mb{L}_\sss{\text{M}}
  - \int_{\mathcal{T}} \mb{d}t \wedge\mb{q}_{\text{extra}}[t]\:.
  \label{definition of extra piece of microcanonical action}
\end{equation}
Under variations between classical solutions, one finds that the
$\mathcal{T}$-boundary contributions to the extra piece of the
microcanonical action are given by
\begin{equation}
  \begin{split}
     \delta I_{\text{extra}} &= \int_{\mathcal{T}} ( -\mb{d}t\wedge
     \mb{q}_{\text{extra}}[t] + \overline{\mb{\varrho}} ) \\
     &= \int \mb{d}t \int_{\mathcal{B}} (
     N\frak{V}_{a_1\cdots a_{p-2}}\,
     \delta\mb{\frak{Q}}^{a_1\cdots a_{p-2}} - N^a\,
     \delta\mb{\frak{J}}_a\\
     &\qquad + N \mb{\frak{K}}^{a_1\cdots a_{p-1}}\,
     \delta \frak{W}_{a_1\cdots a_{p-1}})
  \end{split}
  \label{variation of extra piece of microcanonical action}
\end{equation}
where equation~\eqref{pull-back of boundary term onto T} and
equation~\eqref{decomposition of timelike matter momentum} have been
used to evaluate the~$\overline{\mb{\varrho}}$ term, and
equation~\eqref{on-shell matter Noether current} has been used to find
\begin{equation}
  \mb{q}_{\text{extra}}[t] = N\frak{V}_{a_1\cdots a_{p-2}}
  \mb{\frak{Q}}^{a_1\cdots a_{p-2}} - N^a \mb{\frak{J}}_a
  \label{extra Noether charge on quasilocal surface}
\end{equation}
on the boundary~$\mathcal{T}$.  Notice that in
equation~\eqref{variation of extra piece of microcanonical action},
the extensive variables are varied on the boundary.
Thus, the extra
piece~$I_{\text{extra}}$ is just the form of the matter action
that should be added to the microcanonical action of
equation~\eqref{definition of microcanonical action} in order
to get a microcanonical action containing both the matter and
gravity sectors.

The determination of the entropy and the first law of thermodynamics
proceeds exactly as in the last section of the last chapter.
The additional contribution to the entropy of the black hole
from matter fields comes from the extra piece of the Noether
charge (see equation~\eqref{definition of entropy}), which is
proportional to~$t\cdot\mb{\frak{A}}$ evaluated on the event horizon.
One should use part of the gauge freedom to make this
factor vanish; otherwise, free-falling observers will encounter
a diverging potential~$\mb{\frak{V}}$ in their orthonormal
frame~\cite{gh:77}.  Thus, the matter does not explicitly contribute
to the entropy, which is still given by equation~\eqref{entropy}.
The first law of thermodynamics,
equation~\eqref{first law of thermodynamics}, will pick up extra terms
from the Euclideanization of
equation~\eqref{variation of extra piece of microcanonical action}.
One finds that the first law of thermodynamics with matter fields
present is
\begin{equation}
  \begin{split}
    \delta S &\approx \int_{\mathcal{B}} \beta (
    \delta\mb{\mathcal{E}} - \omega^a\,
    \delta\mb{\tilde{\mathcal{J}}}_a
    + \mb{\mathcal{S}}^{ab}\,\delta\sigma_{ab}
    + \mb{\mu}\,\delta\phi \\
    &\qquad + \frak{V}_{a_1\cdots a_{p-2}}\,
    \delta\mb{\frak{Q}}^{a_1\cdots a_{p-2}} 
    + \mb{\frak{K}}^{a_1\cdots a_{p-1}} \,\delta
    \frak{W}_{a_1\cdots a_{p-1}} )\:.
  \end{split}
  \label{first law of thermodynamics with matter}
\end{equation}
Here~$\mb{\tilde{\mathcal{J}}}_a=\mb{\mathcal{J}}_a
+\mb{\frak{J}}_a$ is the effective quasilocal surface momentum density
that includes the \acro{emf} term.

\section{Special Cases}
\label{s:matter cases}

In this section, I examine some specific types of matter that
are encompassed in the wide class that has been considered.
The special cases are simply particular choices
of the value of~$p$ where the field strength is a~$p$-form.

\begin{enumerate}
\renewcommand{\theenumi}{\Roman{enumi}}
\renewcommand{\labelenumi}{Case~\theenumi.}
\item $p=1$:  In this case, one has~$\mb{\frak{F}}=\mb{d}\varPhi$,
  where I have written~$\varPhi$ as the scalar potential rather
  than~$\frak{A}$.  The Lagrangian density is proportional
  to~$(\nabla\varPhi)^2$, so this case corresponds to a massless
  scalar field with dilaton coupling.  Notice that there is no gauge
  invariance, so one will not have a Gauss law or a conserved charge
  for the scalar field.  The stress energy tensor and dilaton source
  are just as expected for a massless scalar field:
  \begin{equation}
    T_{ab} = W(\phi) \bigl( (\nabla_a\varPhi)(\nabla_b\varPhi)
    - \half(\nabla\varPhi)^2 \bigr)
    \label{stress energy for massless scalar}
  \end{equation}
  and
  \begin{equation}
    U = \frac{dW}{d\phi}\,(\nabla\varPhi)^2 \:.
    \label{dilaton source for massless scalar}
  \end{equation}
  The sourceless equation of motion for the massless scalar field is
  \begin{equation}
    (E_\varPhi) = \nabla_a\bigl(W(\phi)\nabla^a\varPhi\bigr)=0 \:.
    \label{eom for massless scalar}
  \end{equation}

  On the timelike boundary, the momentum conjugate to the scalar
  field is
  \begin{equation}
    \mb{\varPi} = -\overline{\mb{\epsilon}}\,W(\phi)n^a\nabla_a
    \varPhi\,,
    \label{momentum conjugate to scalar field on T}
  \end{equation}
  and one finds
  that~$\underline{\mb{\varPi}}\,\delta\varPhi=
  \mb{\frak{K}}\,\delta\varPhi$ where
  \begin{equation}
    \mb{\frak{K}}=-\underline{\overline{\mb{\epsilon}}}\,W(\phi)
    n^a\nabla_a\varPhi
    \label{potential conjugate to scalar field}
  \end{equation}
  is the potential on the quasilocal surface conjugate to the
  scalar field; the work term appearing in the first law of
  thermodynamics will be~$\mb{\frak{K}}\,\delta\varPhi$.  Notice that
  there are no charge or \acro{emf} densities.

  On the spacelike boundary,
  \begin{equation}
    \mb{P} = \underline{\mb{\epsilon}}\,W(\phi)
    \overset{\circ}{\varPhi}
    \label{momentum conjugate to scalar field on S}
  \end{equation}
  is the momentum conjugate to the scalar field, and
  $\overset{\circ}{\varPhi}=u\cdot\mb{d}\varPhi$~is just the electric
  field scalar.  One can evaluate the Hamiltonian and momentum
  constraints:
  \begin{equation}
    \mathcal{H}_\sss{\text{M}} =
    \frac{W(\phi)}{2}\,\bigl( (\nablas\varPhi)^2
    + \overset{\circ}{\varPhi}{}^2 \bigr)
    \label{Hamiltonian constraint for massless scalar field}
  \end{equation}
  and
  \begin{equation}
    (\mathcal{H}_\sss{\text{M}})_a =
    W(\phi)\overset{\circ}{\varPhi}\nablas_a\varPhi
    \label{momentum constraint for massless scalar field}
  \end{equation}
  respectively.  Notice that there is no Gauss constraint.

\item $p=2$: In this case, the field strength is a two-form, and the
  potential is a one-form.  Thus, this case is just a generalization
  of electromagnetism to an $n$-dimensional spacetime with a dilaton
  coupling.  The stress energy tensor and dilaton source are
  \begin{equation}
    T_{ab} = W(\phi) ( \frak{F}_{ac} \frak{F}_b{}^c - \fourth
    g_{ab} \frak{F}^{cd}\frak{F}_{cd} )
    \label{electromagnetic stress tensor}
  \end{equation}
  and
  \begin{equation}
    U = \frac{1}{2}\,\frac{dW}{d\phi}\, \frak{F}^{ab}\frak{F}_{ab}
    \label{electromagnetic dilaton source}
  \end{equation}
  respectively, and the sourceless equation of motion for the
  electromagnetic field is given by
  \begin{equation}
    (E_{\frak{A}})^a=\nabla_b\bigl(W(\phi)\frak{F}^{ba}\bigr)=0\:.
    \label{electromagnetic eom}
  \end{equation}

  The momentum conjugate to the electromagnetic potential on the
  timelike boundary~$\mathcal{T}$
  is~$\mb{\varPi}^a=
  -\overline{\mb{\epsilon}}\,W(\phi)n_c\frak{F}^{ca}$
  and the quasilocal surface charge density, \acro{emf} density,
  and current are
  \begin{equation}
    \mb{\frak{Q}} = \underline{\overline{\mb{\epsilon}}}\,W(\phi)
    n_a \frak{E}^a\,,
    \label{electromagnetic charge density}
  \end{equation}
  \begin{equation}
    \mb{\frak{J}}_a = -\mb{\frak{Q}}\frak{W}_a\,,
    \label{electromagnetic emf density}
  \end{equation}
  and
  \begin{equation}
    \mb{\frak{K}}^a = \underline{\overline{\mb{\epsilon}}}\,W(\phi)
    \sigma^a{}_b n_c \frak{F}^{bc}
    \label{electromagnetic current density}
  \end{equation}
  respectively.  All three of these contribute work terms in the
  first law of thermodynamics.

  On the spacelike boundary~$\varSigma$, the momentum
  $\mb{P}^a=\underline{\mb{\epsilon}}\,W(\phi)\frak{E}^a$~is
  proportional to the electric field strength.  The contributions
  to the Hamiltonian and momentum constrains are given by
  \begin{equation}
    \mathcal{H}_\sss{\text{M}} = W(\phi) \biggl( \frac{1}{2(n-3)!}\,
    \frak{B}^{a_1\cdots a_{n-3}} \frak{B}_{a_1\cdots a_{n-3}}
    + \frac{1}{2} \frak{E}^a \frak{E}_a \biggr)
    \label{electromagnetic Hamiltonian constraint}
  \end{equation}
  and
  \begin{equation}
    (\mathcal{H}_\sss{\text{M}})_a = W(\phi)\frak{F}_{ab}\frak{E}^b\:.
    \label{electromagnetic momentum constraint}
  \end{equation}
  In addition, there is a Gauss constraint
  \begin{equation}
    \mathcal{G} = -\nablas_a \bigl( W(\phi) \frak{E}^a \bigr)
    \label{electromagnetic Gauss constraint}
  \end{equation}
  conjugate to the Lagrange multiplier~$\frak{A}_t$.  By integrating
  this Gauss constraint over the spacelike hypersurface~$\varSigma$,
  one obtains a conserved quantity on the quasilocal boundary:
  \begin{equation}
    \frak{Q} = \int_{\mathcal{B}} \mb{\frak{Q}} \:.
    \label{conserved electromagnetic charge}
  \end{equation}

\item $p=n$: Because the field strength is an $n$-form, one can define
  a scalar field, $\varLambda$, that is related to the field strength
  by~$\varLambda=-(\ast\mb{\frak{F}})^2/4$.  The Lagrangian density
  can be written in terms of the scalar field.  One finds that
  $\mb{L}=-2\mb{\epsilon}\,W(\phi)\varLambda$.  Thus, when the
  function~$W$ is constant, the scalar
  field~$\varLambda$ assumes the r{\^o}le of a cosmological constant.
  The stress energy tensor is found to
  be~$T_{ab}=2W(\phi)\varLambda g_{ab}$, which is just what one would
  expect for a cosmological constant except for the extra function of
  the dilaton.

  The $n$-form field gives rise to a Gauss constraint in the
  Hamiltonian:
  \begin{equation}
    \mathcal{G}^{a_1\cdots a_{n-2}} = \frac{1}{(n-2)!}\,\nablas_b
    \bigl( W(\phi) (\ast\mb{\frak{F}})
    \underline{\epsilon}^{ba_1\cdots a_{n-2}} \bigr) \:.
    \label{cosmological constant Gauss constraint}
  \end{equation}
  When the Gauss law constraint holds, the quantity
  $\lambda=-\bigl(W(\phi)(\ast\mb{\frak{F}})/2\bigr)^2$ must be
  constant.  One can contract
  equation~\eqref{cosmological constant Gauss constraint} with
  the volume form on the $(n-2)$-dimensional quasilocal surface
  and integrate it over the spacelike hypersurface.  The result is
  the conserved charge,
  \begin{equation}
    \frak{Q}[\underline{\overline{\epsilon}}]=2\sqrt{|\lambda|}\,A\,,
    \label{conserved cosmological constant charge}
  \end{equation}
  when the Gauss constraint holds.  Here, $A$~is the area of the
  quasilocal surface with volume
  form~$\underline{\overline{\mb{\epsilon}}}$;  the constant~$\lambda$
  is related to the charge per unit area on the quasilocal surface:
  $\lambda=-\fourth(Q/A)^2$.  The Lagrangian density
  can be written in terms of~$\lambda$ as
  $\mb{L}=-2\mb{\epsilon}\,w(\phi)\lambda$ where~$w(\phi)=1/W(\phi)$.
  Thus, when the dilaton coupling is constant, the field~$\lambda$
  represents a cosmological constant.  Note, however, that
  $\lambda$~is constant only when the Gauss constraint holds, that is,
  $\lambda$~is a constant of integration rather than a ``physical''
  constant.
\end{enumerate}

\section{Yang-Mills Matter Fields}
\label{s:matter Yang-Mills}

I now consider non-Abelian Yang-Mills gauge fields with an arbitrary
gauge group~$\frak{G}$.  Such fields have been considered in
reference~\cite{cm:95a}.  The quasilocal analysis of Yang-Mills
fields resembles the analysis of electromagnetic
fields, which are the special case of an Abelian~U(1) gauge group;
the conserved charges of the Yang-Mills fields are more complex
because of the fact that Yang-Mills field lines carry charge and,
thus, the charge of a source is gauge-dependent.  The first law of
thermodynamics, however, will remain gauge-independent.

In the following, I denote internal gauge space indices with
Fraktur characters, and I assume the adjoint representation.
The gauge covariant derivative (which is both gauge covariant and
compatible with the metric~$g_{ab}$) is given
by~$(\frak{D}_a)^{\frak{a}}{}_{\frak{b}}=
\nabla_a\delta^{\frak{a}}{}_{\frak{b}}+
\frak{f}^{\frak{a}}{}_{\frak{bc}}\frak{A}_a{}^{\frak{c}}$ where
$\frak{A}_a{}^{\frak{a}}$ is the connection
and~$\frak{f}^{\frak{a}}{}_{\frak{bc}}$ are the structure constants
of the group.  The curvature of this gauge covariant derivative is
the field strength tensor:
$\frak{F}_{ab}{}^{\frak{a}}=2\nabla_{[a}\frak{A}_{b]}{}^{\frak{a}}+
\frak{f}^{\frak{a}}{}_{\frak{bc}}
\frak{A}_a{}^{\frak{b}}\frak{A}_b{}^{\frak{c}}$.  This field strength
is covariant under gauge transformations of the
form~$\frak{A}_a{}^{\frak{a}}\to\frak{A}_a{}^{\frak{a}}+
(\frak{D}_a)^{\frak{a}}{}_{\frak{b}}\frak{x}^{\frak{b}}$
where~$\frak{x}^{\frak{a}}$ is a Lie-algebra-valued scalar field
on~$\mathcal{M}$.  The internal gauge indices are raised and
lowered by means of the Killing
metric~$\frak{g}_{\frak{ab}}=
\half\frak{f}^{\frak{c}}{}_{\frak{da}}
\frak{f}^{\frak{d}}{}_{\frak{cb}}$.  The field strength can be
decomposed into electric and magnetic pieces on a spacelike
hypersurface~$\varSigma$ as before; the gauge covariant derivative
compatible with the metric~$h_{ab}$
is~$(\frak{d}_a)^\frak{a}{}_{\frak{b}}$.

The matter action for the Yang-Mills field with coupling to a
dilaton is
\begin{equation}
  I_\sss{\text{M}} = -\int_{\mathcal{M}}\mb{\epsilon}\,\fourth
  W(\phi)\frak{F}^{ab}{}_{\frak{a}}\frak{F}_{ab}{}^{\frak{a}}\:.
  \label{Yang-Mills action}
\end{equation}
The quasilocal analysis proceeds in a similar manner to the analysis
of an electromagnetic field with dilaton coupling.  I find that
the matter provides gravitational source terms for the metric and
dilaton field equations; these source terms are
\begin{gather}
  T_{ab} = W(\phi) ( \frak{F}_{ac}{}^{\frak{a}}
  \frak{F}_b{}^c{}_{\frak{a}} - \fourth
  g_{ab} \frak{F}^{cd}{}_{\frak{a}}\frak{F}_{cd}{}^{\frak{a}} )
  \label{Yang-Mills stress tensor}\\
  \intertext{and}
  U = \frac{1}{2}\,\frac{dW}{d\phi}\, \frak{F}^{ab}{}_{\frak{a}}
  \frak{F}_{ab}{}^{\frak{a}}
  \label{Yang-Mills dilaton source}
\end{gather}
respectively.  The Yang-Mills field equations are
\begin{equation}
  (E_{\frak{A}})^a{}_{\frak{a}}=(\frak{D}_b)^{\frak{b}}{}_{\frak{a}}
  \bigl(W(\phi)\frak{F}^{ba}{}_{\frak{b}}\bigr)=
  \frak{j}^a{}_{\frak{a}}
  \label{Yang-Mills eom}
\end{equation}
where $\frak{j}^a{}_{\frak{a}}$ is a source current.

On the~$\mathcal{T}$-boundary, the momentum conjugate to the
Yang-Mills potential
is~$\mb{\varPi}^a{}_{\frak{a}}=-\overline{\mb{\epsilon}}\,W(\phi)
n_c\frak{F}^{ca}{}_{\frak{a}}$; notice that this momentum is
Lie-algebra-valued.  The quasilocal surface charge and current
densities are also Lie-algebra-valued; they are
\begin{gather}
  \mb{\frak{Q}}_{\frak{a}} = \underline{\overline{\mb{\epsilon}}}\,
  W(\phi) n_a \frak{E}^a{}_{\frak{a}}
  \label{Yang-Mills charge density}\\
  \intertext{and}
  \mb{\frak{K}}^a{}_{\frak{a}}=\underline{\overline{\mb{\epsilon}}}\,
  W(\phi) \sigma^a{}_b n_c \frak{F}^{bc}{}_{\frak{a}}
  \label{Yang-Mills current density}
\end{gather}
respectively.  These densities contribute the work terms
$\frak{V}^{\frak{a}}\,\delta\mb{\frak{Q}}_{\frak{a}}$
and~$\mb{\frak{K}}^a{}_{\frak{a}}\,\delta\frak{W}_a{}^{\frak{a}}$
to the first law of thermodynamics.  The quasilocal surface \acro{emf}
density~$\mb{\frak{J}}_a = -\mb{\frak{Q}}_{\frak{a}}
\frak{W}_a{}^{\frak{a}}$, however, is a gauge scalar; it modifies
the gravitational angular momentum work term in the usual manner.

The Yang-Mills field provides the following contributions to the
Hamiltonian and momentum constraints:
\begin{gather}
  \mathcal{H}_\sss{\text{M}} = W(\phi) \biggl( \frac{1}{2(n-3)!}\,
  \frak{B}^{a_1\cdots a_{n-3}}{}_{\frak{a}}
  \frak{B}_{a_1\cdots a_{n-3}}{}^{\frak{a}} + \frac{1}{2}
  \frak{E}^a{}_{\frak{a}} \frak{E}_a{}^{\frak{a}} \biggr)
  \label{Yang-Mills Hamiltonian constraint}\\
  \intertext{and}
  (\mathcal{H}_\sss{\text{M}})_a = W(\phi)\frak{F}_{ab}{}^{\frak{a}}
  \frak{E}^b{}_{\frak{a}}\:.
  \label{Yang-Mills momentum constraint}
\end{gather}
respectively.  The Gauss constraint,
\begin{equation}
  \mathcal{G}_{\frak{a}} = -(\frak{d}_a)^{\frak{b}}{}_{\frak{a}}\bigl(
  W(\phi) \frak{E}^a{}_{\frak{b}} \bigr)\,,
  \label{Yang-Mills Gauss constraint}
\end{equation}
is conjugate to the Lagrange multiplier~$\frak{A}_t{}^{\frak{a}}$.
Because the Gauss constraint is Lie-algebra-valued, it is not
possible to generate a gauge independent charge in general.
This problem arises because the equations of motion for the
Yang-Mills field~\eqref{Yang-Mills eom} are gauge covariant rather
than gauge invariant, so the separation of the charge contained in
the source from the charge contained in the field is gauge dependent.

If a solution to the equation of motion for the Yang-Mills
field~\eqref{Yang-Mills eom} possesses certain properties, it
will be possible to construct a conserved charge~\cite{ad:82,it:95}.
Suppose that the solution possesses a gauge Killing
scalar~$\frak{k}^{\frak{a}}$, i.e., a Lie-algebra-valued scalar
field on~$\mathcal{M}$ that is covariantly constant:
$(\frak{D}_a)^{\frak{a}}{}_{\frak{b}}\frak{k}^{\frak{b}}=0$.  Then
the quantity~$\frak{k}^{\frak{a}}\frak{j}^a{}_{\frak{a}}$ is gauge
invariant and divergenceless:
$\nabla_a(\frak{k}^{\frak{a}}\frak{j}^a{}_{\frak{a}})=0$.  One then
finds that the charge
\begin{equation}
  \frak{Q}[\frak{k}]=\int_{\varSigma}\underline{\mb{\epsilon}}\,
  u_a\frak{k}^{\frak{a}}\frak{j}^a{}_{\frak{a}}
  =-\int_{\varSigma} \mb{\mathcal{G}}_{\frak{a}}\frak{k}^{\frak{a}}
  =\int_{\mathcal{B}} \mb{\frak{Q}}_{\frak{a}}\frak{k}^{\frak{a}}
  \label{conserved Yang-Mills charge}
\end{equation}
is conserved.  Therefore, the quantity~$\mb{\frak{Q}}_{\frak{a}}$
can be interpreted as a Yang-Mills charge density provided that
a gauge-fixing scalar~$\frak{k}^{\frak{a}}$ can be found.

\chapter{Black Hole Spacetimes in General Relativity}
\label{c:GR}

I now examine the thermodynamics of specific black
hole spacetimes using the formalism that I have derived.  In this
chapter, I consider two asymptotically anti-de\thinspace Sitter
({\ads}) black hole
spacetimes in General Relativity.  Because the quasilocal surface
is of finite size, the particular asymptotic fall-off conditions
do not affect the applicability of the methods I use.
The solutions I examine in this section are solutions to the
field equations of General Relativity: they do not contain a dilaton
field.  Thus, in this chapter, I take $D=(2\kappa)^{-1}$
where~$\kappa$ is the gravitational coupling constant, $H=0$ since
there is no dilaton kinetic energy, and~$V=-\varLambda/\kappa$
where~$\varLambda$ is the cosmological constant.\footnote{%
$\varLambda$ is a \emph{real} cosmological constant, not the constant
of integration discussed at the end of the previous chapter.}

\section[Reissner-Nordstr{\"o}m-Anti-de\thinspace Sitter Spacetimes]%
  {Reissner-Nordstr{\"o}m-Anti-de\thinspace Sitter\\ Spacetimes}
\label{s:GR RNADS}

The thermodynamics of the
Reissner-Nordstr{\"o}m-Anti-de\thinspace Sitter ({\rnads}) has been
studied recently by Louko and Winters-Hilt~\cite{lw:96}.  Their
motivation was to use the negative cosmological constant instead
of a finite box to cause the heat capacity of a black hole spacetime
to be positive.  Louko and Winters-Hilt analyzed the thermodynamics
using variables obtained from the Hamiltonian evaluated at spacelike
infinity, so they needed to be very careful about the {\ads} fall-off
conditions.  In this section, I obtain the thermodynamic
variables for the {\rnads} spacetime using the quasilocal method.

\begin{figure}[t]
\renewcommand{\baselinestretch}{1}\small
\begin{center}
\setlength{\unitlength}{0.75mm}
\begin{picture}(160,50)
\multiput(0,0)(80,0){2}{%
  \begin{picture}(80,50)(0,25)
  \put(20,50){\line(-1,1){20}}
  \put(20,50){\line(-1,-1){20}}
  \put(20,50){\line(1,1){20}}
  \put(20,50){\line(1,-1){20}}
  \put(60,50){\line(-1,1){20}}
  \put(60,50){\line(-1,-1){20}}
  \put(60,50){\line(1,1){20}}
  \put(60,50){\line(1,-1){20}}
  \put(40,70){\line(1,0){40}}
  \put(40,30){\line(1,0){40}}
  \multiput(2.22,70)(8.88,0){5}{\qbezier(-2.22,0)(0,2.22)(2.22,0)}
  \multiput(6.66,70)(8.88,0){4}{\qbezier(-2.22,0)(0,-2.22)(2.22,0)}
  \multiput(2.22,30)(8.88,0){5}{\qbezier(-2.22,0)(0,-2.22)(2.22,0)}
  \multiput(6.66,30)(8.88,0){4}{\qbezier(-2.22,0)(0,2.22)(2.22,0)}
  \put(20,73){\raisebox{\depth}{\makebox[0pt]{\hss$r=0$\hss}}}
  \put(20,27){\raisebox{-\height}{\makebox[0pt]{\hss$r=0$\hss}}}
  \put(60,73){\raisebox{\depth}{\makebox[0pt]{\hss$r=\infty$\hss}}}
  \put(60,27){\raisebox{-\height}{\makebox[0pt]{\hss$r=\infty$\hss}}}
  \put(60,65){\raisebox{-.5\height}{\makebox[0pt]{\hss I\hss}}}
  \put(60,35){\raisebox{-.5\height}{\makebox[0pt]{\hss I\hss}}}
  \put(40,50){\raisebox{-.5\height}{\makebox[0pt]{\hss II\hss}}}
  \put(0,50){\raisebox{-.5\height}{\makebox[0pt]{\hss II\hss}}}
  \put(20,65){\raisebox{-.5\height}{\makebox[0pt]{\hss III\hss}}}
  \put(20,35){\raisebox{-.5\height}{\makebox[0pt]{\hss III\hss}}}
  \put(10,60){\makebox[0pt]{\hss$r=r_-$\hss}}
  \put(10,40){\makebox[0pt]{\hss$r=r_-$\hss}}
  \put(30,60){\makebox[0pt]{\hss$r=r_-$\hss}}
  \put(30,40){\makebox[0pt]{\hss$r=r_-$\hss}}
  \put(50,60){\makebox[0pt]{\hss$r=r_+$\hss}}
  \put(50,40){\makebox[0pt]{\hss$r=r_+$\hss}}
  \put(70,60){\makebox[0pt]{\hss$r=r_+$\hss}}
  \put(70,40){\makebox[0pt]{\hss$r=r_+$\hss}}
  \end{picture}}
\end{picture}
\end{center}
\begin{quote}\leavevmode
\caption[The extended {\rnads} spacetime]{\small The extended,
  non-extremal {\rnads} spacetime~\cite{l:79}:
  there are two horizons, $r_\pm$ with~$r_+>r_-$,
  and a curvature singularity at~$r=0$.  Notice that~$r=\infty$ is
  timelike.  There are three regions: the outer region~I has~$r>r_+$,
  the intermediate region~II has~$r_-<r<r_+$, and the inner region~III
  has~$r<r_-$.  In this diagram, causal future is to the right while
  causal past is to the left.}
\label{fig:RNADS}
\end{quote}
\end{figure}
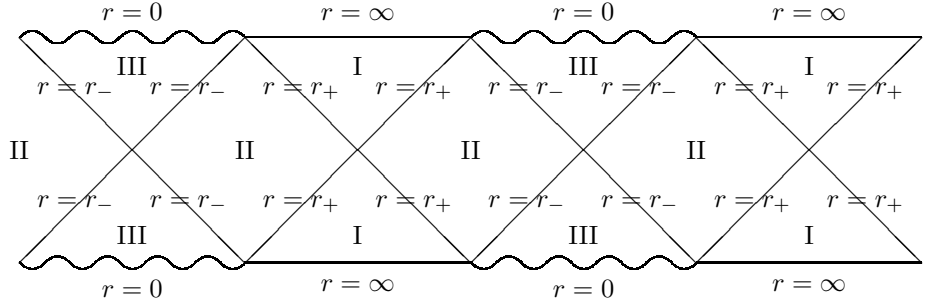

The Lagrangian density for the Einstein-Maxwell theory with a negative
cosmological constant is given by
\begin{equation}
  \mb{L} = \frac{\mb{\epsilon}}{16\pi} ( R[g] + 6\ell^{-2}
  - \frak{F}^{ab} \frak{F}_{ab} )
  \label{Einstein Maxwell Anti-de Sitter Lagrangian}
\end{equation}
where the cosmological constant
is~$\varLambda=-3/\ell^2$ with~$\ell$ being a positive
constant.  The gravitational coupling constant is~$\kappa=8\pi$, while
the Maxwell field has a coupling constant~$W=(4\pi)^{-1}$.  The
spherically symmetric and static black hole solution in four
dimensions has the line element
\begin{equation}
  ds^2 = -N^2(r)\,dt^2 + \frac{dr^2}{f^2(r)} + r^2\,d\omega^2
  \label{sss metric}
\end{equation}
where $d\omega^2$ is the line element on a two-sphere and the metric
components in the $t$-$r$ sector are
\begin{equation}
  N^2(r) = f^2(r) = \frac{r^2}{\ell^2} + 1 - \frac{2m}{r}
  + \frac{q^2}{r^2} \:.
  \label{RNAdS lapse}
\end{equation}
In addition, the electric {\rnads} solution has the electromagnetic
field strength
\begin{equation}
  \mb{\frak{F}} = \frac{q}{r^2}\,\mb{d}t \wedge \mb{d}r \:.
  \label{electric RNAdS Maxwell field}
\end{equation}
The parameters $m$ and~$q$ are constants of integration, and I
require that $m>m_{\text{crit}}$ with
\begin{equation}
  m_{\text{crit}} = \frac{\ell}{3\sqrt{6}}\,\Bigl(
  \sqrt{1+12(q/\ell)^2} + 2 \Bigr) \Bigl( \sqrt{1+12(q/\ell)^2}
  -1 \Bigr)^{1/2} \:.
  \label{RNADS critical mass}
\end{equation}
The case $m=m_{\text{crit}}$ represents an extremal black hole.
I am not interested in such black holes since they should
not represent physically attainable spacetimes according to the third
law of thermodynamics.  The lapse function will have two positive
roots, $r_\pm$ with~$r_+>r_-$, when~$q\ne0$---these are coordinate
singularities.  The event horizon is
at a radius equal to the larger of these two roots: 
$r_\sss{\text{H}}=r_+$;  when~$q=0$, there will only be a single
positive root of the lapse function which will be taken as the
event horizon radius.  There is a curvature singularity at~$r=0$.
A diagram showing the extended {\rnads} spacetime is given in
figure~\ref{fig:RNADS}.

For the spherically symmetric and static (\acro{sss}) metric of
equation~\eqref{sss metric}, the quasilocal quantities can be readily
evaluated.  I take advantage of the spherical symmetry by
choosing the quasilocal surface to be a surface of constant
radius~$r_\sss{\text{B}}>r_\sss{\text{H}}$ and constant time~$t$.
The quasilocal energy is given by equation~\eqref{quasilocal energy}.
With the aid of equation~\eqref{quasilocal energy density}, one finds
that the quasilocal energy is given by
\begin{equation}
  E = - \frac{f(r_\sss{\text{B}})}{\kappa}\, \frac{d}{dr}
  \bigl[ A_{n-2}(r) \bigr]_{r=r_\sss{\text{B}}} - E_\sss0
  \label{sss GR quasilocal energy}
\end{equation}
for a \acro{sss} solution to the field equation of General Relativity.
Here, $A_m(r)$ is the area of an $m$-sphere:
\begin{equation}
  A_m(r) = \frac{(4\pi)^{m/2}\varGamma(m/2)}{\varGamma(m)}\,r^m \:.
  \label{area of m-sphere}
\end{equation}
Because of the spherical symmetry, the quasilocal momentum density
vanishes.  Also, I only consider the variations in the metric
that preserve spherical symmetry, so I need only consider the
surface tension part of the quasilocal surface stress density.
The surface tension is evaluated from
equation~\eqref{surface tension}:
\begin{equation}
  \mathcal{S} = \frac{f(r_\sss{\text{B}})}{\kappa}\, \biggl(
  \frac{d}{dr} \bigl[ \log N(r) \bigr]_{r=r_\sss{\text{B}}}
  + \frac{(n-3)}{r_\sss{\text{B}}} \biggr) - \mathcal{S}_\sss0 \:.
  \label{sss GR surface tension}
\end{equation}
The entropy of the black hole is evaluated from
equation~\eqref{entropy} and is found to be
\begin{equation}
  S = \frac{1}{4}\,A_{n-2}(r_\sss{\text{H}})
  \label{sss GR entropy}
\end{equation}
while the temperature is found to be
\begin{equation}
  T = \beta^{-1} = \frac{1}{N(r_\sss{\text{B}})}\,
  \frac{\varkappa_\sss{\text{H}}}{2\pi}
  \label{stationary temperature}
\end{equation}
from equation~\eqref{temperature and velocity}.  I also need to
calculate the conserved electromagnetic charge and the electromagnetic
potential.  Because of the \acro{sss} symmetries, the \acro{emf}
density and the surface current densities will vanish.
Taking~$W=2/\kappa$, one finds that the conserved charge is
\begin{equation}
  \frak{Q} = \frac{1}{\kappa}\,A_{n-2}(r_\sss{\text{B}})
  \bigl[ n^{ab} \frak{F}_{ab} \bigr]_{r=r_\sss{\text{B}}}
  \label{sss Maxwell charge}
\end{equation}
where equations \eqref{conserved electromagnetic charge}
and~\eqref{electromagnetic charge density} have been used.
Here, $n^{ab}$ is the bi-normal~$2u^{[a}n^{b]}$ to the quasilocal
surface;
$n^{ab}\frak{F}_{ab}=2(f/N)\frak{F}_{tr}$.  The scalar potential
is~$\frak{V}=u\cdot\mb{\frak{A}}=N^{-1}\frak{A}_t$ where
$\mb{\frak{A}}$~is the electromagnetic potential in a gauge for
which the scalar potential is finite on the event horizon.
Since~$\frak{F}_{tr}=-\partial_r\frak{A}_t$, one finds that
\begin{equation}
  \frak{V} = -\frac{1}{N(r_\sss{\text{B}})}\,
  \int_{r_\sss{\text{H}}}^{r_\sss{\text{B}}} \frak{F}_{tr}
  \label{sss electromagnetic potential}
\end{equation}
where the lower limit guarantees that~$\frak{V}$ will be finite
on the event horizon.

The \acro{sss} symmetries are very useful in evaluating the first
law of thermodynamics because they will allow one to convert the
integral form of this law,
equation~\eqref{first law of thermodynamics with matter},
into the differential form:
\begin{equation}
  T\, \delta S = \delta E + \mathcal{S}\, \delta A + \frak{V}\,
  \delta\frak{Q}
  \label{sss Einstein-Maxwell first law of thermodynamics}
\end{equation}
where $A$~is the area of the quasilocal boundary.
Equation~\eqref{sss Einstein-Maxwell first law of thermodynamics}
is the usual expression of the
first law of thermodynamics for a system with electromagnetic and
surface tension work terms.

Now I evaluate the thermodynamic variables for the {\rnads}
solution.  The quantities
\begin{gather}
  E = -r_\sss{\text{B}} f(r_\sss{\text{B}}) - E_\sss0\,,
  \label{RNADS quasilocal energy} \\
  \mathcal{S} = \frac{f(r_\sss{\text{B}})}{8\pi}\,\biggl(
  \frac{1}{r_\sss{\text{B}}} + \frac{d}{dr} \bigl[ \log N(r)
  \bigr]_{r=r_\sss{\text{B}}} \biggr) - \mathcal{S}_\sss0\,,
  \label{RNADS surface tension} \\
  S = \pi r_\sss{\text{H}}^2\,,
  \label{RNADS entropy} \\
  T = \frac{1}{N(r_\sss{\text{B}})}\,\frac{4(r_\sss{\text{H}}/\ell)^2
  + 2 + 2m/r_\sss{\text{H}} }{4\pi r_\sss{\text{H}}}\,,
  \label{RNADS temperature} \\
  \frak{Q}=q \,,
  \label{RNADS charge} \\
  \intertext{and}
  \frak{V} = \frac{q}{N(r_\sss{\text{B}})} \biggl(
  \frac{1}{r_\sss{\text{B}}} - \frac{1}{r_\sss{\text{H}}} \biggr)
  \label{RNADS electric potential}
\end{gather}
are the thermodynamic internal energy, surface tension, entropy,
temperature, charge, and electric potential respectively.
The quasilocal energy can be expressed in terms of the extensive
thermodynamic variables $S$, $A$, and~$\frak{Q}$.  To do so, one
must express the
parameter~$m$ as a function of these variables.  Recall that, on the
event horizon, the lapse function vanishes so
\begin{equation}
  \begin{split}
    2m(S,A,\frak{Q}) &= \frac{1}{r_\sss{\text{H}}}\, (
    r_\sss{\text{H}}^4/\ell^2 + r_\sss{\text{H}}^2 + q^2 ) \\
    &= \sqrt{\frac{\pi}{S}}\,\biggl( \frac{S^2}{\pi^2\ell^2}
    + \frac{S}{\pi} + \frak{Q}^2 \biggr)
  \end{split}
  \label{RNADS mass parameter}
\end{equation}
where equation~\eqref{RNADS entropy} was used.  Notice that the
parameter~$m$ does not depend on the size, $A$, of the quasilocal
system.  Now, one can explicitly show that
\begin{equation}
  T = \frac{\partial E}{\partial S}\,, \quad
  \mathcal{S} = -\frac{\partial E}{\partial A}\,, \quad\text{and}\quad
  \frak{V} = -\frac{\partial E}{\partial\frak{Q}}
  \label{RNADS intensive variables from extensive}
\end{equation}
confirming the first law of thermodynamics~\eqref{sss
Einstein-Maxwell first law of thermodynamics}.  Although I have
not yet specified a reference spacetime, the
relations~\eqref{RNADS intensive variables from extensive} will
hold for any choice of the reference spacetime because of
the functional relationship given in
equation~\eqref{reference functional relationship}.

I shall now examine the asymptotic behaviour of the thermodynamic
variables as $r_\sss{\text{B}}\to\infty$, or, equivalently, for
small values of~$u=1/r_\sss{\text{B}}$.  The entropy and the
electromagnetic charge are both independent of the size.  The
quasilocal energy has the following behaviour:
\begin{equation}
  E = \bigl\{ -\ell^{-1}u^{-2} - (\ell/2) + m\ell u + O(u^2) \bigr\}
  - E_\sss0 \:;
  \label{asymptotic RNADS quasilocal energy}
\end{equation}
unless~$E_\sss0$ is suitably chosen, the quasilocal energy will
diverge as~$r_\sss{\text{B}}\to\infty$.  The temperature and
scalar electromagnetic potential both behave
as~$O(u)$, so they will redshift to zero as the quasilocal
region grows:  this is expected in an {\ads} spacetime.
The surface tension has the
asymptotic behaviour
\begin{equation}
  \mathcal{S} = \frac{1}{8\pi}\,\bigl\{ 2\ell^{-1} + m\ell u^3
  + O(u^4) \bigr\} + \parenfrac{\partial E_\sss0}{\partial A} \:.
  \label{asymptotic RNADS surface tension}
\end{equation}
A natural choice for the reference spacetime is the usual {\ads}
spacetime, which can be obtained by setting the constants of
integration, $m$ and~$q$, to zero.  With this choice, one finds
\begin{equation}
  E_\sss0 = -r_\sss{\text{B}} \sqrt{(r_\sss{\text{B}}/\ell)^2 + 1}\:.
  \label{RNADS reference energy}
\end{equation}
The quasilocal energy, then, has the asymptotic
behaviour~$E=m\ell u+O(u^2)$ while the surface tension has
the behaviour~$\mathcal{S}=(8\pi)^{-1}m\ell u^3+O(u^4)$.
The quasilocal energy now vanishes
as~$r_\sss{\text{B}}\to\infty$;  however, the conserved mass on the
quasilocal surface, defined by equation~\eqref{mass}, is finite
and approaches the value~$m$ as~$r_\sss{\text{B}}\to\infty$.

The sourceless four-dimensional Einstein-Maxwell theory is invariant under the
duality rotation~$\mb{\frak{F}}\to\ast\mb{\frak{F}}$.  Thus, in
addition to the electric {\rnads} solution, there is also a magnetic
{\rnads} solution; the only difference being in the electromagnetic
field strength, which now takes the form
\begin{equation}
  \mb{\frak{F}} = -q\, \mb{d}\vartheta \wedge \sin\vartheta\,
  \mb{d}\varphi \:.
  \label{magnetic RNADS Maxwell field}
\end{equation}
Although the field equations are invariant under the duality rotation,
the quasilocal quantities relating to the electromagnetic field are not.
Nevertheless, one would expect that the
thermodynamics of the magnetic {\rnads} solution should be the same
as for the electric solution.

Since the only difference between the electric and the magnetic
{\rnads} solutions is in the form of the electromagnetic field strength,
the quasilocal energy, entropy, surface tension, and temperature
will all be the same for both solutions.  Thus, I need only compute
the quasilocal quantities arising from the electromagnetic part of
the action.  There is a difficulty, however, in evaluating the
electromagnetic work term.  It is not possible to produce the
electromagnetic field strength of
equation~\eqref{magnetic RNADS Maxwell field} with a non-singular
potential.  For example, the
potential~$\mb{\frak{A}}=-q(1-\cos\vartheta)\mb{d}\varphi$ possesses
a Dirac string singularity at~$\vartheta=\pi$.\footnote{%
The form~$\mb{d}\varphi$ is not defined for $\vartheta=0$
and~$\vartheta=\pi$ but, in the former position, the electromagnetic potential
vanishes so there is no singularity.}  This singularity
does not affect the calculation of the quasilocal quantities mentioned
above because they do not depend on the electromagnetic potential.
However, the quasilocal quantities arising from the electromagnetic
part of the action are affected because they involve a decomposition
of the action into projections of the potential~$\mb{\frak{A}}$.
I use the following trick to avoid the Dirac string singularity
in order to calculate the electromagnetic work term.
Let~$\frak{U}\,\delta\frak{P}$ denote the electromagnetic work term
for the magnetic solution with $\frak{P}$ being the magnetic charge
and~$\frak{U}$ the magnetic scalar potential; from
equation~\eqref{first law of thermodynamics with matter}, one has
\begin{equation}
  \beta\frak{U}\,\delta\frak{P} = \int_{\hat{\mathcal{T}}} N\mb{d}t
  \wedge \mb{\frak{K}}^a\,\delta\frak{W}_a
  \label{definition of RNADS magnetic work term}
\end{equation}
where the~$\frak{V}\,\delta\mb{\frak{Q}}$ term does not appear
because~$\mb{\frak{Q}}$ vanishes for purely spatial electromagnetic
field strength two-forms.  The \acro{emf} density also does not
appear because the magnetic solution is still a \acro{sss} solution.
To prevent the surface~$\hat{\mathcal{T}}$ from crossing the Dirac
singularity, I consider the Euclidean
manifold~$\check{\mathcal{M}}$ with
topology~$D^2\times S^2_\theta$ where $D^2$~is a two-dimensional disk
and $S^2_\theta$~is the fragment of a sphere
with~$0\le\vartheta\le\theta<\pi$.  The boundary
of~$\check{\mathcal{M}}$, $\check{\mathcal{T}}$, consists of the
(periodic) history of the sphere fragment~$S^2_\theta$ of
radius~$r_\sss{\text{B}}$, and the (periodic) history of the cone,
$C^2_\theta$, of constant~$\vartheta=\theta$ and radius running
from the event horizon~$r_\sss{\text{H}}$ to the
boundary~$r_\sss{\text{B}}$.  Because the solution is static,
\begin{equation}
  \beta(\frak{U}\,\delta\frak{P})\spcheck = \Delta\tau \biggl(
  \int_{S^2_\theta} N \mb{\frak{K}}^a\,\delta\frak{W}_a
  + \int_{C^2_\theta} N \mb{\frak{K}}^a\,\delta\frak{W}_a \biggr) \:.
  \label{definition of RNADS magnetic work on check manifold}
\end{equation}

Using equation~\eqref{electromagnetic current density}, one finds
that~$\mb{\frak{K}}^a=0$ on~$S^2_\theta$.  On~$C^2_\theta$, however,
one has the unit normal~$n^a=r^{-1}(\partial/\partial\vartheta)^a$;
the surface current density and surface potential, respectively, are
\begin{gather}
  \mb{\frak{K}}^\varphi = \underline{\overline{\mb{\epsilon}}}\,
  \frac{q}{4\pi r^3\sin\vartheta}
  \label{magnetic RNADS cone surface current density}\\
  \intertext{and}
  \frak{W}_\varphi = -q(1-\cos\vartheta) \:.
  \label{magnetic RNADS surface potential}
\end{gather}
I can now evaluate
equation~\eqref{definition of RNADS magnetic work on check manifold}:
\begin{equation}
  \beta(\frak{U}\,\delta\frak{P})\spcheck = \beta
  \parenfrac{1-\cos\theta}{2} \Biggl( \frac{q}{N(r_\sss{\text{B}})}
  \biggl( \frac{1}{r_\sss{\text{B}}} - \frac{1}{r_\sss{\text{H}}}
  \biggr) \Biggr) \,\delta q \:.
  \label{RNADS magnetic work on check manifold}
\end{equation}
I assume that~$\frak{U}\,\delta\frak{P}=\lim_{\theta\to\pi}(\frak{U}
\,\delta\frak{P})\spcheck$; one is then motivated to define
\begin{gather}
  \frak{P} = q
  \label{RNADS magnetic charge}\\
  \intertext{and}
  \frak{U} = \frac{q}{N(r_\sss{\text{B}})} \biggl(
  \frac{1}{r_\sss{\text{B}}} - \frac{1}{r_\sss{\text{H}}} \biggr) \:.
  \label{RNADS magnetic potential}
\end{gather}
Notice that equations \eqref{RNADS magnetic charge}
and~\eqref{RNADS magnetic potential} have the same form as
\eqref{RNADS charge} and~\eqref{RNADS electric potential}, so the
thermodynamics of the magnetic solution is no different from the
thermodynamics of the electric solution.

The heat capacity of a thermodynamic system is a useful indicator
of the stability of the system: if a system has a negative heat
capacity, then the temperature of the system will \emph{decrease}
as heat is added, and thus it cannot come to thermal equilibrium
with a system that initially has a higher temperature.  Similarly,
if a system with a negative heat capacity is in thermal contact
with a reservoir of a lower temperature, the temperature of the
system will \emph{increase} as heat is lost to the reservoir.
It is important to determine the conditions under which the thermal
equilibrium of a quasilocal system containing a black hole is stable.
In general, the heat capacity of a system at constant extensive
variables is computed via the formula
\begin{equation}
  C_{\text{extensive variables}} = T \,
  \parenfrac{\partial T}{\partial S}^{-1}_{\text{extensive variables}}
  \label{heat capacity}
\end{equation}
where the partial derivative is computed holding the extensive
variables fixed.  This heat capacity is appropriate for systems in
which the temperature and the extensive variables are fixed on the
surface, i.e., for a system in the canonical ensemble.

In the case of the {\rnads} spacetime, the extensive variables to be
fixed are the electromagnetic charge and the area of the quasilocal
boundary (or, alternately, the value of~$r_\sss{\text{B}}$).
The heat capacity is
\begin{equation}
  C_{A,\frak{Q}} = 2\pi r_\sss{\text{H}}^4\varkappa_\sss{\text{H}}\,
  \frac{\bigl(\ell r_\sss{\text{B}} N(r_\sss{\text{B}})\bigr)^2}%
  {ap(r_\sss{\text{B}})}
  \label{RNAdS heat capacity}
\end{equation}
where $a=3m-2r_\sss{\text{H}}$ and the polynomial~$p(r)$ is
\begin{gather}
  p(r) = r^4 + \ell^2 r^2 + b r + \ell^2 q^2
  \label{RNAdS heat capacity denom}\\
  \intertext{with}
  ab = 4mr_\sss{\text{H}}^3 + (\ell^2-4q^2)r_\sss{\text{H}}^2
  + 2m\ell^2r_\sss{\text{H}} - 5m^2\ell^2 \:.
  \label{RNAdS heat capacity denom coeff}
\end{gather}
When $a\ne0$, the qualitative relationship between the heat capacity
and the size of the quasilocal system depends upon the roots of the
polynomial~$p(r)$.  The polynomial may have zero, one, or two
positive real roots depending on the value of~$b$.  If~$b$ is
positive, then the polynomial~$p(r)$ is positive for all values
of~$r>0$.  If~$b$ is negative, however, then there may be a single
root~$r=r_{\text{c}}$, or two roots~$r=r_{1,2}$.  In the case of
a single root, the polynomial is positive everywhere except at the
root, and, in the case of two roots, the polynomial is negative
only between the roots.  The overall sign of the heat capacity
depends on the sign of the polynomial as well as the sign of~$a$;
the heat capacity diverges for sizes corresponding to the roots
of the polynomial~$p(r)$.
The heat capacity vanishes as the system shrinks to the event
horizon, and, for systems with a large radius~$r_\sss{\text{B}}$,
the heat capacity approaches the
value~$2\pi r_\sss{\text{H}}^4\varkappa_\sss{\text{H}}/a$.
However, when~$a=0$, the heat capacity is a monotonically increasing
or decreasing (depending on the sign of~$ab$) function of the system
size.  The heat capacity diverges for large systems in this case.

For many {\rnads} spacetimes, it is possible to set the size of the
quasilocal surface surrounding the black hole such that the heat
capacity of the system is positive.  In particular,
when~$2r_\sss{\text{H}}<3m$, a large system will always have a
positive heat capacity.

\section{The (2+1)-Dimensional Black Hole}
\label{s:GR BTZ}

The (2+1)-dimensional asymptotically {\ads} black hole was originally
discovered by Ba{\~n}ados, Teitelboim, and Zanelli
(\acro{btz})~\cite{btz:92} and has been studied in more detail
in reference~\cite{bhtz:93}.  The quasilocal thermodynamics of this
solution was analyzed in~\cite{z:94,bcm:94}.
The \acro{btz} black hole solution is a
rotating, circularly symmetric, and stationary spacetime solution
to the vacuum Einstein field equations with a cosmological constant.
I am interested in this
spacetime because the black hole is rotating.  In higher dimensions
it is very difficult to calculate the thermodynamic variables for
rotating black holes: Martinez~\cite{m:94} was only able to get
an approximate form of the quasilocal energy.  However, I can
get exact results for the (2+1)-dimensional
black hole.

In this section, I shall set~$\kappa=\pi$ and~$\varLambda=-1/\ell^2$.
The stationary line element for the \acro{btz} solution is
\begin{gather}
  ds^2 = -N^2(r)\,dt^2 + \frac{dr^2}{f^2(r)} + r^2 \bigl(
  N^\varphi(r)\,dt + d\varphi \bigr)^2
  \label{BTZ metric}\\
  \intertext{with}
  N^2(r) = f^2(r) = - m + \parenfrac{r}{\ell}^2 + \parenfrac{j}{2r}^2
  \label{BTZ lapse}\\
  \intertext{and}
  N^\varphi(r) = - \frac{j}{2r^2} \:.
  \label{BTZ shift}
\end{gather}
The parameters $m$ and~$j$ are constants of integration that will
eventually be associated with the mass and the angular momentum
of the black hole spacetime.  When $m>0$ and~$|j|<m\ell$, the black
hole has two horizons---where the lapse function vanishes---at the
radii
\begin{equation}
  r_\pm = \Bigl( \half\ell\bigl( m\ell \pm \sqrt{(m\ell)^2 - j^2}\,
  \bigr) \Bigr )^{1/2}
  \label{BTZ horizons}
\end{equation}
with $r_+>r_-$.  When~$|j|=m\ell$, these two horizons coincide.
The event horizon is at the larger of these two
radii~$r_\sss{\text{H}}=r_+$.  In addition, there is a ``causal''
singularity at~$r=0$~\cite{bhtz:93}.

Although the \acro{btz} solution is not static, it is circularly
symmetric; I can use this symmetry, as I did with
the {\rnads} solution, to integrate the integral form of the first law
of thermodynamics to obtain a differential form.\footnote{%
This treatment would not be possible in higher dimensional
axially-symmetric
solutions, such as the Kerr solution, since the surfaces of constant
temperature would not be the same as the surfaces of constant
observer angular momentum.}
I take the quasilocal boundary~$\mathcal{B}$ to be a circle of
constant time~$t$ and radius~$r_\sss{\text{B}}$.  A calculation
of the thermodynamic quantities yields~\cite{bcm:94}
\begin{gather}
  E = -2 f(r_\sss{\text{B}}) - E_\sss0 \,,
  \label{BTZ energy} \\
  \mathcal{S} = \frac{1}{\pi r_\sss{\text{B}}N(r_\sss{\text{B}})}\,
  \bigl( (r_\sss{\text{B}}/\ell)^2 - (j/2r_\sss{\text{B}})^2 \bigr)
  - \mathcal{S}_\sss0 \,,
  \label{BTZ surface tension} \\
  S = 4\pi r_\sss{\text{H}} \,,
  \label{BTZ entropy} \\
  T = \frac{1}{N(r_\sss{\text{B}})} \,
  \parenfrac{(r_\sss{\text{H}}/\ell)^2-(j/2r_\sss{\text{H}})^2}%
  {4\pi r_\sss{\text{H}}} \,,
  \label{BTZ temperature}\\
  J = j \,,
  \label{BTZ angular momentum} \\
  \intertext{and}
  \omega = \frac{1}{N(r_\sss{\text{B}})} \, \bigl(
  N^\varphi(r_\sss{\text{B}}) - N^\varphi(r_\sss{\text{H}}) \bigr)
  \label{BTZ angular velocity}
\end{gather}
for the quasilocal energy, surface tension, entropy, temperature,
angular momentum and angular velocity.  The angular momentum is the
conserved charge associated with the rotational Killing vector, and
the angular velocity is calculated from
equation~\eqref{temperature and velocity}.  Notice that the angular
velocity is the angular velocity of the quasilocal surface relative
to that at the event horizon,\footnote{%
The angular velocity is finite for observers on the event horizon.
This property is similar to the requirement that the gauge fields
potentials be finite on the event horizon.}
i.e., the relative velocity of the zero
angular momentum observers with respect to the rigidly rotating
radiation fluid that is in equilibrium with the black
hole~\cite{bmy:91,bcm:94}.  I assume that the reference
spacetime does not rotate.  The first law of thermodynamics for
the \acro{btz} solution is
\begin{equation}
  T\,\delta S = \delta E + \omega\,\delta J + \mathcal{S}\,\delta A
  \label{BTZ first law of thermodynamics}
\end{equation}
where $A$ is the ``area'' (circumference) of the quasilocal surface.
One can check that the first law of thermodynamics of
equation~\eqref{BTZ first law of thermodynamics} is accurate for
the thermodynamic variables of equations
\eqref{BTZ energy}--\eqref{BTZ angular velocity} by showing that
the following relations hold:
\begin{equation}
  T = \frac{\partial E}{\partial S}\,, \quad
  \mathcal{S} = -\frac{\partial E}{\partial A}\,, \quad\text{and}\quad
  \omega = -\frac{\partial E}{\partial J} \:.
  \label{BTZ intensive variables from extensive}
\end{equation}

The entropy and angular momentum of the \acro{btz} solution are
independent of the size of the thermodynamic system.  One can
determine
the asymptotic behaviour of the remaining thermodynamic variables.
The quasilocal energy has the following behaviour for small values
of~$u=1/r_\sss{\text{B}}$:
\begin{equation}
  E = \bigl\{ -2\ell^{-1}u^{-1} + m\ell u + O(u^3) \bigr\}-E_\sss0\,,
  \label{asymptotic BTZ quasilocal energy}
\end{equation}
so the quasilocal energy will diverge as~$r_\sss{\text{B}}\to\infty$
unless a suitable reference spacetime is chosen.  The temperature
has the behaviour~$T=O(u)$, so it will redshift to zero as the
radius of the quasilocal boundary approaches infinity.  The angular
velocity, $\omega=O(u)$, has a similar behaviour for large
values of~$r_\sss{\text{B}}$.  The surface tension is
\begin{equation}
  \mathcal{S} = \frac{1}{2\pi}\, \bigl\{ \ell^{-1} + m\ell u^2
  + O(u^4) \bigr\} + \parenfrac{\partial E_\sss0}{\partial A}
  \label{asymptotic BTZ surface tension}
\end{equation}
for large values of~$r_\sss{\text{B}}$.  One possible choice for the
reference spacetime would be the \acro{btz} spacetime with the
parameters $m$ and~$j$ both set to zero.  Note that this choice
differs
from the {\ads} spacetime, which would correspond to $m=-1$ and~$j=0$.
However, the {\ads} spacetime is not obtainable as a limit of black
hole spacetimes by smoothly changing the parameter~$m$ because the
solutions
corresponding to~$0<m<1$ have naked singularities.  The contribution
to the quasilocal energy from such a reference spacetime
is~$E_\sss0=-2r_\sss{\text{B}}/\ell$.  The quasilocal energy and
surface tension then have the asymptotic behaviour
$E=m\ell u+O(u^3)$ and~$\mathcal{S}=(2\pi)^{-1}m\ell u^2+O(u^4)$.
Now the quasilocal energy vanishes
as~$r_\sss{\text{B}}\to\infty$.  One can also calculate the
quasilocal mass associated with the timelike Killing vector~$t^a$
from equation~\eqref{conserved charge}.  (Note that one may not use
equation~\eqref{mass} because the spacetime is not static and, thus,
the vector~$t^a$ is not hypersurface orthogonal.)  One finds that
the mass of the \acro{btz} black hole is
\begin{equation}
  M = -Q_t = N(r_\sss{\text{B}}) E - N^\varphi(r_\sss{\text{B}})J\:.
  \label{BTZ mass}
\end{equation}
As $r_\sss{\text{B}}\to\infty$, the mass approaches the value~$m$,
so the parameter~$m$ can be interpreted as the asymptotic mass.

The heat capacity, at constant $r_\sss{\text{B}}$ and~$J$,
of the \acro{btz} black hole can be computed via
equation~\eqref{heat capacity}.  The result is
\begin{equation}
  C_{r_{\text{B}},J} =
  \frac{8\pi\ell^2r_\sss{\text{H}}^4\varkappa_\sss{\text{H}}%
      N^2(r_\sss{\text{B}})}%
    {\ell^2(2mr_\sss{\text{H}}^2+j^2)N^2(r_\sss{\text{B}})%
      +2(m^2\ell^2-j^2)r_\sss{\text{H}}^2} \:.
  \label{BTZ heat capacity}
\end{equation}
The denominator of equation~\eqref{BTZ heat capacity} is a
monotonically increasing function of~$r_\sss{\text{B}}$ and is
positive everywhere outside of the event horizon.  Thus, the
heat capacity is positive for any system
with~$r_\sss{\text{B}}>r_\sss{\text{H}}$.  The heat capacity
vanishes at the event horizon, while, for large~$r_\sss{\text{B}}$,
the heat capacity approaches the value
\begin{equation}
  C_{r_{\text{B}},J} \asymp 4\pi r_\sss{\text{H}}\,
  \parenfrac{1-x}{1+x}
  \label{BTZ asymp heat capacity}
\end{equation}
where $x=j^2/(2mr_\sss{\text{H}}^2)$.  Although the heat capacity
is finite and positive everywhere outside of the event horizon, the
quasilocal surface should be chosen at some radius less than
the inverse of the angular velocity of the black hole so that the
equilibrating radiation rotates with a velocity less than the speed
of light.

\chapter{Black Hole Spacetimes in Dilaton Gravity}
\label{c:dilaton}

Although General Relativity in four dimensions has been very
successful in describing classical gravitation, there has been a
growing interest in dilaton gravity.  It is likely that quantum
gravity will require some modification to General Relativity, and
one candidate, string theory, indicates that a particular form of
dilaton gravity
represents the low-energy effective field theory of gravitation.
In addition, in order to study gravitation in two-dimensional
spacetimes, one needs to supplement General Relativity with additional
structure, such as a dilaton field, because
the Riemann curvature tensor has only one degree of freedom, which
would be set to zero by the Einstein field equations.  Thus,
two-dimensional General Relativity is trivial, though there
are non-trivial two-dimensional dilaton theories.
One can relate a given dilaton theory
of gravity to other dilaton theories of gravity
by performing a conformal transformation of the theory.  In
particular, a dilaton theory of gravity can possibly be related to
General Relativity by a conformal transformation.\footnote{%
In order for such a transformed theory to represent a physically
different theory, one would have to assume that particles follow
geodesics in the conformally transformed theory rather than the
conformal transform of the geodesics in the original theory.}

In this chapter I consider four black hole solutions to
dilaton gravity.  Three of these solutions are solutions to the
string theory inspired dilaton gravity theories, two in four
dimensions and the third in two dimensions; the fourth is
a solution to an alternate theory of two-dimensional dilaton
gravity.  However, I first examine the circumstances under
which two theories of dilaton gravity will be conformally related,
and I demonstrate how conserved quantities that are preserved
under such conformal transformations can be defined.

\section[Conformally Related Theories of Dilaton Gravity]%
  {Conformally Related Theories\\ of Dilaton Gravity}
\label{s:dilaton conform}

Consider two theories of dilaton gravity with the Lagrangian densities
of the gravitational sectors given by
\begin{gather}
  I_\sss{\text{G}} = \int_{\mathcal{M}}\mb{\epsilon} \, \bigl(
  D(\phi) R[g] + H(\phi)(\nabla\phi)^2 + V(\phi) \bigr)
  -2\int_{\partial\mathcal{M}} \overline{\mb{\epsilon}} \,
  D(\phi) \varTheta - I^0
  \label{action of gravitational sector} \\
  \intertext{and}
  I_\sss{\text{G}} = \int_{\mathcal{M}}\mb{\tilde{\epsilon}} \, \bigl(
  \tilde{D}(\tilde{\phi}) \tilde{R}[\tilde{g}]
  + \tilde{H}(\tilde{\phi})(\nabla\tilde{\phi})^2
  + \tilde{V}(\tilde{\phi}) \bigr)
  -2\int_{\partial\mathcal{M}} \overline{\mb{\tilde{\epsilon}}} \,
  \tilde{D}(\tilde{\phi}) \tilde{\varTheta} - \tilde{I}^0
  \label{tilde action of gravitational sector}
\end{gather}
In reference~\cite{ccm:96}, it was shown that these theories are
related by a conformal transformation of the form
\begin{equation}
  \tilde{g}_{ab} = \varOmega^2(\phi) g_{ab} \quad \text{and} \quad
  \tilde{\phi} = \varUpsilon(\phi)
  \label{conformal transformation}
\end{equation}
provided that the following conditions hold:
\begin{equation}
  \tilde{D}(\tilde{\phi}) = \varOmega^{2-n}(\phi) D(\phi)\,,
  \label{condition 1}
\end{equation}
\begin{multline}
  \Biggl( \frac{\tilde{H}(\tilde{\phi})}{\tilde{D}(\tilde{\phi})}
  - \frac{n-1}{n-2}\biggl( \frac{d}{d\tilde{\phi}}\,
  \log\tilde{D}(\tilde{\phi}) \biggr)^2 \Biggr)
  \parenfrac{d\varUpsilon}{d\phi}^2 \\
  = \Biggl( \frac{H(\phi)}{D(\phi)} - \frac{n-1}{n-2} \biggl(
  \frac{d}{d\phi}\,\log D(\phi) \biggr)^2 \Biggr)\,,
  \label{condition 2}
\end{multline}
and
\begin{equation}
  \tilde{D}^{n/(2-n)}(\tilde{\phi}) \tilde{V}(\tilde{\phi})
  = D^{n/(2-n)}(\phi) V(\phi) \:.
  \label{condition 3}
\end{equation}
In addition to these three conditions, one must also require that the
reference action functionals transform into each other under the
conformal transformation~\eqref{conformal transformation}.
(The~$n=2$ version of these relations are given in
reference~\cite{mann:93}.)
Clearly the background action functional~$I^0=0$ is conformally
invariant.  However, one would prefer to have a non-trivial
conformally invariant background action functional that would
cancel divergences in the conserved quasilocal mass for asymptotically
flat spacetimes (at least).

For a \acro{sss} spacetime, one can
show that equation~\eqref{boundary action functional 1}, obtained
by the first prescription of section~\ref{s:quasi ref}, will have
the desired properties when the reference spacetime is also a
\acro{sss} spacetime.  This can be seen explicitly by considering a
\acro{sss} solution possessing the line element of
equation~\eqref{sss metric}.  When the quasilocal boundary is
chosen to be the~$r=r_\sss{\text{B}}$ boundary of a spacelike
surface of constant time~$t$, one can compute the extrinsic curvature
of the quasilocal boundary as embedded in a constant time slice of
a \acro{sss} reference spacetime with the line element
\begin{equation}
  d\sigma_\sss0^2=\frac{dr^2}{f_\sss0^2(r)}+r^2 d\omega^2\:.
  \label{spatial slice of reference spacetime}
\end{equation}
One finds
that~$k_\sss0=-(n-2)f_\sss0(r_\sss{\text{B}})/r_\sss{\text{B}}$.
Similarly, one can compute the extrinsic curvature of the quasilocal
surface of the conformally related spacetime,
\begin{equation}
  d\tilde{s}^2 = -\varOmega^2 N^2(r)\,dt^2
  + \frac{dr^2}{\varOmega^{-2}f^2(r)} + \varOmega^2r^2\,d\omega^2\,,
  \label{tilde sss metric}
\end{equation}
in the reference
spacetime~\eqref{spatial slice of reference spacetime}; one
finds~$\tilde{k}_\sss0=-(n-2)f_\sss0(r_\sss{\text{B}})/
(\varOmega r_\sss{\text{B}})$.  The
relationship~$\tilde{k}_\sss0=\varOmega k_\sss0$ is just what is
needed to ensure the conformal invariance of the background action
functional of equation~\eqref{boundary action functional 1}.
However, the reference action functional obtained by
the second prescription of section~\ref{s:quasi ref} will not be
conformally invariant.  The reference action functional will
necessarily satisfy
the three conditions \eqref{condition 1}--\eqref{condition 3}, but
for \emph{conformally related} reference solutions.
Thus, one would find
that the second prescription is \emph{not} conformally invariant
if the \emph{same} reference solution is used in both theories.

Even if the actions of two theories are conformally related, the
quantities constructed from the actions need not be conformally
invariant.  While the quasilocal mass is conformally invariant,
the quasilocal energy is not.  For a \acro{sss} spacetime of the
form~\eqref{sss metric}, one can show that the quasilocal energy is
\begin{equation}
  E = 2\biggl( f_\sss0(r_\sss{\text{B}})\,
  \frac{(n-2)A_{n-2}(r_\sss{\text{B}})D(\phi)}{r_\sss{\text{B}}}
  - f(r_\sss{\text{B}})\,\frac{d}{dr} \bigl[ A_{n-2}(r)D(\phi)
  \bigr]_{r=r_\sss{\text{B}}} \biggr)
  \label{sss quasilocal energy}
\end{equation}
where the background action functional was chosen by the first
prescription of section~\ref{s:quasi ref}.  The quasilocal energy of
the conformally related solution~\eqref{tilde sss metric} is
\begin{equation}
  \tilde{E} = 2\biggl( f_\sss0(r_\sss{\text{B}})\,
  \frac{(n-2)\tilde{A}_{n-2}(r_\sss{\text{B}})%
  \tilde{D}(\tilde{\phi})}{\varOmega r_\sss{\text{B}}}
  - \tilde{f}(r_\sss{\text{B}})\,\frac{d}{dr} \bigl[
  \tilde{A}_{n-2}(r)\tilde{D}(\tilde{\phi})
  \bigr]_{r=r_\sss{\text{B}}} \biggr) \:.
  \label{tilde sss quasilocal energy}
\end{equation}
Because the area has the conformal transformation
$\tilde{A}_{n-2}(r)=\varOmega^{n-2}A_{n-2}(r)$, one can verify that
the quasilocal energy is not conformally invariant but
rather~$\tilde{E}=E/\varOmega$.  However, the conserved masses are
$M=N(r_\sss{\text{B}})E$
and~$\tilde{M}=\tilde{N}(r_\sss{\text{B}})\tilde{E}$, so the
conserved mass is conformally invariant.

One can also compute the mass of the \acro{sss}
spacetime~\eqref{sss metric} with the background action functional
obtained by the second prescription of section~\ref{s:quasi ref}.
Given that
a constant time slice of the reference metric has the line element
of equation~\eqref{spatial slice of reference spacetime}, one finds
\begin{equation}
  M = 2N(r_\sss{\text{B}}) \bigl( f_\sss0(r_\sss{\text{B}})
  - f(r_\sss{\text{B}}) \bigr) \,\frac{d}{dr} \bigl[
  A_{n-2}(r)D(\phi) \bigr]_{r=r_\sss{\text{B}}} \:.
  \label{alt sss mass}
\end{equation}
In order for this mass to be conformally invariant,
$f_\sss0(r_\sss{\text{B}})$ must have the same properties under
a conformal transformation as~$f(r_\sss{\text{B}})$.  Thus,
one cannot use the same
reference solution for the conformally transformed theory.

Although the quasilocal energy is not conformally invariant, the
first law of thermodynamics should be.
In fact, one can show that the quasilocal energy has just the
properties that one would expect.  The
entropy of a black hole spacetime, which is given by
equation~\eqref{entropy}, is conformally invariant because the
conformal transformation of the area of the horizon cancels
the conformal transformation of~$D(\phi)$.  Furthermore, the
surface gravity of the black hole is conformally
invariant~\cite{jk:93}.
However, the temperature contains a redshift factor and,
thus, $\tilde{T}=T/\varOmega$.  The quasilocal energy is
\emph{covariant} under a conformal transformation, and this is
precisely the form
that is needed in order for the first law of thermodynamics,
$T\,dS=dE+\text{work terms}$, to be conformally invariant.

\section{Garfinkle-Horowitz-Strominger Spacetimes}
\label{s:dilaton GHS}

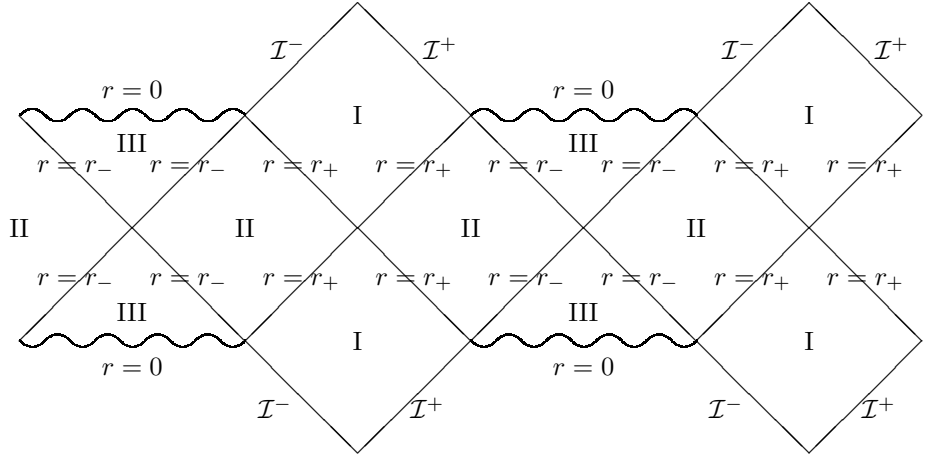
\begin{figure}[t]
\renewcommand{\baselinestretch}{1}\small
\begin{center}
\setlength{\unitlength}{0.75mm}
\begin{picture}(160,100)
\multiput(0,25)(80,0){2}{%
  \begin{picture}(80,100)(0,25)
  \put(20,50){\line(-1,1){20}}
  \put(20,50){\line(-1,-1){20}}
  \put(20,50){\line(1,1){20}}
  \put(20,50){\line(1,-1){20}}
  \put(60,50){\line(-1,1){20}}
  \put(60,50){\line(-1,-1){20}}
  \put(60,50){\line(1,1){20}}
  \put(60,50){\line(1,-1){20}}
  \put(40,70){\line(1,1){20}}
  \put(40,30){\line(1,-1){20}}
  \put(80,70){\line(-1,1){20}}
  \put(80,30){\line(-1,-1){20}}
  \multiput(2.22,70)(8.88,0){5}{\qbezier(-2.22,0)(0,2.22)(2.22,0)}
  \multiput(6.66,70)(8.88,0){4}{\qbezier(-2.22,0)(0,-2.22)(2.22,0)}
  \multiput(2.22,30)(8.88,0){5}{\qbezier(-2.22,0)(0,-2.22)(2.22,0)}
  \multiput(6.66,30)(8.88,0){4}{\qbezier(-2.22,0)(0,2.22)(2.22,0)}
  \put(20,73){\raisebox{\depth}{\makebox[0pt]{\hss$r=0$\hss}}}
  \put(20,27){\raisebox{-\height}{\makebox[0pt]{\hss$r=0$\hss}}}
  \put(50,80){%
    \raisebox{\depth}{\makebox[0pt]{\hss$\scri^-\quad$\hss}}}
  \put(70,80){%
    \raisebox{\depth}{\makebox[0pt]{\hss$\qquad\scri^+$\hss}}}
  \put(50,20){%
    \raisebox{-\height}{\makebox[0pt]{\hss$\scri^-\qquad$\hss}}}
  \put(70,20){%
    \raisebox{-\height}{\makebox[0pt]{\hss$\quad\scri^+$\hss}}}
  \put(60,70){\raisebox{-.5\height}{\makebox[0pt]{\hss I\hss}}}
  \put(60,30){\raisebox{-.5\height}{\makebox[0pt]{\hss I\hss}}}
  \put(40,50){\raisebox{-.5\height}{\makebox[0pt]{\hss II\hss}}}
  \put(0,50){\raisebox{-.5\height}{\makebox[0pt]{\hss II\hss}}}
  \put(20,65){\raisebox{-.5\height}{\makebox[0pt]{\hss III\hss}}}
  \put(20,35){\raisebox{-.5\height}{\makebox[0pt]{\hss III\hss}}}
  \put(10,60){\makebox[0pt]{\hss$r=r_-$\hss}}
  \put(10,40){\makebox[0pt]{\hss$r=r_-$\hss}}
  \put(30,60){\makebox[0pt]{\hss$r=r_-$\hss}}
  \put(30,40){\makebox[0pt]{\hss$r=r_-$\hss}}
  \put(50,60){\makebox[0pt]{\hss$r=r_+$\hss}}
  \put(50,40){\makebox[0pt]{\hss$r=r_+$\hss}}
  \put(70,60){\makebox[0pt]{\hss$r=r_+$\hss}}
  \put(70,40){\makebox[0pt]{\hss$r=r_+$\hss}}
  \end{picture}}
\end{picture}
\end{center}
\begin{quote}\leavevmode
\caption[The extended \acro{ghs} spacetime]{\small The extended,
  non-extremal \acro{ghs} spacetime: there are two horizons, $r_\pm$
  with~$r_+>r_-$, and a curvature singularity at~$r=0$.  There are
  three regions: the outer region~I has~$r>r_+$, the intermediate
  region~II has~$r_-<r<r_+$, and the inner region~III has~$r<r_-$.
  In this diagram, causal future is to the right while causal past is
  to the left.}
\label{fig:GHS}
\end{quote}
\end{figure}

The Garfinkle-Horowitz-Strominger solutions are presented in
references~\cite{ghs:91,h:92}.  These solutions are \acro{sss}
solutions
to the string-inspired dilaton theory of gravity with Lagrangian
density
\begin{equation}
  \mb{L} = \frac{1}{16\pi}\,\mb{\epsilon}\,\e^{-2\phi} \bigl(
  R[g] + 4(\nabla\phi)^2 - \frak{F}^{ab}\frak{F}_{ab} \bigr) \:.
  \label{string theory}
\end{equation}
This theory of gravity is conformally related to General Relativity
by the conformal transformation
\begin{equation}
  g^\sss{\text{S}}_{ab} = \e^{2\phi} g^\sss{\text{E}}_{ab}
  \label{string conformal transformation}
\end{equation}
where $g^\sss{\text{S}}_{ab}$ is the metric of a solution in
the ``string frame,'' i.e., a solution to the field equations
obtained from equation~\eqref{string theory}, and
$g^\sss{\text{E}}_{ab}$ is the metric in the ``Einstein frame,'' which
is a solution to the field equations of General Relativity with
additional scalar and gauge matter fields:
\begin{equation}
  \mb{L} = \frac{1}{16\pi}\,\mb{\epsilon}\, \bigl(
  R[g] - 2(\nabla\phi)^2 - \e^{-2\phi}\frak{F}^{ab}\frak{F}_{ab}
  \bigr) \:.
  \label{Einstein-Maxwell-scalar theory}
\end{equation}
In the string frame, the magnetically charged \acro{ghs} solution is
given by~\cite{ghs:91}
\begin{gather}
  \e^{-2\phi} = \e^{-2\phi_0} \biggl( 1 - \frac{q^2}{ar}
  \biggr) \,,
  \label{GHS magnetic phi}\\
  \mb{\frak{F}} = -q\,\mb{d}\vartheta \wedge \sin\vartheta\,
  \mb{d}\varphi \,,
  \label{GHS magnetic F}\\
  f^2(r) = \biggl( 1 - \frac{2a}{r} \biggr) \biggl( 1 - \frac{q^2}{ar}
  \biggl) \,,
  \label{GHS magnetic f2}\\
  \intertext{and}
  N(r) = f(r)\,\e^{2(\phi-\phi_0)}
  \label{GHS magnetic N}
\end{gather}
where the line element is given by equation~\eqref{sss metric}.
Here, $\phi_\sss0$, $q$, and~$a$ are constants of
integration.%
\footnote{The constant of integration, $a$, is the combination of
constants $m\e^{\phi_0}$ used in references~\cite{ghs:91,h:92}.}
In the non-extremal case, $q^2<2a^2$, this black hole has two
horizons; the outer horizon at~$r_\sss{\text{H}}=2a$
is taken to be the event horizon.
The electric string black hole solution is~\cite{h:92}
\begin{gather}
  \e^{-2\phi} = \e^{2\phi_0} + \frac{q^2}{ar} \,,
  \label{GHS electric phi}\\
  \mb{\frak{F}} = \frac{q\e^{2\phi}}{r^2}\,
  N(r)\,\mb{d}t \wedge f^{-1}(r)\,\mb{d}r \,,
  \label{GHS electric F} \\
  f^2(r) = 1 + \e^{-2\phi_0} \parenfrac{q^2-2a^2}{ar} \,,
  \label{GHS electric f2}\\
  \intertext{and}
  N(r) = f(r)\,\e^{2(\phi+\phi_0)} \;.
  \label{GHS electric N}
\end{gather}
Again, when $q^2<2a^2$, there are two horizons with the event horizon
at~$r_\sss{\text{H}}=\e^{-2\phi_0}(2a-q^2/a)$.  The causal structure
of these \acro{ghs} spacetimes are the same as the asymptotically
flat Reissner-Nordstr{\"o}m spacetime: see figure~\ref{fig:GHS}.

One can compute the conserved mass for the two \acro{ghs} solutions
in the string frame via equation~\eqref{sss quasilocal energy};
one finds~$\lim_{r\to\infty}M=a$ for the electric solution
and~$a\e^{-2\phi_0}$ for the magnetic solution.  In addition, one
can compute the mass in the Einstein frame; here, the solution
is given by the line element
\begin{equation}
  ds^2 = -N^2\e^{-2\phi}\,dt^2
  + \frac{d\varrho^2}{f^2\e^{2\phi}(d\varrho/dr)^2}
  + \varrho^2\,d\omega^2
  \label{Einstein GHS metric}
\end{equation}
where $\varrho=r\e^{-\phi}$.  One finds~$\lim_{\varrho\to\infty}M=a$
for the electric solution and~$a\e^{-2\phi_0}$ for the magnetic
solution: the mass is
conformally invariant as expected.  However,
when~$\phi_\sss0\ne0$, the conformal transformation
of equation~\eqref{string conformal transformation}
does not approach unity at spatial infinity, and, thus, the lapse
function~$N\e^{-\phi}$ of equation~\eqref{Einstein GHS metric}
also fails to approach unity at spatial infinity.  Since the solution
is static, one can rescale the time coordinate:
$t\to t_{\text{new}}=\e^{\pm\phi_0}$ where the
plus (minus) applies to the magnetic (electric) solution.  The new
lapse function in the Einstein frame
is~$N\e^{-(\phi\mp\phi_0)}$.  The mass rescales by a constant
factor, and the result~$M_\sss{\text{ADM}}=a\e^{-\phi_0}$ is
obtained for both the magnetic and electric solutions in agreement
with references~\cite{ghs:91,h:92}.  Note that such a rescaling is
always possible for static solutions and, unless there is some
required asymptotic behaviour for the lapse function, the definition
of mass contains an arbitrary constant factor.

Consider now the mass defined by equation~\eqref{alt sss mass} with
the reference solution of Minkowski spacetime with a constant dilaton.
In the Einstein frame, the mass evaluated above and the mass
of equation~\eqref{alt sss mass} agree because~$D$ is constant.
However, in the string frame, equation~\eqref{alt sss mass} yields
$M=\bigl(a-q^2/(2a)\bigr)$ for the electric solution
and~$M=\e^{-2\phi_0}\bigl(a+q^2/(2a)\bigr)$ for the magnetic
solution.  The mass is not conformally invariant when
equation~\eqref{alt sss mass} is used.

I first consider the electrically charged \acro{ghs} solution.
The intensive variables can be evaluated on a quasilocal surface
of constant radius~$r=r_\sss{\text{B}}$.  One finds
\begin{equation}
  S = 2\pi a r_\sss{\text{B}} \biggl(
  \frac{2a^2-q^2}%
    {ar_\sss{\text{B}}\e^{-2\phi}-q^2} \biggr)
  \label{GHS electric entropy}
\end{equation}
is the entropy of the black hole,
\begin{equation}
  E = -f(r_\sss{\text{B}})\,\biggl(
  r_\sss{\text{B}}\e^{-2\phi} - \frac{q^2}{2a}
  \biggr) - E_{\sss0}
  \label{GHS electric energy}
\end{equation}
is the quasilocal energy, and
\begin{equation}
  \frak{Q} = q
  \label{GHS electric Maxwell charge}
\end{equation}
is the Maxwell charge.
The extensive variables include the temperature, the surface tension,
the dilaton potential, and the Maxwell potential.  The temperature
is~$T=\varkappa_\sss{\text{H}}/(2\pi N)$ with
$\kappa_\sss{\text{H}}=f(dN/dr)$ evaluated on the event horizon:
\begin{equation}
  T = \frac{1}{8\pi a N(r_\sss{\text{B}})}\, \biggl(
  \e^{-2\phi} - \frac{q^2}{ar_\sss{\text{B}}} \biggr) \;.
  \label{GHS electric temperature}
\end{equation}
The surface tension is
\begin{equation}
  \mathcal{S} = \frac{1}{8\pi r_\sss{\text{B}}^2}\, \Biggl(
   f(r_\sss{\text{B}})\, \biggl( r_\sss{\text{B}}\e^{-2\phi}
   - \frac{q^2}{2a} \biggr) + \frac{a}{f(r_\sss{\text{B}})} \Biggr)
  - \mathcal{S}_{\sss0}
  \label{GHS electric surface tension}
\end{equation}
and the dilaton potential is
\begin{equation}
  \mu = -f(r_\sss{\text{B}})\, \biggl(
  2r_\sss{\text{B}}\e^{-2\phi} - \frac{q^2}{2a}
  \biggr) - \frac{a}{f(r_\sss{\text{B}})} - \mu_{\sss0} \;.
  \label{GHS electric dilaton potential}
\end{equation}
The contributions of the reference spacetime to the surface tension
and dilaton potential, $\mathcal{S}_{\sss0}$ and~$\mu_{\sss0}$,
are determined in terms of~$E_{\sss0}$ by
equation~\eqref{reference functional relationship}.
Finally, the Maxwell potential, $\frak V$, is given by
\begin{equation}
  \frak{V} = -\frac{1}{N(r_\sss{\text{B}})}
  \int^{r_{\text{B}}}_{r_{\text{H}}} \frak{F}_{tr}\, dr
  = - \frac{q}{2a}\,f(r_\sss{\text{B}}) \;.
  \label{GHS electric Maxwell potential}
\end{equation}

Because the \acro{ghs} solution is spherically symmetric, the first
law of thermodynamics can be written in the differential form
\begin{equation}
  T\,dS = dE + \mathcal{S}\,dA + \mu\,d\phi + \frak{V}\,d\frak{Q}\:.
  \label{GHS first law of thermodynamics}
\end{equation}
Equation~\eqref{GHS electric entropy} can be solved for the
parameter~$a$ to obtain an expression for~$a$ in terms of the
extensive variables $S$, $\frak{Q}$, $\phi$,
and~$A=4\pi r_\sss{\text{B}}^2$.  One can write~$f(r_\sss{\text{B}})$
in terms of these extensive variables and, thus, one
has~$E=E(S,A,\phi,\frak{Q})$.  It can then be shown that
\begin{equation}
  T = \frac{\partial E}{\partial S}, \quad
  \mathcal{S} = -\frac{\partial E}{\partial A}, \quad
  \mu = -\frac{\partial E}{\partial\phi}, \quad\text{and}\quad
  \frak{V} = -\frac{\partial E}{\partial\frak{Q}}\;.
  \label{GHS intensive variables}
\end{equation}
Thus, the quasilocal energy is the appropriate choice for the
thermodynamic internal energy of the gravitating system contained
within the quasilocal boundary; I have explicitly demonstrated the
first law of thermodynamics given by
equation~\eqref{GHS first law of thermodynamics}.  This
thermodynamic relation is independent of the choice of the background
action functional~$I^0$ and, thus, independent of the choice
of~$E_{\sss0}$.

I now examine the asymptotic behaviour of the thermodynamic
variables for large values of~$r_\sss{\text{B}}$
or, equivalently, for small values of~$u=1/r_\sss{\text{B}}$.
The Maxwell charge, $\frak{Q}=q$, and the entropy,
$S=2\pi\e^{-2\phi_0}(2a^2-q^2)$, are both independent of the
size of the quasilocal system.  The quasilocal energy has the
following behaviour:
\begin{equation}
  E = \biggl\{ -\e^{2\phi_0} u^{-1} + \biggl( a - \frac{q^2}{a}
  \biggr) + O(u) \biggr\} - E_{\sss0} \;.
  \label{GHS asymp electric energy}
\end{equation}
Unless $E_{\sss0}$ is suitably chosen, the quasilocal
energy will diverge as~$r_\sss{\text{B}}\to\infty$.
The temperature,
\begin{equation}
  T = \frac{1}{8\pi a}\, \biggl\{ \e^{2\phi_0} + \biggl(
  a + \frac{q^2}{2a} \biggr)\,u + O(u^2) \biggr\},
  \label{GHS asymp electric temperature}
\end{equation}
approaches the value~$\e^{2\phi_0}/(8\pi a)$
as~$r_\sss{\text{B}}\to\infty$, and the Maxwell
potential,
\begin{equation}
  \frak{V} = -\frac{q}{2a}\, \biggl\{ 1 + \e^{-2\phi_0}
  ( q^2/2 - a^2 ) u + O(u^2) \biggr\},
  \label{GHS asymp Maxwell potential}
\end{equation}
approaches~$-q/(2a)$.  The surface tension and the dilaton potential
have the following behaviour:
\begin{equation}
  \mathcal{S} = \frac{1}{8\pi}\, \biggl\{ \e^{2\phi_0}u
  + \frac{q^2}{a}\,u^2 + O(u^3) \biggr\} + \biggl(
  \frac{\partial E_{\sss0}}{\partial A} \biggr)
  \label{GHS asymp surface tension}
\end{equation}
and
\begin{equation}
  \mu = \biggl\{ -2\e^{2\phi_0}u^{-1} + \biggl(
  a - \frac{5q^2}{2a} \biggr) + O(u) \biggr\} + \biggl(
  \frac{\partial E_\sss0}{\partial\phi} \biggr)
  \label{GHS asymp dilaton potential}
\end{equation}
respectively.  As with the energy, the asymptotic behaviour of these
functions depends upon the choice of background action functional;
in particular, the dilaton potential will diverge
as~$r_\sss{\text{B}}\to\infty$ unless a suitable choice
for~$E_{\sss0}$ is made.  I adopt the procedure that was found
to be conformally invariant in section~\ref{s:dilaton conform}.
When the reference spacetime is just Minkowski spacetime, one finds
$E_{\sss0}=-r_\sss{\text{B}}\e^{-2\phi}$,
$\mu_\sss0=-2r_\sss{\text{B}}\e^{-2\phi}$,
and~$\mathcal{S}_\sss0=\e^{-2\phi}/(8\pi r_\sss{\text{B}})$.
The quasilocal energy is finite in the
limit~$r_\sss{\text{B}}\to\infty$ and approaches the value~$a$;
the dilaton potential is also finite and approaches~$q^2/(2a)$ in
this limit; the surface tension behaves as~$\mathcal{S}=O(u^3)$
for large values of~$r_\sss{\text{B}}$.

Recall now the magnetic string solution of
equations~\eqref{GHS magnetic phi}--\eqref{GHS magnetic N}.
I wish to construct the thermodynamic variables in the same way
as I have just done for the electric string solution.  The
entropy and the energy can both be calculated as before.  They are
\begin{equation}
  S = 2\pi ar_\sss{\text{B}}\e^{-2\phi}\, \biggl(
  \frac{2a^2-q^2}{ar_\sss{\text{B}}-q^2} \biggr)
  \label{GHS magnetic entropy}
\end{equation}
and
\begin{equation}
  E = -\e^{-2\phi}\,\biggl(
  f(r_\sss{\text{B}})r_\sss{\text{B}}
  +N(r_\sss{\text{B}})\frac{q^2}{2a} \biggr)
  - E_\sss0
  \label{GHS magnetic energy}
\end{equation}
respectively.  I can also calculate the following intensive
variables as before.  The temperature is
\begin{equation}
  T = \frac{1}{8\pi aN(r_\sss{\text{B}})},
  \label{GHS magnetic temperature}
\end{equation}
the surface tension is
\begin{equation}
  \mathcal{S} = \frac{1}{8\pi r_\sss{\text{B}}}
  \e^{-2\phi}\,\biggl(
  f(r_\sss{\text{B}})r_\sss{\text{B}}
  +N(r_\sss{\text{B}})\frac{q^2}{2a}
  +\frac{a}{N(r_\sss{\text{B}})} \biggr)
  - \mathcal{S}_\sss0,
  \label{GHS magnetic surface tension}
\end{equation}
and the dilaton potential is
\begin{equation}
  \mu = -\e^{-2\phi}\,\biggl(
  2f(r_\sss{\text{B}})r_\sss{\text{B}}
  +N(r_\sss{\text{B}})\frac{q^2}{2a}
  +\frac{a}{N(r_\sss{\text{B}})} \biggr)
  - \mu_\sss0\;.
  \label{GHS magnetic dilaton potential}
\end{equation}

Just as with the {\rnads} magnetic solution of section~\ref{s:GR RNADS},
I cannot globally define a non-singular electromagnetic potential
that gives rise to the field strength of the magnetic \acro{ghs}
solution.  The remedy is to use the same procedure as was used in
section~\ref{s:GR RNADS}: when
these steps are repeated, one finds that~$\frak{U}\,\delta\frak{P}$
represents the electromagnetic work term with
\begin{gather}
  \frak{P}=q
  \label{GHS magnetic Maxwell charge}\\
  \intertext{and}
  \frak{U} = \e^{-2\phi}\,
  \frac{q}{f(r_\sss{\text{B}})} \biggl(
  \frac{1}{r_\sss{\text{B}}} -
  \frac{1}{r_\sss{\text{H}}} \biggr) \;.
  \label{GHS magnetic Maxwell potential}
\end{gather}

The quasilocal energy can be written in terms of the extensive
variables $S$, $A$, $\phi$, and~$\frak{P}$;  then, the relations
of equation~\eqref{GHS intensive variables} can be shown to hold.
Thus, the definitions of the thermodynamic variables are consistent
with the first law of
thermodynamics~\eqref{GHS first law of thermodynamics}, where the
$\frak{V}\,\delta\frak{Q}$ work term is to be replaced
with~$\frak{U}\,\delta\frak{P}$.

The entropy, $S=2\pi\e^{-2\phi_0}(2a^2-q^2)$, and
Maxwell charge, $\frak{P}=q$, are independent of the size,
$r_\sss{\text{B}}$, of the quasilocal region.
For~$r_\sss{\text{B}}\to\infty$, $T\to(8\pi a)^{-1}$,
and~$\frak{U}\to-\e^{-2\phi_0}q/(2a)$.
Finally, I shall obtain the asymptotic behaviour of the energy, the
surface tension, and the dilaton force for small values
of~$u=1/r_\sss{\text{B}}$.  I find
\begin{equation}
  E = \e^{-2\phi_0}\,\biggl\{ u^{-1} + \biggl( a + \frac{q^2}{a}
  \biggr) + O(u) \biggr\} - E_\sss0\,,
  \label{GHS asymp magnetic energy}
\end{equation}
\begin{equation}
  \mathcal{S} = \frac{1}{8\pi}\,\e^{-2\phi_0}\, \biggl\{ u -
  \frac{q^2}{a}\,u^2 + O(u^3) \biggr\}
  + \biggl(\frac{\partial E_\sss0}{\partial A}\biggr)\,,
  \label{GHS asymp magnetic surface tension}
\end{equation}
and
\begin{equation}
  \mu = \e^{-2\phi_0}\,\biggl\{ -2u^{-1} + \biggl( a +
  \frac{5q^2}{2a} \biggr) + O(u) \biggr\}
  +\biggl(\frac{\partial E_\sss0}{\partial\phi}\biggr)
  \label{GHS asymp magnetic dilaton potential}
\end{equation}
respectively.  For the background action functional as given by
the first prescription of section~\ref{s:quasi ref}, one
has~$E_\sss0=-\e^{-2\phi}u^{-1}$.  In this case, the
quasilocal energy approaches the value~$\e^{-2\phi_0}a$
as~$r_\sss{\text{B}}\to\infty$, and the dilaton
potential approaches~$\e^{-2\phi_0}\bigl(a+q^2/(2a)\bigr)$.  For
large~$r_\sss{\text{B}}$, $\mathcal{S}=O(u^3)$.

One can compute the heat capacities of the \acro{ghs} spacetimes
using equation~\eqref{heat capacity}.  For the magnetic and
electric solutions, the heat capacities at constant system size,
dilaton field, and electromagnetic charge are
\begin{gather}
  C_{\phi,A,\frak{P}} = -8\pi a^2\e^{-2\phi_0}\,
  \frac{(x-1)(2x-3\alpha+\alpha^2)}{2x^2-3(1+\alpha)x+4\alpha}
  \label{GHS magnetic heat capacity}\\
  \intertext{and}
  C_{\phi,A,\frak{Q}} = -8\pi a^2\e^{-2\phi_0}\,
  \frac{(1-\alpha)(x-1)(2x-\alpha)}%
    {2(1-\alpha)x^2-(3-\alpha)x+\alpha}
  \label{GHS electric heat capacity}
\end{gather}
respectively.  Here, $x=r/r_\sss{\text{H}}$ is the radius of the
quasilocal surface in units of the radius of the event horizon
and $\alpha=q^2/(2a^2)$; these must satisfy $x>1$ and~$0\le\alpha<1$.
The denominators of equations \eqref{GHS magnetic heat capacity}
and~\eqref{GHS electric heat capacity} both have a single root
and are negative for~$x$ less than this root and positive
for~$x$ greater than the root.  The heat capacities of the magnetic
and electric black holes have
similar behaviour: the heat capacities are zero near the event
horizon, increase without bound until some critical radius is
reached, and then increase from an infinitely negative value
just beyond the critical radius to the value~$-8\pi a^2\e^{-2\phi_0}$
at~$r_\sss{\text{B}}=\infty$.  Therefore, stable thermodynamic
equilibria can be reached if the quasilocal boundaries are placed
within the critical radii about the black holes.

\section{Two-Dimensional Spacetimes}
\label{s:dilaton 2d}

Two-dimensional black hole solutions have been of considerable
interest recently because they have been used as tractable toy
models of quantum gravity.  In two dimensions, gravitational theories
are simplified greatly; in fact, General Relativity becomes trivial
because the complete Riemann curvature tensor is uniquely specified
by the Ricci scalar.  Thus, one requires the extra complexity of
dilaton gravity in order to have a non-trivial theory.

The expressions for the thermodynamic variables are also simplified 
when one goes to two dimensions.  For example, the extrinsic
curvature of the one-dimensional boundary~$\mathcal{T}$ is
just~$\varTheta_{ab}=\gamma_{ab}\varTheta=-\gamma_{ab}n^ca_c$.
There will be no quasilocal surface momentum or stress
densities.  Similarly, a Maxwell field is dual to a scalar:
$\frak{F}_{ab}=f\epsilon_{ab}$
where~$f=-\epsilon^{ab}\nabla_a\frak{A}_b$.  This implies that
there is no surface electromotive force or current
densities created by the Maxwell field.  In fact, because the
quasilocal surface is just a point, the integral form of the
first law of thermodynamics~\eqref{first law of thermodynamics}
automatically becomes a differential equation.

A static two-dimensional spacetime can always be expressed in the form
\begin{equation}
  ds^2 = -N^2(x)\,dt^2 + N^{-2}(x)\,dx^2 \:.
  \label{2d static metric}
\end{equation}
A black hole has an event horizon at the point where the lapse
function vanishes, and a curvature singularity when the
Ricci scalar, $R=-d^2(N^2)/dx^2$, diverges.
One can then write the thermodynamic variables in terms of the
lapse function and the dilaton alone~\cite{cm:95c}.  The entropy is
\begin{equation}
  S = 4\pi D(\phi_\sss{\text{H}})
  \label{2d entropy}
\end{equation}
where $\phi_\sss{\text{H}}$ is the value of the dilaton on the
event horizon.  The quasilocal energy is the only other extensive
variable:
\begin{equation}
  E = -2 N(x_\sss{\text{B}}) \, \frac{dD}{d\phi}\,
  \frac{d\phi}{dx}\biggr|_{x=x_\sss{\text{B}}} - E_\sss0 \:.
  \label{2d energy}
\end{equation}
The two intensive variables are the temperature,
\begin{equation}
  T = \frac{1}{4\pi N(x_\sss{\text{B}})} \, \frac{d}{dx} \bigl[
  N^2(x) \bigr]_{x=x_\sss{\text{B}}} \,,
  \label{2d temperature}
\end{equation}
and the dilaton potential
\begin{equation}
  \mu = 2 \biggl[ N(x) H(\phi)\, \frac{d\phi}{dx} + \frac{dD}{d\phi}\,
  \frac{dN}{dx} \biggr]_{x=x_\sss{\text{B}}} -\mu_\sss0 \:.
  \label{2d dilaton force}
\end{equation}
When a Maxwell field is present, the quasilocal charge is
\begin{equation}
  \frak{Q} = W(\phi) f(x_\sss{\text{B}})
  \label{2d charge}
\end{equation}
where $f=\frak{A}_t$, and the quasilocal Maxwell potential is
\begin{equation}
  \frak{V} = -\frac{1}{N(x_\sss{\text{B}})}\,
  \int^{x_\sss{\text{B}}}_{x_\sss{\text{H}}} f(x)\,dx \:.
  \label{2d Maxwell potential}
\end{equation}
The first law of thermodynamics is
\begin{equation}
  T\,dS = dE + \mu\,d\phi + \frak{V}\,d\frak{Q}
  \label{2d first law of thermodynamics}
\end{equation}
where the electromagnetic work term has been included.

The first example of a two-dimensional black hole arises from
the string-inspired theory~\cite{hs:92}
\begin{equation}
  \mb{L} = \frac{1}{2\kappa}\,\mb{\epsilon}\,
  \e^{-2\phi}\bigl( R[g] + 4(\nabla\phi)^2
  - \fourth \, \frak{F}^{ab}\frak{F}_{ab} + a^2 \bigr)
  \label{2d string theory}
\end{equation}
where $a$ is a constant.  The field equations of this theory possess
the following solution~\cite{mny:92,np:92}
in the Schwarzschild-like coordinates of
equation~\eqref{2d static metric}:
\begin{gather}
  f = q\e^{2\phi}\,,
  \label{2d string Maxwell field} \\
  N^2 = 1 - \frac{2m}{a}\,\e^{2\phi} + \frac{q^2}{2a^2}\,\e^{4\phi}\,,
  \label{2d string lapse2} \\
  \intertext{and}
  \phi = -\half ax
  \label{2d string dilaton}
\end{gather}
where $q$ and~$m$ are constants of integration.  Note that a third
constant of integration, $x_\sss0$, has been absorbed into the
definition of the origin.  The dilaton field is proportional to the
coordinate~$x$, so one can view the various quantities as functions
of the dilaton value on the quasilocal boundary rather than as
functions of the coordinate value of the quasilocal boundary.

The solution admits inner and outer event horizons given
by $\exp(-2\phi_\pm)=(1\pm\varDelta)m/a$
where~$\varDelta^2=1-q^2/2m^2$.  I will consider only the
case for which~$0<\varDelta\le1$, that is, $2m^2>q^2\ge0$.
(When~$q=0$ there is only a single horizon.)  The outer event horizon
is the horizon with the larger \emph{coordinate} value,
$\phi_\sss{\text{H}}=\phi_+$.  The Ricci scalar diverges
as~$\phi\to\infty$ ($x\to-\infty$) and vanishes for~$\phi\to -\infty$
($x\to\infty$);  thus I position the quasilocal boundary such
that~$\phi_\sss{\text{B}}<\phi_\sss{\text{H}}=\phi_+\le\phi_-<\infty$.
A natural choice for the reference spacetime is the solution
with~$m=q=0$ for which $N=1$ and~$f=0$ everywhere.  For the reference
solution, there is no event horizon and the Ricci scalar is zero
everywhere.  The quasilocal thermodynamic quantities can now be
evaluated.  The extensive variables are
\begin{gather}
  S = \frac{4\pi m}{a}\,(1+\varDelta)\,,
  \label{2d string entropy} \\
  E = 2a\e^{-2\phi}(1-N)\,,
  \label{2d string energy} \\
  \intertext{and}
  \frak{Q}=q
  \label{2d string charge}
\end{gather}
where I have set the gravitational coupling constant
to~$\kappa=\half$.
These functions are to be evaluated at~$\phi=\phi(x_\sss{\text{B}})$.
In the asymptotically flat regime, $\phi\to-\infty$, the quasilocal
energy approaches the value~$2m$ so the parameter~$m$ can be
interpreted as a measure of the black hole mass.  Similarly, one can
interpret the parameter~$q$ as the charge of the black hole.
Equation~\eqref{2d string energy} can be solved for the mass
parameter; using~\eqref{2d string charge}, one finds
\begin{equation}
  m = \frac{\pi\frak{Q}^2}{aS} + \frac{aS}{8\pi} \:.
  \label{2d string mass parameter}
\end{equation}
Using this equation, one can express the quasilocal energy in terms
of $\frak{Q}$ and~$S$.

The intensive variables can be calculated either by equations
\eqref{2d temperature}, \eqref{2d dilaton force},
and~\eqref{2d Maxwell potential} or by the
first law of thermodynamics:
\begin{equation}
  T = \frac{\partial E}{\partial S}, \quad
  \mu = -\frac{\partial E}{\partial\phi}, \quad \text{and} \quad
  \frak{V} = -\frac{\partial E}{\partial\frak{Q}} \:.
\end{equation}
Both methods yield the same results.  The temperature, dilaton
force, and Maxwell potential are
\begin{gather}
  T = \frac{a}{2\pi N}\,\parenfrac{1}{1+1/\varDelta}\,,
  \label{2d string temperature}\\
  \mu = 4a\e^{-2\phi}(1-N) - \frac{4}{N} \biggl( m - \frac{q^2}{2a}\,
  \e^{2\phi} \biggr)\,,
  \label{2d string dilaton force}\\
  \intertext{and}
  \frak{V} = \frac{q}{aN}\,( \e^{2\phi} - \e^{2\phi_{\text{H}}} )
  \label{2d string Maxwell potential}
\end{gather}
respectively where the quantities are to be evaluated on the
quasilocal surface.
The temperature is always positive outside of the black hole.
The dilaton potential~$\mu$ can be interpreted as a force because the
dilaton can be used as the coordinate value of the quasilocal surface;
thus, $\mu$~is conjugate to the size of the system.

The heat capacity at constant electric charge and dilaton value
is obtained from equation~\eqref{heat capacity}; one obtains
\begin{equation}
  C_{\phi,\frak{Q}} = \frac{NT}{T^2\e^{2\phi}+4\pi\frak{Q}^2/aS^3}\:.
  \label{2d string heat capacity}
\end{equation}
Outside of the event horizon, the temperature, entropy, and lapse
are all positive, so the heat capacity is also positive.

Another important two-dimensional theory of gravity is
the~``$R=\kappa T$'' theory~\cite{m:91,sm:91,mmss:91}.  In this
theory, the Lagrangian density of the gravitational sector is
\begin{equation}
  \mb{L}_\sss{\text{G}} = \frac{1}{2\kappa}\,\mb{\epsilon}\,\bigl(
  \phi R[g] + \half (\nabla\phi)^2 \bigr) \:.
  \label{R=T gravitational Lagrangian}
\end{equation}
When the matter fields do not couple to the dilaton, the field
equation for the metric takes the form~$R=\kappa T$ where~$T$ is
the trace of the divergenceless stress energy tensor.  The matter
field of interest here is the Liouville field, $\psi$, for which
the Lagrangian density is~\cite{mann:94}
\begin{equation}
  \mb{L}_\sss{\text{L}} = \mb{\epsilon}\,\bigl(
  \varLambda\e^{-2a\psi} - b(\nabla\psi)^2 - c\psi R[g] \bigr) \:.
  \label{Liouville Lagrangian}
\end{equation}
Notice that the Liouville field is not minimally coupled to the
metric because of the~$-c\psi R[g]$ term.  The presence of this
term raises some concerns about the applicability of the formulation
in chapter~2, which required that all matter fields be minimally
coupled to the metric.  However, the Liouville field is easily
accommodated if one groups the Liouville field together with the
dilaton in the function~$D$ wherever necessary~\cite{cm:95c}.
In the following, I require $a>0$, $b=a^2/\kappa$,
and~$c>a/\kappa$.
These restrictions allow one to obtain a solution with physically
reasonable thermodynamics.
For convenience, I define the quantities~$d=c/a-1/\kappa$
(always positive) and~$\lambda^2=-\varLambda/2d$.

The field equations generated from the Lagrangian
density~$\mb{L}=\mb{L}_\sss{\text{G}}+\mb{L}_\sss{\text{L}}$
possess the following solution in the two-dimensional line element
of equation~\eqref{2d static metric}:
\begin{gather}
  N^2 = 1 - \frac{\lambda^2}{m^2}\,\e^{-2a\psi}\,,
  \label{2d Liouville lapse} \\
  \phi = 2a(\psi-\psi_\sss0) + \phi_\sss0\,,
  \label{2d Liouville dilation} \\
  \intertext{and}
  \psi = \frac{m}{a}\,x + \psi_\sss0
  \label{2d Liouville Louville}
\end{gather}
where $\psi_\sss0=-mx_\sss0/a$, $\phi_\sss0$, and~$m$ are constants
of integration; I take~$m$ to be positive.  Since the Liouville
field is proportional to the spatial coordinate~$x$, one can use the
value of the Liouville field instead of the spatial coordinate.
The Ricci scalar is~$R=(2\lambda)^2\e^{-2a\psi}$, so the solution is
singular in the limit~$\psi\to-\infty$ ($x\to-\infty$) and is
asymptotically flat in the limit~$\psi\to\infty$ ($x\to\infty$).
The curvature is finite in the~$m\to0$ limit.  The solution possesses
a single event horizon
at~$\psi_\sss{\text{H}}=a^{-1}\log(\lambda/m)$.  The quasilocal
boundary should be chosen outside of the event horizon:
$\psi_\sss{\text{B}}>\psi_\sss{\text{H}}$.

The thermodynamic variables can be calculated for this solution.
The entropy and quasilocal energy have contributions arising from
both the dilaton and the Liouville fields; equations
\eqref{2d entropy} and~\eqref{2d energy} must be changed to include
the extra contributions of the Liouville field.  One finds that the
energy and entropy are
\begin{equation}
  E = (2md)\,N
  \label{2d Liouville energy}
\end{equation}
(evaluated on the quasilocal boundary) and
\begin{equation}
  S = (4\pi d) \log(m/\lambda) + S_\sss0
  \label{2d Liouville entropy}
\end{equation}
respectively with
\begin{displaymath}
  S_\sss0 = \frac{2\pi}{\kappa}\,\bigl( \phi_\sss0 - 2a\psi_\sss0
  \bigr) \:.
\end{displaymath}
Notice that the parameter~$m$ can be written in terms of the
extensive thermodynamic variables:
\begin{equation}
  m = \lambda \exp \bigl( d^{-1} ( -\phi/2\kappa + a\psi/\kappa
  + S/4\pi ) \bigr)
  \label{2d Liouville mass parameter}
\end{equation}
and the lapse function on the quasilocal boundary is given by
\begin{equation}
  N^2 = 1 - \exp\bigl( d^{-1}(\phi/\kappa - 2c\psi - S/2\pi)\bigr)\:.
  \label{2d Liouville lapse on B}
\end{equation}
Therefore, the energy can be written as a function of extensive
variables alone.

The intensive variables are calculated according to equations
\eqref{2d temperature} and~\eqref{2d dilaton force}.  In addition,
there is a force, $\nu$, conjugate to the Liouville field, which can
be calculated from the analog of~\eqref{2d dilaton force} with the
Liouville field replacing the dilaton field.  The temperature,
dilaton force, and Liouville force are
\begin{gather}
  T = \frac{m}{2\pi N}\,,
  \label{2d Liouville temperature} \\
  \mu = \frac{m}{\kappa N} - \mu_\sss0\,,
  \label{2d Liouville dilaton force} \\
  \intertext{and}
  \nu = \frac{2ma}{N}\,(N^2d-c/a) - \nu_\sss0
  \label{2d Liouville Liouville force}
\end{gather}
respectively.  Because the energy can be written in terms of the
extensive variables, one can also obtain the intensive variables from
the quasilocal energy with the aid of the first law of thermodynamics:
\begin{equation}
  T\,dS = dE + \mu\,d\phi + \nu\,d\psi \:.
  \label{2d Liouville first law of thermodynamics}
\end{equation}
This method gives the same expressions as in
equations~\eqref{2d Liouville temperature}--%
\eqref{2d Liouville Liouville force}.

A natural choice for the background action functional
is~$I^0=0$ which implies that $E_\sss0$, $\mu_\sss0$, and~$\nu_\sss0$
are all zero.  In the limit~$\phi\to\infty$, the quasilocal energy
is finite and approaches the value~$2md$.  Therefore, the
parameter~$m$ is related to the mass of the black hole.

One can compute the heat capacity at constant dilaton and Liouville
fields using equation~\eqref{heat capacity}.  One finds
\begin{equation}
  C_{\phi,\psi} = \frac{4\pi N^2 d}{2N^2-1}\:.
  \label{2d Liouville heat capacity}
\end{equation}
The heat capacity diverges
when~$\psi_{\text{crit}}=\psi_\sss{\text{H}}+a^{-1}\log2$
(i.e., $N^2=\half$).  For $\psi<\psi_{\text{crit}}$, the heat capacity
is negative but it approaches zero as~$\psi\to\psi_\sss{\text{H}}$.
For $\psi>\psi_{\text{crit}}$, the heat capacity is positive and it
approaches the value~$4\pi d$ as~$\psi\to\infty$.  This qualitative
behaviour is opposite to what one finds for the four-dimensional
Schwarzschild solution.

\chapter{Summary}
\label{c:summary}

I have developed a quasilocal formalism for analyzing the
thermodynamics of stationary solutions to a general class of
dilaton theories of
gravity with Abelian gauge fields and Yang-Mills matter fields.
This formalism is quite
robust: I have applied it to black hole spacetimes in two,
three, and four dimensions, to rotating black holes, and to
black holes possessing electric and magnetic charges.

The quasilocal method is useful for many reasons.  First, because
the quasilocal surface can be taken to be at any finite radius,
the particular asymptotic behaviour of the solution does not limit
the applicability of the formalism.  I have illustrated this by
calculating the thermodynamics of the {\rnads}
and the \acro{btz} black holes, which are both solutions
to the Einstein field equations with a negative cosmological constant.
Second, by putting a black hole in a box, certain technical
difficulties of statistical mechanics can be alleviated.
In particular, with a judicious choice of the position of the
quasilocal surface, the heat capacity of the system can be made
positive, and the system will be stable.  Also, for a rotating
black hole in equilibrium with a radiation fluid, the radiation
must be confined in a region small enough so that it can rotate
rigidly at the angular velocity of the black hole without travelling
faster than the speed of light.  Third, a quasilocal
analysis is generally preferable to one at spacelike infinity
because any physically realistic experiment must be performed on a
quasilocal surface.

Several interesting features of the quasilocal formalism have been
found.  The quasilocal thermodynamic variables are observer-dependent
in that they depend on how spacetime is foliated.  In other words,
the quasilocal variables depend on the motion of the observer;
furthermore, they also depend on an arbitrary reference spacetime,
so an observer may choose an arbitrary zero point for the
thermodynamic internal energy.  However, the laws of thermodynamics
are independent of these freedoms as one would expect.  In particular,
the first law of thermodynamics depends only on differences between
equilibrium systems, and thus the arbitrary reference value does
not appear.

When the quasilocal system possesses certain symmetries,
conserved charges that are not foliation dependent can be constructed.
Thus, when the quasilocal system is stationary, a conserved mass for
the system is present, and when the system has some sort of axial
symmetry, a conserved angular momentum is present.  These ``charges''
are conserved in the following way: if one is given a timelike history
of the
quasilocal boundary that contains the orbits of the Killing vector
field
associated with a symmetry, the value of the conserved charge
does not depend on the manner in which this boundary is foliated into
quasilocal surfaces.  In particular, the value of the charge will
be the same no matter which quasilocal surface is used to evaluate
the charge; the charge is conserved.  When gauge fields are
present, conserved charges can be associated with them too.  In the
case of electromagnetism, the familiar Gauss law is the means by
which the electromagnetic charge of the system is found.

The quasilocal energy at spacelike infinity is equivalent to the
\acro{adm} mass for asymptotically flat solutions in General
Relativity.  More generally, the quasilocal energy is equal to
the on-shell value of the gravitational Hamiltonian on a quasilocal
surface for which the lapse is unity and the shift vanishes.  More
importantly, though, the quasilocal energy is the
thermodynamic internal energy for a quasilocal system.  The conserved
mass parameter is more closely associated with the value of the
on-shell Hamiltonian and, thus, with the notion of the total energy
of the spacetime.  However, it is \emph{not} the thermodynamic
internal energy.  The conserved mass is equal to the quasilocal
energy at spacelike infinity for asymptotically flat spacetimes.

Under certain conditions, the solutions of two theories of dilaton
gravity will be related by a conformal transformation.  When this
is the case, it is of interest to find quantities that are invariant
under the conformal transformation.  In order to construct
such a quantity, further restrictions on the conformal properties
of the reference action functional need to be imposed.  I have
shown that for spherically symmetric spacetimes there is a
prescription for choosing the reference action functional that
yields a conformally invariant definition of the mass.  The entropy
of a black hole is also conformally invariant.  However, because
the redshift factor of the temperature is not conformally invariant,
the thermodynamic internal energy cannot be conformally invariant
either.

The notion of extensive and intensive thermodynamic variables
has been formally defined to refer to variables that, respectively,
can and cannot be
constructed from the phase-space variables on the quasilocal surface.
Actions of various statistical ensembles can be
constructed for which special combinations of extensive and intensive
variables must be held fixed on the quasilocal surface in order that
the equations of motion are generated from variations of the action.
The microcanonical action is defined as the action for which the
extensive variables alone must be fixed in generating the field
equations.  Thus the variation among classical solutions
of the microcanonical density of states depends on the variations
of the extensive thermodynamic variables alone.  Since the entropy
is the logarithm of the microcanonical density of states, the
variation of the entropy between nearby classical solutions yields
the first law of thermodynamics.

Unfortunately, the first law of thermodynamics is an integral equation
relating the variation of the entropy with the variations of the
extensive thermodynamic variables on the quasilocal surface.
Because the varied extensive variables within the integral over the
quasilocal surface contain intensive coefficients (such as the
inverse temperature), one cannot express the first law of
thermodynamics in the usual differential form unless these intensive
factors are constant on the quasilocal surface.  However, for
gravitating systems, the temperature will not be constant even
if the system is in ``thermodynamic equilibrium'' because
the temperature is blue-shifted by the gravitational field.
Isothermal quasilocal surfaces will not necessarily be the same
as isopotential surfaces, so one cannot in general simplify the
integral form of the first law of thermodynamics.

When a solution possesses very strong symmetries, the integral
form of the first law of thermodynamics can be reduced to a
differential form.  For example, spherically symmetric spacetimes
will always have a differential form if the quasilocal surfaces
are chosen to respect the spherical symmetries since the temperature
and all the potentials must also be spherically symmetric.
Furthermore, in three dimensions, the solution need only possess
circular symmetry, so rotating solutions such as the \acro{btz}
solution can be analyzed.  In two dimensions, the quasilocal
surface is just a point and, thus, the first law of thermodynamics is
automatically in a differential form.  All the black hole spacetimes
that I have examined possess the necessary symmetries to write
the first law of thermodynamics in a differential form.

I have examined two solutions to the Einstein field equations
with a negative cosmological constant.  The first solution was
the {\rnads} family of black hole spacetimes, which has
electromagnetic charge.  In the case of the electrically charged
{\rnads} solution, I was able to compute the thermodynamic variables
and to show explicitly that the first law of thermodynamics holds.
The magnetically charged {\rnads} solution was more difficult to
analyze because of the Dirac string singularity in the electromagnetic
potential.  In order to evaluate the electromagnetic work term,
I devised a trick to avoid the Dirac string singularity.  Once
again I found that the thermodynamic variables were consistent
with the first law of thermodynamics.  The natural choice of
reference spacetime for the {\rnads} solution is the {\ads} spacetime.
With this choice, I found that the quasilocal energy goes to zero
as the quasilocal surface approaches spacelike infinity.  However,
the quasilocal mass approaches a constant value equal to the
``mass parameter'' of the {\rnads} spacetimes.  The behaviour of
the heat capacity with the quasilocal system size depends on the
parameters of the solution.  In many cases, it is possible to find
a system size for which the heat capacity is positive.

The second asymptotically {\ads} solution that I examined was the
rotating \acro{btz} solution in three dimensions.  The thermodynamic
variables, including the rotational work term, were found to be
consistent with the first law of thermodynamics.  In this case,
the natural reference spacetime was taken to be the spacetime
in which the parameters associated with the mass and angular
momentum are set to zero.  One again, the quasilocal energy of
the black hole vanishes at spacelike infinity while the conserved mass
is equal to the ``mass parameter'' at spacelike infinity.  The
angular momentum, which is a conserved charge, is equal to the
``angular momentum parameter.''  The heat capacity of the \acro{btz}
black hole is positive for all quasilocal boundaries outside of the
event horizon; if the equilibrating radiation surrounding a rotating
black hole is considered, however, the system should be small enough
that the radiation fluid can rotate at less than the speed of light.

I analyzed four asymptotically flat solutions to dilaton gravity.
The two \acro{ghs} solutions are spherically symmetric solutions to a
string-inspired theory of dilaton gravity in four dimensions
possessing an electromagnetic charge.  This theory is conformally
related to General Relativity with a scalar field coupled to
a gauge field.  I showed explicitly that the masses of these
spacetimes are the same in the string frame and in the Einstein frame.
I computed the thermodynamic variables in the string frame and
obtained consistent thermodynamics.  For the solution possessing
a magnetic charge, I used the same trick as I used for the {\rnads}
magnetic solution.  The thermodynamic variables have well-behaved
asymptotic properties when Minkowski spacetime is chosen as the
reference.  The electric and magnetic solutions have heat
capacities that behave in a qualitatively similar manner: near
the black hole, the heat capacity is positive and diverges at
a critical radius.  Beyond the critical radius, the heat capacity
is negative and approaches a constant value for large system sizes.

In two-dimensional dilaton gravity, the quasilocal variables are
simplified considerably.  I calculated these variables for solutions
of two theories: a string-inspired two-dimensional theory and
the ``$R=\kappa T$'' theory with a Liouville field.  A consistent
thermodynamics was obtained for both of these solutions.  In the
case of the Liouville black hole solution, I modified the quasilocal
formalism slightly to include the non-minimally coupled Liouville
field.  The heat capacity of the string-inspired black hole is
positive everywhere outside the event horizon, while the Liouville
black hole has a heat capacity that has a similar behaviour to the
\acro{ghs} black holes, except that the sign is reversed.

The quasilocal formalism that I have presented is a
powerful tool in analyzing the thermodynamics of solutions to
a wide class of gravitational theories.  The primary purpose of
extending the quasilocal formalism to include a larger class of
gravitational theories is so that the thermodynamic properties
of spacetimes in these theories can be more easily compared to
the familiar thermodynamics of black holes in General Relativity.
Hopefully, when a better understanding of the quantum nature of
black holes in the alternate theory is obtained, one will gain
insight into the quantum nature of General Relativity.  Additionally,
the quasilocal formalism I have presented encompasses several
theories of gravity that have been proposed to extend General
Relativity.

At present, the quasilocal formalism has only been applied to
a few black hole spacetimes.  It would be interesting to apply
the quasilocal formalism to a charged, rotating solution to illustrate
the effect of the electromotive force work term.  However, it may
not be possible to obtain exact results for the complicated
Kerr-Newman spacetime, and it is not yet clear how physical the
charged rotating $(2+1)$-dimensional black hole spacetimes are.
A detailed analysis of asymptotically de\thinspace Sitter black
holes would also be of interest though it is not clear exactly
where the quasilocal surface should be located.
Future work in quasilocal theories of gravity may include
higher curvature terms in the Lagrangian.  Such terms are often
included in order to renormalize General Relativity, and the
effective action of string theories include such terms at
higher energies.  The study of additional types of matter fields, such
as ``topological matter'' which couples to the connection, would make
an interesting extension to the gauge fields that I have studied here.
Further work should also be conducted in
determining the semi-classical corrections to the quasilocal
variables arising from the inclusion of the radiation field that
surrounds the black hole.  In addition to these ``on-shell''
corrections to the variables that are useful in studying
thermodynamics, one could include ``off-shell'' corrections
that would be useful in understanding the statistical mechanics
of the black hole spacetimes as well as in studying non-equilibrium
processes and evolution of black hole spacetimes.

\appendix

\chapter{Action Principle for a Non-Gravitating System}
\label{a:action}

This appendix serves as an illustration of how one can use
the boundary terms derived from the action of a system to define a
quasilocal energy, and how one can find the entropy of a system
given its action via path integral techniques.
I restrict attention here to non-gravitating systems.
The analysis of chapter~\ref{c:quasi} is
closely related to the Hamilton-Jacobi theory, which I review.
A similar review is given by Brown and York~\cite{by:93a};
more detail can be found in the following references.
The Weiss action principle and Noether's theorem are standard results
in classical mechanics (see, e.g., \cite{ms:91});
a detailed discussion of Jacobi's action and its analog for
gravitating systems is found in reference~\cite{by:89}; the
microcanonical functional integral~\cite{by:93b} is related to
the path-integral definition of the canonical partition
function~\cite{fh:65,gh:77}.

\section{Weiss Action Principle}

Let $\varSigma$ be an $(n-1)$-dimensional spacelike manifold called
configuration space; the coordinates on~$\varSigma$ are $\{x\}$.
Paths on configuration space are parameterized by a time variable,
$t$.  The \emph{Lagrangian}, $L(x,\dot{x},t)$, is defined for a
path~$\gamma=x(t)$.  Here, an overdot indicates a total derivative
with respect to the time parameter.  Suppose that the endpoints
of the path~$\gamma$ are $A=x(t_A)$ and~$B=x(t_B)$.  The \emph{action}
of this path is
\begin{equation}
  I[\gamma] = \int_{t_A}^{t_B} L(x,\dot{x},t)\,dt \qquad
  \text{(along $\gamma$)}\:.
  \label{cl action}
\end{equation}
Suppose that a nearby path, $\gamma'$, is related to~$\gamma$ by
\begin{equation}
  x'(t)=x(t)+\varepsilon\eta(t)
  \label{cl deviation}
\end{equation}
where $\eta(t)$ is some deviation function and $\varepsilon$ is a
small parameter.  The endpoints of~$\gamma'$ are also shifted
relative to those of~$\gamma$:
\begin{equation}
  t_{A'} = t_A + \varepsilon s_A \quad \text{and} \quad
  t_{B'} = t_B + \varepsilon s_B \:.
  \label{cl endpoint deviation}
\end{equation}
The action of the path~$\gamma'$ can be related to the action
of the path~$\gamma$ as follows:
\begin{equation}
  \begin{split}
    I[\gamma'] &= \int_{t_{A'}}^{t_{B'}} L(x',\dot{x}',t)\,dt \\
    &= \int_{t_A}^{t_B} L(x+\varepsilon\eta,\dot{x}
    +\varepsilon\dot{\eta},t)\,dt \\
    &\qquad + \int_{t_B}^{t_B+\varepsilon s_B}
    L(x,\dot{x},t)\,dt - \int_{t_A}^{t_A+\varepsilon s_A}
    L(x,\dot{x},t)\,dt + O(\varepsilon^2) \\
    &= I[\gamma] + \varepsilon\int_{t_A}^{t_B} \biggl(
    \frac{\partial L}{\partial x}\,\eta
    + \frac{\partial L}{\partial\dot{x}}\,\dot{\eta} \biggr)\,dt
    + \varepsilon[Ls]_{t_A}^{t_B} + O(\varepsilon^2)\:.
  \end{split}
  \label{cl action prime}
\end{equation}
The second term in the integral is integrated by parts; then, the
difference, $\delta I=I[\gamma']-I[\gamma]$, between the actions
along the path~$\gamma'$ and the path~$\gamma$ is given by
\begin{equation}
  \delta I = \varepsilon\int_{t_A}^{t_B} \biggl(
  \frac{\partial L}{\partial x} - \frac{d}{dt}
  \frac{\partial L}{\partial\dot{x}} \biggr)\eta\,dt
  + \varepsilon\biggl[ \frac{\partial L}{\partial\dot{x}}\,\eta
  + Ls \biggr]_{t_A}^{t_B} + O(\varepsilon^2) \:.
  \label{cl var action 1}
\end{equation}
The coordinate values of the endpoints of the two paths differ
by the following amounts:
\begin{equation}
  \begin{split}
     \delta x(t_A) &= x'(t_{A'}) - x(t_A) \\
     &= \varepsilon\bigl( \dot{x}(t_A)s_A + \eta(t_A) \bigr)
     + O(\varepsilon^2)
  \end{split}
  \label{cl delta x A}
\end{equation}
for the initial endpoint and, similarly,
\begin{equation}
   \delta x(t_B) = \varepsilon\bigl( \dot{x}(t_B)s_B + \eta(t_B)
   \bigr) + O(\varepsilon^2)
   \label{cl delta x B}
\end{equation}
for the final endpoint.  The parameter values of the endpoints
differ by
\begin{align}
  \delta t(t_A) &= t_{A'} - t_A = \varepsilon s_A
  \label{cl delta t A} \\
  \intertext{and}
  \delta t(t_B) &= t_{B'} - t_B = \varepsilon s_B
  \label{cl delta t B}
\end{align}
for the initial and final endpoints respectively.  With these
definitions, the quantity~$\delta I$ can be written, to first
order in the small parameter~$\varepsilon$, as
\begin{equation}
  \delta I = \varepsilon\int_{t_A}^{t_B} \biggl(
  \frac{\partial L}{\partial x} - \frac{d}{dt}
  \frac{\partial L}{\partial\dot{x}} \biggr)\eta\,dt
  + \biggl[ \frac{\partial L}{\partial\dot{x}}\,\delta x
  + \Bigl(L - \frac{\partial L}{\partial\dot{x}}\,\dot{x}\Bigr)
  \delta t \biggr]_{t_A}^{t_B} \:.
  \label{cl var action 2}
\end{equation}

If the endpoints are fixed so that the two paths, $\gamma'$ and
$\gamma$, have the same coordinate and parameter values at their
endpoints, then only the integral will remain in
equation~\eqref{cl var action 2}.  Since the deviation function,
$\eta(x)$, is arbitrary, one finds that the path for which the
action has an extremal value (so that~$\delta I=0$) is the one
for which the Euler-Lagrange equations of motion,
\begin{equation}
  \frac{\partial L}{\partial x} - \frac{d}{dt}
  \frac{\partial L}{\partial\dot{x}} = 0\,,
  \label{cl eom}
\end{equation}
hold.  Conversely, if the two paths, $\gamma'$ and $\gamma$, are
classical trajectories for which the Euler-Lagrange equations of
motion hold, then the only contribution to the difference in the
actions~$\delta I$ is due to the endpoint contributions.  Thus,
the \emph{Weiss action principle} states that a system follows
the path~$\gamma$ for which variations of the action contain only
the endpoint contributions:
\begin{equation}
  \delta I = [ p\,\delta x - E\,\delta t ]_{t_A}^{t_B}
  \label{Wiess action principle}
\end{equation}
where the \emph{momentum} of the system is given by~$p$ and the
\emph{energy} of the system is given by~$E$.

By comparing equation~\eqref{Wiess action principle} with
equation~\eqref{cl var action 2}, one sees that the momentum
and energy of a system are given by
\begin{equation}
  p = \frac{\partial L}{\partial\dot{x}} \quad \text{and} \quad
  E = \frac{\partial L}{\partial\dot{x}}\,\dot{x} - L
  \label{cl momentum and energy}
\end{equation}
respectively; these expressions should be evaluated for a solution
to the Euler-Lagrange equations.  The momentum and energy can also
be written as the functional derivatives $\delta I/\delta x$
and~$-\delta I/\delta t$ respectively.  One obtains the
Hamilton-Jacobi equation:
\begin{equation}
  -\frac{\delta I}{\delta t} = H\Bigl(x,\frac{\delta I}{\delta x},t
  \Bigr)
  \label{Hamilton-Jacobi}
\end{equation}
where $H(x,p,t)=p\dot{x}-L$ is the Hamiltonian.

\section{Noether's Theorem}

Spacetime is defined as the $n$-dimensional manifold~$\mathcal{M}$
that has the topology of the direct product of the topology of
the configuration manifold~$\varSigma$ with a real interval.
The coordinates on spacetime are~$\{t,x\}$.  Suppose that a variation
in the path~$\gamma$ is generated by a spacetime vector~$\xi^a$ via
\begin{equation}
  \delta t = \xi^a\partial_a t \quad \text{and} \quad
  \delta x = \xi^a\partial_a x \:.
  \label{cl var t and x}
\end{equation}
The Lagrangian is invariant under such a variation if
\begin{equation}
  \xi^a\partial_a L = 0 \:.
  \label{cl invariant Lagrangian}
\end{equation}
In this case, the variation in the action must vanish:
\begin{equation}
  0 = \delta I = [ p \xi^a\partial_a x 
  - E \xi^a\partial_a t ]_{t_A}^{t_B}
  \label{cl invariant action}
\end{equation}
so the quantity in brackets has the same value when evaluated
at~$t_A$ as it has when evaluated at~$t_B$.  This quantity, then,
is a conserved quantity associated with the generator~$\xi^a$.
\emph{Noether's Theorem} states that, for every diffeomorphism
generated by~$\xi^a$ under which the Lagrangian is invariant,
there is a conserved charge,
\begin{equation}
  q[\xi] = p \xi^a\partial_a x - E \xi^a\partial_a t,
  \label{cl Noether charge}
\end{equation}
that is a constant of motion.

I present two examples of Noether's theorem for Lagrangians that
are invariant under temporal and spatial translations.
\begin{itemize}
  \item When $\xi^a$ is a temporal vector~$t^a$ so that the Lagrangian
    is invariant if~$\partial L/\partial t=0$, then the conserved
    Noether charge is
    \begin{equation}
       q[t]=-E
       \label{cl t-conserved Noether charge}
    \end{equation}
    since $\delta x=0$ and~$\delta t=1$.  Hence the energy of a
    system is conserved when the Lagrangian is invariant under time
    translations.
  \item When $\xi^a=x^a$ generates a spacelike transformation for
    which $\delta x=1$, $\delta t=0$, and~$\partial L/\partial x=0$,
    then the conserved Noether charge is the momentum:
    \begin{equation}
       q[x]=p\:.
       \label{cl x-conserved Noether charge}
    \end{equation}
    Thus, the momentum is a conserved charge when the Lagrangian is
    invariant under translations along the vector~$x^a$.
\end{itemize}

\section[The Microcanonical Action and Functional Integral]%
  {The Microcanonical Action and\\ Functional Integral}

A stationary system is one for which the Lagrangian density has no
explicit dependence on the time parameter.  Such a system possesses
the conserved Noether charge~$q[t]=-E$.  The microcanonical action,
$I_{\text{m}}$, is related to the action~$I$ via the canonical
transformation
\begin{equation}
  \begin{split}
    I_{\text{m}}[\gamma] &= I[\gamma] - q[t]\,t |_{t_A}^{t_B} \\
    &= \int_{t_A}^{t_B} L(x,\dot{x})\,dt + [Et]_{t_A}^{t_B} \:.
  \end{split}
  \label{cl micro action}
\end{equation}
Under variations about a path~$\gamma$ corresponding to a classical
solution, the variation in the microcanonical action is
\begin{equation}
  \delta I_{\text{m}}[\gamma] = [ p\,\delta x + t\,\delta E 
  ]_{t_A}^{t_B} \:.
  \label{cl var micro action}
\end{equation}
Therefore, the microcanonical action is a functional of the energy
of the path~$\gamma$ rather than the time of the endpoints
of~$\gamma$; the microcanonical action is extremized for variations
of fixed energy (and endpoint coordinate values) about a classical
solution.  Notice that the microcanonical action has the same
functional dependence on the coordinate values of the endpoints
of the path as the action~$I$:
$\delta I_{\text{m}}/\delta x=\delta I/\delta x$.

For a stationary system, the Hamilton-Jacobi equation has the
simplified form
\begin{equation}
  E = H\Bigl(x,\frac{\delta I_{\text{m}}}{\delta x}\Bigr) \:.
  \label{stationary Hamilton-Jacobi}
\end{equation}
Suppose that this equation can be solved
for~$\delta I_{\text{m}}/\delta x$; then, the Jacobi action is
defined by
\begin{equation}
  W(x,E) = \int^x \frac{\delta I_{\text{m}}}{\delta x}(x,E)\,dx \:.
  \label{cl Jacobi action}
\end{equation}
The microcanonical action and the Jacobi action are closely related,
but the microcanonical action is a functional of the position and
the energy of a system whereas the Jacobi action is an ordinary
function of the position and energy.  Thus,
$\partial W/\partial E=\delta I_{\text{m}}/\delta E=t_B-t_A$ and
$\partial W/\partial x=\delta I_{\text{m}}/\delta x=p$.

The microcanonical action can be written in canonical form as
follows.  Let the path~$\gamma$ be parameterized by
the parameter~$\sigma$:
\begin{equation}
  \begin{split}
    I_{\text{m}}[\gamma] &= \int_{\sigma_A}^{\sigma_B} \Bigl(
    p\,\frac{dx}{d\sigma} - H(x,p) \Bigr)\,d\sigma
    - [Et]_{t(\sigma_A)}^{t(\sigma_B)} \\
    &= \int_{\sigma_A}^{\sigma_B} \Bigl(
    p\,\frac{dx}{d\sigma} - N\mathcal{H}(x,p) \Bigr)\,d\sigma
  \end{split}
  \label{cl canon micro action}
\end{equation}
where $\mathcal{H}=H-E$ is the \emph{Hamiltonian constraint}
and~$N$, which is defined by~$N\,d\sigma=dt$, measures the lapse
of time with parameter~$\sigma$.  There is a gauge freedom in the
choice of the parameterization of the path and, thus, in the choice
of the lapse function~$N(\sigma)$.  However, this gauge freedom
can be removed by redefining the lapse function as the constant
\begin{equation}
  N = \frac{\tau}{\sigma_B-\sigma_A}
  \label{cl constant lapse}
\end{equation}
where $\tau$ is the integral of the original lapse function over
the path.  The gauge-fixed microcanonical action in canonical form
is a functional of $x$ and~$p$ and is an ordinary function of~$\tau$:
\begin{equation}
  I_{\text{m}}[x,p;\tau) = \int_{\sigma_A}^{\sigma_B} \Bigl(
  p\,\frac{dx}{d\sigma} - \frac{\tau}{\sigma_B-\sigma_A}\,
  \mathcal{H}(x,p) \Bigr)\,d\sigma \:.
  \label{cl canon micro action gauge fixed}
\end{equation}

The microcanonical density matrix is defined by the functional
integral
\begin{equation}
  \rho(x_B,x_A) = \frac{1}{2\pi\hbar} \int d\tau \int [\varGamma]\,
  \exp\biggl\{ \frac{i}{\hbar} \int_{\sigma_A}^{\sigma_B} \Bigl(
  p\,\frac{dx}{d\sigma} - \frac{\tau}{\sigma_B-\sigma_A}\,
  \mathcal{H}(x,p) \Bigr)\,d\sigma \biggr\}
  \label{cl micro density matrix}
\end{equation}
where $[\varGamma]$ is the measure of paths in phase-space with
endpoints $x_A$ and~$x_B$.  The \emph{density of states} is defined
as the trace of the microcanonical density matrix; thus, the
density of states is given by the microcanonical functional integral
\begin{equation}
  \begin{split}
    \nu(E) &= \int \rho(x,x)\,dx \\
    &= \frac{1}{2\pi\hbar} \int d\tau \oint [\varGamma]\,
    \exp\biggl\{ \frac{i}{\hbar} \int_{\sigma_A}^{\sigma_B} \Bigl(
    p\,\frac{dx}{d\sigma} - \frac{\tau}{\sigma_B-\sigma_A}\,
    \mathcal{H}(x,p) \Bigr)\,d\sigma \biggr\}
    \end{split}
  \label{cl micro density of states}
\end{equation}
where the symbol~$\oint$ is used to indicate that the path integral
is taken over all \emph{periodic} paths
with~$x(\sigma_B)=x(\sigma_A)$.  The density of states can also
be written in terms of the the evolution
operator~$\exp(-i\tau\hat{\mathcal{H}}/\hbar)$
as~$\nu(E)=\Tr\bigl(\delta(\hat{\mathcal{H}})\bigr)
=\Tr\bigl(\delta(E-\hat{H})\bigr)$.\footnote{This result is
explicitly calculated from the microcanonical functional
integral~\eqref{cl micro density of states} for the harmonic
oscillator in reference~\cite{by:93c}.}

\chapter{Boundaries of the Manifold}
\label{a:man}

Let spacetime be an $n$-dimensional Lorentzian manifold equipped with a
metric~$g_{ab}$.  I consider a region of this manifold,
$\mathcal{M}$, and I define various tensors
on the boundary.  I require that the region~$\mathcal{M}$ have
the topology of the direct product of a spacelike hypersurface,
$\varSigma$, with a real (timelike) interval.  This requirement allows
me to foliate the manifold into leaves~$\varSigma_t$ of constant
foliation parameter~$t$.  The vector~$t^a$ is defined by the
condition~$t\cdot\mb{d}t=1$; $t^a$ is not generally equal to the
normal vector, $u^a$, to the surfaces~$\varSigma_t$.

\begin{figure}[t]
\renewcommand{\baselinestretch}{1}\small
\begin{center}
\setlength{\unitlength}{.75mm}
\begin{picture}(100,100)
  \put(10,10){\line(1,0){80}}
  \put(10,10){\line(0,1){80}}
  \put(10,90){\line(1,0){40}}
  \qbezier(50,90)(50,70)(70,50)
  \qbezier(70,50)(90,30)(90,10)
  \qbezier(10,40)(60,40)(70,50)
  \put(70,50){\vector(1,1){20}}
  \put(70,50){\vector(-1,1){20}}
  \put(70,50){\circle*{2}}
  \put(90,70){$n^a$}
  \put(44,70){$u^a$}
  \put(20,33){$\varSigma_t$,$\underline{\mb{\epsilon}}$}
  \put(20,60){$\mathcal{M}$,$\mb{\epsilon}$}
  \put(52,80){$\mathcal{T}$,$\overline{\mb{\epsilon}}$}
  \put(73,48){$\mathcal{B}_t$,$\underline{\overline{\mb{\epsilon}}}$}
\end{picture}
\end{center}
\begin{quote}\leavevmode
\caption[Manifolds, boundaries, and normal vectors]{\small
  The manifold,
  its boundaries, and the normal vectors to these boundaries.}
\label{fig:man}
\end{quote}
\end{figure}
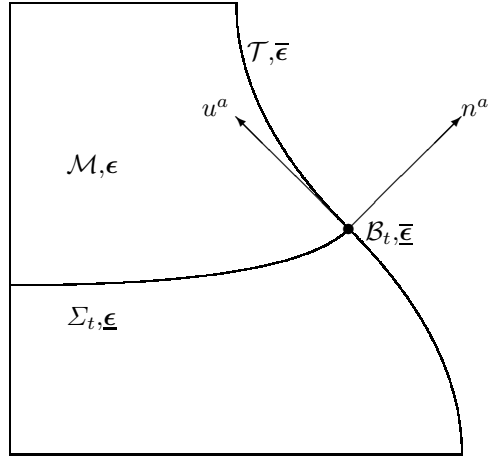

The boundary of the region~$\mathcal{M}$
is the union of ``initial'' and ``final'' hypersurfaces,
$\varSigma_{\text{i}}$ and~$\varSigma_{\text{f}}$, and the timelike
hypersurface~$\mathcal{T}$.  This timelike hypersurface can also
be foliated into the spacelike \emph{quasilocal}
surfaces~$\mathcal{B}_t$.  The (spacelike) normal vector to the
boundary element~$\mathcal{T}$ is~$n^a$ while the bi-normal
to the quasilocal surface~$\mathcal{B}$ is~$n^{ab}=2u^{[a}n^{b]}$.
I orient the normal vectors to be directed outwards on the timelike
boundary elements and future-directed on the spacelike boundary
elements.
For simplicity, I enforce the condition~$g_{ab}u^an^b=0$.
The volume element on the manifold~$\mathcal{M}$
is~$\mb{\epsilon}=\sqrt{-g}\,\mb{d}x^1\wedge\cdots\wedge\mb{d}x^n$
where~$\{x\}$ is a local coordinate system and~$\sqrt{-g}$ is the
square-root of the negative of the determinant of the metric in
these local coordinates.  The volume element on the spacelike
manifold~$\varSigma$
is~$\underline{\mb{\epsilon}}=\underline{u\cdot\mb{\epsilon}}$;
on the timelike manifold~$\mathcal{T}$, the volume element
is~$\overline{\mb{\epsilon}}=\overline{n\cdot\mb{\epsilon}}$;
the quasilocal surface~$\mathcal{B}$ has the volume
element~$\underline{\overline{\mb{\epsilon}}}=
\overline{n\cdot\underline{\mb{\epsilon}}}$.  These various
manifolds are illustrated in figure~\ref{fig:man}.

If $\mb{\omega}=\omega\cdot\mb{\epsilon}$ is an $(n-1)$-form, then
the pull back onto~$\varSigma$
is~$\underline{\mb{\omega}}=-\omega^au_a\underline{\mb{\epsilon}}$,
while the pull back onto~$\mathcal{T}$
is~$\overline{\mb{\omega}}=\omega^an_a\overline{\mb{\epsilon}}$.
Furthermore, if~$\mb{\alpha}=\alpha\cdot\underline{\mb{\epsilon}}$
is an $(n-2)$-form on~$\varSigma$, then the pull back
onto~$\mathcal{B}=\partial\varSigma$
is~$\overline{\mb{\alpha}}=\omega^an_a
\underline{\overline{\mb{\epsilon}}}$.  Finally,
if~$\mb{\alpha}=\alpha\cdot\overline{\mb{\epsilon}}$ is an
$(n-2)$-form on~$\mathcal{T}$, then the pull back
onto~$\mathcal{B}=\partial\mathcal{T}$
is~$\underline{\mb{\alpha}}=\omega^au_a
\underline{\overline{\mb{\epsilon}}}$.  This last expression
follows because~$\underline{\overline{\mb{\epsilon}}}=
-\underline{u\cdot\overline{\mb{\epsilon}}}$.

\section{The Timelike Boundary}
\label{s:man time}

I now examine the geometry on the timelike
boundary~$\mathcal{T}$.
The two \emph{fundamental forms} on~$\mathcal{T}$ are the
\emph{induced metric}~$\gamma_{ab}=g_{ab}-n_an_b$, and the
\emph{extrinsic curvature}~$\varTheta_{ab}=-\half\Lie_n\gamma_{ab}$.
The restriction of the induced metric to the boundary~$\mathcal{T}$
can be viewed as the physical metric on the $(n-1)$-dimensional
manifold~$\mathcal{T}$.  Alternately, one can view the
operator~$\gamma^a{}_b$ as a projection operator that will take a
vector on the tangent space of~$\mathcal{M}$ to a vector on the
tangent space of~$\mathcal{T}$.  The derivative operator compatible
with the metric~$\gamma_{ab}$ is~$\nablat$.

The second fundamental form can be thought of as the failure of the
vectors~$n^a$ to coincide when parallel-transported along the
boundary~$\mathcal{T}$.  The definition of extrinsic curvature yields
\begin{equation}
  \begin{split}
    \varTheta_{ab} &= -\half\Lie_n \gamma_{ab} \\
                   &= -\half\Lie_n ( g_{ab} - n_a n_b ) \\
                   &= -( \nabla_a n_b - n^c n_a \nabla_c n_b ) \\
                   &= -\gamma^c{}_a \nabla_c n_b\,,
  \end{split}
  \label{forms of extrinsic curvature}
\end{equation}
which is the difference between the normal vector and the parallel
transport (along $\mathcal{T}$) of a nearby normal.  In the third
line I have used the hypersurface orthogonality of the unit vector,
the definition of the Lie derivative, and the compatibility of the
metric with the derivative operator~$\nabla$.  From this viewpoint,
the second fundamental form describes how curved
the surface~$\mathcal{T}$ is.  However, this description requires the
embedding of~$\mathcal{T}$ in some higher dimensional space since
the description is based on the normal vector.

\section{The Spacelike Boundary}
\label{s:man space}

On the spacelike boundary~$\varSigma$, the induced metric
is~$h_{ab}=g_{ab}+u_au_b$ while the extrinsic curvature
of~$\varSigma$ embedded in~$\mathcal{M}$
is~$K_{ab}=-\half\Lie_u h_{ab}=-h^c{}_a\nabla_cu_b$.  Here, the
operator~$h^a{}_b$ is a projection operator onto the tangent space
of~$\varSigma$.
Define the \emph{lapse function} as the normalization of the
unit normal~$u^a$ relative to the vector~$t^a$:
$N=(u\cdot\mb{d}t)^{-1}$.  The \emph{shift vector}~$N^a=h^a{}_bt^b$
is the projection of the vector~$t^a$ onto the surface~$\varSigma$.
Then the vector~$t^a$ can be decomposed into a portion normal
to~$\varSigma$ and a portion tangent to~$\varSigma$: $t^a=Nu^a+N^a$.

Let~$\nablas$ be the derivative operator compatible with the
metric~$h_{ab}$.  It is straightforward to show
that~$\nablas_aT^{b\cdots c}{}_{d\cdots e}=\perp\nabla_a
T^{b\cdots c}{}_{d\cdots e}$ where~$\perp$ indicates that all indices
are projected onto~$\varSigma$ using~$h^a{}_b$.  The curvature of
the derivative operator~$\nablas$ is written~$R_{abcd}[h]$.  It
is related to the curvature of the derivative operator~$\nabla$,
$R_{abcd}[g]$, by the \emph{Gauss-Codacci} relations:
\begin{gather}
  R_{abcd}[h]=\perp R_{abcd}[g] + 2 K_{d[a} K_{b]c}
  \label{Gauss Codacci a}\\
  \intertext{and}
  2\nablas_{[b} K^b{}_{a]} = n^c h^d{}_a R_{cd}[g]\:.
  \label{Gauss Codacci b}
\end{gather}
An immediate consequence of~\eqref{Gauss Codacci a} is
\begin{equation}
  \begin{split}
    2 u^a u^b G_{ab}[g] &= h^{ac} h^{bd} R_{abcd}[g] \\
                        &= R[h] + K^2 - K^{ab} K_{ab}
  \end{split}
  \label{uu Einstein}
\end{equation}
where $K=h^{ab}K_{ab}$ and~$G_{ab}[g]$ is the Einstein tensor
calculated for the metric~$g_{ab}$.  Furthermore, using the definition
of the Riemann tensor, one finds
\begin{equation}
  u^a u^b R_{ab}[g] = K^2 - K^{ab} K_{ab} + \nabla_b ( u^b K + a^b )
  \label{uu Ricci}
\end{equation}
where $a^b=u^a\nabla_au^b$ is the \emph{acceleration} of the unit
normal~$u^a$.  Combining \eqref{uu Einstein} and~\eqref{uu Ricci},
one finds
\begin{equation}
  \begin{split}
    R[g] &= 2 u^a u^b ( G_{ab}[g] - R_{ab}[g] ) \\
      &= R[h] + K^{ab} K_{ab} - K^2 - \nabla_b ( u^b K + a^b )\:.
  \end{split}
  \label{Ricci scalar}
\end{equation}
Finally, the extrinsic curvature~$K_{ab}$ can be expressed in terms
of the time derivative of the induced metric~$h_{ab}$:
\begin{equation}
  K_{ab} = -\half N^{-1}( \Lie_t h_{ab} - 2\nablas_{(a}N_{b)} )\:.
  \label{extrinsic curvature as velocity}
\end{equation}
The derivations of equations
\eqref{Gauss Codacci a}--\eqref{extrinsic curvature as velocity} can
be found in, e.g., Wald~\cite{w:84}.

\section{The Quasilocal Surface}
\label{s:man quasi}

Because $u^a$ and~$n^a$ are taken to be orthogonal, the quasilocal
surface~$\mathcal{B}$ can be viewed either as the boundary
of~$\varSigma$ or as a leaf in the foliation of~$\mathcal{T}$.
The induced metric on~$\mathcal{B}$
is~$\sigma_{ab}=g_{ab}+u_au_b-n_an_b$.  The extrinsic curvature
of~$\mathcal{B}$ embedded in~$\varSigma$
is~$k_{ab}=-\sigma^c{}_a\nablas_cn_b$.  A straightforward analysis
yields~\cite{by:93a} the following relationship between the various
extrinsic curvatures:
\begin{equation}
  \varTheta_{ab} = k_{ab} + u_a u_b n_c a^c
    + 2 \sigma^c{}_{(a} u_{b)} n^d K_{cd} \:.
  \label{extrinsic curvature relationship}
\end{equation}
Thus, the projection of~$\varTheta_{ab}$ onto~$\mathcal{B}$
is the extrinsic curvature~$k_{ab}$.  Furthermore, the projection
of~$\varTheta_{ab}$ along the normal~$u^a$ is~$n_ca^c$.  The
quantities $\varTheta_{ab}$ and~$K_{ab}$ are related
by~$\sigma^a{}_cu^b\varTheta_{ab}=-\sigma^a{}_cn^bK_{ab}$.  Finally,
the trace of~\eqref{extrinsic curvature relationship} yields
\begin{equation}
  \varTheta = k - n_c a^c \:.
  \label{trace extrinsic curvature relationship}
\end{equation}
where $\varTheta$ and~$k$ are the traces of the extrinsic curvatures
$\varTheta_{ab}$ and~$k_{ab}$ respectively.

Variations of the induced metric, $\gamma_{ab}$, on the timelike
boundary~$\mathcal{T}$ can be decomposed into pieces that are
normal-normal,
normal-tangential, and tangential-tangential to the quasilocal
surface~$\mathcal{B}$:
\begin{equation}
  \delta\gamma_{ab} = \sigma^c{}_a \sigma^d{}_b\,\delta\sigma_{cd}
  - \frac{2}{N}\,u_a u_b\,\delta N - \frac{2}{N}\,u_{(a} \sigma_{b)c}
  \,\delta N^c\:.
  \label{decomposition of timelike metric}
\end{equation}
The variation of the metric on the quasilocal surface can also be
decomposed into a variation of the square-root of the determinant,
$\sqrt{\sigma}$, plus a variation of the conformally invariant
part of the
metric~$\varsigma_{ab}=(\sqrt{\sigma})^{-2/(n-2)}\sigma_{ab}$:
\begin{equation}
  \delta\sigma_{ab} = \frac{2}{n-2}\,\biggl(
  \frac{\sigma_{ab}}{\sqrt{\sigma}} \biggr)\,\delta\sqrt{\sigma}
  + (\sqrt{\sigma})^{2/(n-2)}\,\delta\varsigma_{ab}\:.
  \label{decomposition of quasilocal metric}
\end{equation}
The second term in
equation~\eqref{decomposition of quasilocal metric} represents
changes in the ``shape'' of the quasilocal surface that preserve
the determinant whereas the changes in the determinant---given
by the first term---reflect a change in the ``size'' of the
quasilocal surface while maintaining the same shape.

A summary of the definition of the above quantities appears in
table~\ref{tab:not}.
\begin{table}[b]
\renewcommand{\baselinestretch}{1}\small
\begin{quote}\leavevmode
\caption[Summary of manifold variables]{\small A summary of the tensor
  variables defined on the various manifolds.  For the
  boundary~$\mathcal{B}$ there is a normal
  bi-vector~$n^{ab}=2u^{[a}n^{b]}$ (I require that $u^a$
  and~$n^a$ are orthogonal) and~$k_{ab}$ is the extrinsic
  curvature of~$\mathcal{B}$ embedded in~$\varSigma$; the other
  extrinsic curvatures are as embedded in~$\mathcal{M}$.}
\label{tab:not}
\end{quote}
\begin{center}
\newlength{\longest}
\settowidth{\longest}{Compatible derivative \ldots}
\begin{tabular}{p{\longest}*{4}{c}}
  \hline
   & \multicolumn{4}{c}{Manifold}\\
  \cline{2-5}
   & $\mathcal{M}$ & $\mathcal{T}$ & $\varSigma$ & $\mathcal{B}$ \\
  \hline\hline
  Volume form \dotfill& $\mb{\epsilon}$ & $\overline{\mb{\epsilon}}$ &
    $\underline{\mb{\epsilon}}$ & $\underline{\overline{\mb{\epsilon}}}$\\
  Normal (bi-)vector \dotfill& -- & $n^a$ & $u^a$ & $n^{ab}$ \\
  Metric \dotfill& $g_{ab}$ & $\gamma_{ab}$ & $h_{ab}$ & $\sigma_{ab}$ \\
  Compatible derivative \dotfill& $\nabla_a$ & $\nablas_a$ & $\nablat_a$ & \\
  Intrinsic curvature \dotfill& $R_{abcd}[g]$ & & $R_{abcd}[h]$ & \\
  Extrinsic curvature \dotfill& -- & $\varTheta_{ab}$ & $K_{ab}$ & $k_{ab}$ \\
  \hline
\end{tabular}
\end{center}
\end{table}


\begin{thebibliography}{10}
\newcommand{\enquote}[1]{``#1''}

\bibitem{ad:82}
Abbott, L.~F. and Deser, S. (1982) \enquote{Charge Definition in Nonabelian
  Gauge Theories,} \emph{Phys. Lett. B}, \textbf{116}, 259.

\bibitem{adm:62}
Arnowitt, R., Deser, S., and Misner, C.~W. (1962) \enquote{The Dynamics of
  General Relativity,} in \emph{Gravitation: An Introduction to Current
  Research}, ed. L.~Witten, (New York: Wiley).

\bibitem{bhtz:93}
Ba{\~n}ados, M., Henneaux, M., Teitelboim, C., and Zanelli, J. (1993)
  \enquote{Geometry of the (2+1) Black Hole,} \emph{Phys. Rev. D}, \textbf{48},
  1506--1525.

\bibitem{btz:92}
Ba{\~n}ados, M., Teitelboim, C., and Zanelli, J. (1992) \enquote{The Black Hole
  in Three-Dimensional Space-Time,} \emph{Phys. Rev. Lett.}, \textbf{69},
  1849--1851.

\bibitem{bch:73}
Bardeen, J.~M., Carter, B., and Hawking, S.~W. (1973) \enquote{The Four Laws of
  Black Hole Mechanics,} \emph{Comm. Math. Phys.}, \textbf{31}, 161--170.

\bibitem{b:72}
Bekenstein, J.~D. (1972) \enquote{Black Holes and the Second Law,} \emph{Nuovo
  Cimento Lett.}, \textbf{4}, 737--740.

\bibitem{b:80}
Bekenstein, J.~D. (1980) \enquote{Black-Hole Thermodynamics,} \emph{Physics
  Today}, \textbf{33}, 24--31.

\bibitem{b:94}
Bekenstein, J.~D. (1994) \enquote{Do We Understand Black Hole Entropy?}
  preprint: \texttt{gr-qc/9409015}.

\bibitem{bd:61}
Brans, C. and Dicke, R.~H. (1961) \enquote{Mach's Principle and a Relativistic
  Theory of Gravitation,} \emph{Phys. Rev.}, \textbf{42}, 925--935.

\bibitem{bcm:94}
Brown, J.~D., Creighton, J. D.~E., and Mann, R.~B. (1994) \enquote{Temperature,
  Energy, and Heat Capacity of Asymptotically Anti-de\thinspace Sitter Black
  Holes,} \emph{Phys. Rev. D}, \textbf{50}, 6394--6403.

\bibitem{bmy:91}
Brown, J.~D., Martinez, E.~A., and York, Jr., J.~W. (1991) \enquote{Complex
  Kerr-Newman Geometry and Black-Hole Thermodynamics,} \emph{Phys. Rev. Lett.},
  \textbf{66}, 2281--2284.

\bibitem{by:89}
Brown, J.~D. and York, Jr., J.~W. (1989) \enquote{Jacobi's Action and the
  Recovery of Time in General Relativity,} \emph{Phys. Rev. D}, \textbf{40},
  3312--3318.

\bibitem{by:93c}
Brown, J.~D. and York, Jr., J.~W. (1993) \enquote{Jacobi's Action and the
  Density of States,} preprint: \texttt{gr-qc/9301018}.

\bibitem{by:93b}
Brown, J.~D. and York, Jr., J.~W. (1993) \enquote{Microcanonical Functional
  Integral for the Gravitational Field,} \emph{Phys. Rev. D}, \textbf{47},
  1420--1431.

\bibitem{by:93a}
Brown, J.~D. and York, Jr., J.~W. (1993) \enquote{Quasilocal Energy and
  Conserved Charges Derived from the Gravitaional Action,} \emph{Phys. Rev. D},
  \textbf{47}, 1407--1419.

\bibitem{bw:90}
Burnett, G.~A. and Wald, R.~M. (1990) \enquote{A Conserved Current for
  Perturbations of Einstein-Maxwell Space-Times,} \textbf{430}, 57--67.

\bibitem{ccm:96}
Chan, K. C.~K., Creighton, J. D.~E., and Mann, R.~B. (1996) \enquote{Conserved
  Masses in {\sc ghs} Einstein and String Black Holes and Consistent
  Thermodynamics,} \emph{Phys. Rev. D}, \textbf{54}, 3892--3899.

\bibitem{cm:95a}
Creighton, J. D.~E. and Mann, R.~B. (1995) \enquote{Quasilocal Thermodynamics
  of Dilaton Gravity Coupled to Gauge Fields,} \emph{Phys. Rev. D},
  \textbf{52}, 4569--4587.

\bibitem{cm:95c}
Creighton, J. D.~E. and Mann, R.~B. (1995) \enquote{Thermodynamics of Dilatonic
  Black Holes in $n$~Dimensions,} preprint: \texttt{gr-qc/9511012}.

\bibitem{fh:65}
Feynman, R.~P. and Hibbs, A.~R. (1965) \emph{Quantum Mechanics and Path
  Integrals}, (New York: McGraw-Hill).

\bibitem{f:95}
Frolov, V.~P. (1995) \enquote{Black Hole Entropy and Physics at Planckian
  Scales,} preprint: \texttt{hep-th/9510156}.

\bibitem{ft:89}
Frolov, V.~P. and Thorne, K.~S. (1989) \enquote{Renormalized Stress-Energy
  Tensor near the Horizon of a Slowly Evolving, Rotating Black Hole,}
  \emph{Phys. Rev. D}, \textbf{39}, 2125--2154.

\bibitem{ghs:91}
Garfinkle, D., Horowitz, G.~T., and Strominger, A. (1991) \enquote{Charged
  Black Holes in String Theory,} \emph{Phys. Rev. D}, \textbf{43}, 3140--3143,
  erratum: \emph{ibid}. \textbf{45}, 3888 (1992).

\bibitem{gh:77}
Gibbons, G.~W. and Hawking, S.~W. (1977) \enquote{Action Integrals and
  Partition Functions in Quantum Gravity,} \emph{Phys. Rev. D}, \textbf{15},
  2752--2756.

\bibitem{gm:88}
Gibbons, G.~W. and Maeda, K. (1988) \enquote{Black Holes and Membranes in
  Higher Dimensional Theories with Dilaton Fields,} \emph{Nucl. Phys. B},
  \textbf{298}, 741--775.

\bibitem{hs:92}
Harvey, J.~A. and Strominger, A. (1992) \enquote{Quantum Aspects of Black
  Holes,} preprint: \texttt{hep-th/9209055}.

\bibitem{h:71}
Hawking, S.~W. (1971) \enquote{Gravitational Radiation from Colliding Black
  Holes,} \emph{Phys. Rev. Lett.}, \textbf{26}, 1344--1346.

\bibitem{h:75}
Hawking, S.~W. (1975) \enquote{Particle Creation by Black Holes,} \emph{Comm.
  Math. Phys.}, \textbf{43}, 199--220.

\bibitem{hh:95}
Hawking, S.~W. and Horowitz, G.~T. (1996) \enquote{The Gravitational
  Hamiltonian, Action, Entropy and Surface Terms,} \emph{Class. Quant. Grav.},
  \textbf{13}, 1487--1498.

\bibitem{h:94}
Hayward, G. (1994) \enquote{Quasilocal Energy Conditions,} \emph{Phys. Rev. D},
  \textbf{52}, 2001--2006.

\bibitem{h:92}
Horowitz, G.~T. (1992) \enquote{The Dark Side of String Theory: Black Holes and
  Black Strings,} in \emph{Proceedings of String Theory and Quantum Gravity
  '92, Trieste, 1992}, pp. 55--99, preprint: \texttt{hep-th/9210119}.

\bibitem{iw:94}
Iyer, V. and Wald, R.~M. (1994) \enquote{Some Properties of Noether Charge and
  a Proposal for Dynamical Black Hole Entropy,} \emph{Phys. Rev. D},
  \textbf{50}, 846--864.

\bibitem{iw:95}
Iyer, V. and Wald, R.~M. (1995) \enquote{A Comparison of Noether Charge and
  Euclidean Methods for Computing the Entropy of Stationary Black Holes,}
  \emph{Phys. Rev. D}, \textbf{52}, 4430--4439.

\bibitem{it:95}
Izquierdo, J.~M. and Townsend, P.~K. (1995) \enquote{Supersymmetric Space-Times
  in (2+1) ADS Supergravity Models,} \emph{Class. Quant. Grav.}, \textbf{12},
  895--924.

\bibitem{jk:93}
Jacobson, T. and Kang, G. (1993) \enquote{Conformal Invariance of Black Hole
  Temperature,} \emph{Class. Quant. Grav.}, \textbf{10}, L201--L206.

\bibitem{l:79}
Lake, K. (1979) \enquote{Riessner-Nordstr{\"o}m-de\thinspace Sitter metric, the
  Third Law, and Cosmic Censorship,} \emph{Phys. Rev. D}, \textbf{19},
  421--429.

\bibitem{lw:96}
Louko, J. and Winters-Hilt, S.~N. (1996) \enquote{Hamiltonian Thermodynamics of
  the Reissner-Nordstr{\"o}m-Anti-de\thinspace Sitter Black Hole,} \emph{Phys.
  Rev. D}, \textbf{54}, 2647--2663.

\bibitem{m:91}
Mann, R.~B. (1991) \enquote{The Simplest Black Holes,} \emph{Found. Phys.
  Lett.}, \textbf{4}, 425--449.

\bibitem{mann:93}
Mann, R.~B. (1993) \enquote{Conservation Laws and 2-D Black Holes in Dilaton
  Gravity,} \emph{Phys. Rev. D}, \textbf{47}, 4438--4442.

\bibitem{mann:94}
Mann, R.~B. (1994) \enquote{Liouville Black Holes,} \emph{Nucl. Phys. B},
  \textbf{418}, 231--256.

\bibitem{mmss:91}
Mann, R.~B., Morsink, S.~M., Sikkema, A.~E., and Steele, T.~G. (1991)
  \enquote{Semiclassical Gravity in (1+1)-Dimensions,} \emph{Phys. Rev. D},
  \textbf{43}, 3948--3957.

\bibitem{m:94}
Martinez, E.~A. (1994) \enquote{Quasilocal Energy for a Kerr Black Hole,}
  \emph{Phys. Rev. D}, \textbf{50}, 4920--4928.

\bibitem{ms:91}
Matzner, R.~A. and Shepley, L.~C. (1991) \emph{Classical Mechanics}, (New
  Jersey: Prentice Hall).

\bibitem{mny:92}
McGuigan, M.~D., Nappi, C.~R., and Yost, S.~A. (1992) \enquote{Charged Black
  Holes in Two-Dimensional String Theory,} \emph{Nucl. Phys. B}, \textbf{375},
  421--452.

\bibitem{mtw:73}
Misner, C.~W., Thorne, K.~S., and Wheeler, J.~A. (1973) \emph{Gravitation},
  (San Francisco: Freeman).

\bibitem{np:92}
Nappi, C.~R. and Pasquinucci, A. (1992) \enquote{Thermodynamics of
  Two-Dimensional Black Holes,} \emph{Mod. Phys. Lett. A}, \textbf{7},
  3337--3346.

\bibitem{rw:96}
R{\'a}cz, I. and Wald, R.~M. (1996) \enquote{Global Extensions of Spacetimes
  Describing Asymptotic Final States of Black Holes,} \emph{Class. Quant.
  Grav.}, \textbf{13}, 539--552.

\bibitem{rt:74}
Regge, T. and Teitelboim, C. (1974) \enquote{Role of Surface Integrals in the
  Hamiltonian Formulation of General Relativity,} \emph{Ann. Phys. (NY)},
  \textbf{88}, 286--318.

\bibitem{sm:91}
Sikkema, A.~E. and Mann, R.~B. (1991) \enquote{Gravitation and Cosmology in
  Two-Dimensions,} \emph{Class. Quant. Grav.}, \textbf{8}, 219--236.

\bibitem{tpmsz:86}
Thorne, K.~S., Price, R.~H., Macdonald, D.~A., Suen, W.-M., and Zhang, X.-H.
  (1986) \enquote{Rapidly Rotating Holes,} in \emph{Black Holes: The Membrane
  Paradigm}, eds. K.~S. Thorne, R.~H. Price, and D.~A. Macdonald, pp. 67--120,
  (New Haven: Yale University Press).

\bibitem{w:74}
Wald, R.~M. (1974) \enquote{Gedanken Experiments to Destroy a Black Hole,}
  \emph{Ann. Phys. (NY)}, \textbf{82}, 548--556.

\bibitem{w:84}
Wald, R.~M. (1984) \emph{General Relativity}, (Chicago: University of Chicago
  Press).

\bibitem{w:93}
Wald, R.~M. (1993) \enquote{Black Hole Entropy is Noether Charge,} \emph{Phys.
  Rev. D}, \textbf{48}, 3427--3431.

\bibitem{w:95}
Wald, R.~M. (1995) \emph{Quantum Field Theory in Curved Space-Time and Black
  Hole Thermodynamics}, (Chicago: University of Chicago Press).

\bibitem{will:93}
Will, C.~M. (1993) \emph{Theory and Experiment in Gravitational Physics},
  (Cambridge: Cambridge University Press), revised edn.

\bibitem{y:86}
York, Jr., J.~W. (1986) \enquote{Black-Hole Thermodynamics and the Euclidean
  Einstein Action,} \emph{Phys. Rev. D}, \textbf{33}, 2092--2099.

\bibitem{z:94}
Zaslavskii, O.~B. (1994) \enquote{Thermodynamics of $2+1$ Black Holes,}
  \emph{Class. Quant. Grav.}, \textbf{11}, L33--L38.

\end{thebibliography}
\end{document}